\documentclass[aps,prb,twocolumn,superscriptaddress,nofootinbib,floatfix]{revtex4-2}

\pdfoutput = 1

\usepackage{amsmath}
\usepackage{amssymb}
\usepackage{amsthm}
\usepackage{bm}
\usepackage{braket}
\usepackage[svgnames,psnames]{xcolor}
\usepackage{graphicx}
\usepackage[caption=false]{subfig}
\usepackage{adjustbox}
\usepackage[colorlinks, linktocpage, hypertexnames = false]{hyperref}

\usepackage[status=draft,inline,nomargin]{fixme}
\fxusetheme{color}
\FXRegisterAuthor{ki}{aki}{KI}

\hypersetup{linkcolor = Crimson, citecolor = MediumBlue, urlcolor = MediumBlue}

\definecolor{red}{rgb}{1,0,0}
\definecolor{blue}{rgb}{0,0,1}
\definecolor{dblue}{rgb}{0,0,0.4}
\definecolor{green}{rgb}{0,1,0}
\definecolor{dgreen}{rgb}{0,0.4,0}
\definecolor{black}{rgb}{0,0,0}
\definecolor{white}{rgb}{1,1,1}

\definecolor{brn}{rgb}{.8,.4,.0}
\definecolor{redo}{rgb}{1,.5,.0}
\definecolor{ddgrn}{rgb}{0,0.4,0}
\definecolor{dgrn}{rgb}{0,0.55,0}
\definecolor{dbl}{rgb}{0,0,0.5}
\definecolor{grey}{rgb}{0.5,0.5,0.5}

\usepackage[bbgreekl]{mathbbol}

\newcommand{\Z}{\mathbb{Z}}

\renewcommand{\t}[1]{\widetilde{#1}}

\newcommand{\Rf}[1]{Ref.~\onlinecite{#1}}

\newcommand{\bpm}{\begin{pmatrix}}
\newcommand{\epm}{\end{pmatrix}}
\newcommand{\bmm}{\begin{matrix}}
\newcommand{\emm}{\end{matrix}}

\newcommand{\cC}{ {\cal C} }

\newcommand{\cR}{ {\cal R} }

\newcommand{\cV}{ {\cal V} }

\usepackage{euscript}

\newcommand\eM          {\EuScript{M}}

\newcommand\eZ         {\EuScript{Z}}

\newcommand{\veps}{\varepsilon}

\begin{document}

\flushbottom

\title{2+1D symmetry-topological-order from local symmetric operators in 1+1D}

\author{Kansei Inamura}
\affiliation{Institute for Solid State Physics, University of Tokyo, Kashiwa, Chiba 277-8581, Japan}

\author{Xiao-Gang Wen}
\affiliation{Department of Physics, Massachusetts Institute of Technology, Cambridge, Massachusetts 02139, USA}

\date{\today}

\begin{abstract}

A generalized symmetry (defined by the algebra of local symmetric operators)
can go beyond group or higher group description.  A theory of generalized
symmetry (up to holo-equivalence) was developed in terms of symmetry-TO -- a
bosonic topological order (TO) with gappable boundary in one higher dimension.
We propose a general method to compute the 2+1D symmetry-TO from the local
symmetric operators in 1+1D systems. 
Our theory is based on the commutant patch operators, which are extended
operators constructed as products and sums of local symmetric operators.  A
commutant patch operator commutes with all local symmetric operators away from
its boundary.  We argue that topological invariants associated with anyon
diagrams in 2+1D can be computed as contracted products of commutant patch
operators in 1+1D.  In particular, we give concrete formulae for several
topological invariants in terms of commutant patch operators.  Topological
invariants computed from patch operators include those beyond modular data,
such as the link invariants associated with the Borromean rings and the
Whitehead link.  These results suggest that the algebra of commutant patch
operators is described by 2+1D symmetry-TO.  Based on our analysis, we also
argue briefly that the commutant patch operators would serve as order
parameters for gapped phases with finite symmetries.

\end{abstract}

\maketitle

\tableofcontents

\section{Introduction}
\label{sec: Introduction}

Topological order \cite{Wen:1989iv} is a collection of low energy universal
properties of gapped liquid \cite{ZW1490,SM1403} phases of matter, which are
captured by topological quantum field theories \cite{Atiyah1988,Witten:1988hf},
or by non-degenerate braided fusion higher categories
\cite{Kong:2014qka}.  These phases are physically characterized by the algebraic
properties of various topological excitations.  For example, topological orders
in 2+1 dimensions are classified by the fusion and braiding of topological
point-like excitations known as anyons \cite{LM7701,W8257,H8483,ASW8422,W8413},
which are generally described by modular tensor categories
\cite{MS8977,Witten:1988hf} (see, e.g., \cite{Wen:2015qgn, Kitaev:2005hzj} for
a review).  Categorical descriptions and classifications of topological orders
in higher dimensions are also developed recently in \cite{Kong:2014qka,
LW170404221, LW180108530, Johnson-Freyd:2020usu, KZ201102859}.

After the systematic understanding of gapped liquid phases of matter, we like
to gain a systematic understanding of gapless liquid phases of matter with
linear dispersion, which is a long-standing challenge in theoretical physics.
One way to make progress \cite{Chatterjee:2022jll} is to study a key invariant of gapless liquid phases
-- the low energy emergent symmetry.  The  emergent symmetry can be a
generalized symmetry, which is a combination of ordinary group-like symmetries,
anomalous symmetries \cite{H8035}, higher-form symmetries
\cite{NOc0605316,NOc0702377,GKSW2015}, higher-group symmetries
\cite{Kapustin:2013uxa, Barkeshli:2014cna, Cordova:2018cvg, Benini:2018reh},
and more general non-invertible 0-symmetries \cite{Verlinde:1988sn,
Petkova:2000ip, Fuchs:2002cm, Frohlich:2004ef, Frohlich:2006ch, Fuchs:2007tx,
Frohlich:2009gb, DKR2011, Carqueville:2012dk, Brunner:2013ota, Brunner:2014lua,
BT2018, CLSWY2019, Thorngren:2019iar} and non-invertible higher symmetries
\cite{KZ200308898,Kong:2020cie,Freed:2022qnc}.\footnote{Non-invertible symmetries
in higher dimensions have been studied intensively over the past few years following the seminal work of \cite{Choi:2021kmx, Kaidi:2021xfk}, see, e.g.,
\cite{McGreevy:2022oyu, Cordova:2022ruw, Schafer-Nameki:2023jdn,
Brennan:2023mmt, Shao:2023gho} for reviews.} In fact,  we can use non-invertible
gravitational anomalies
\cite{KW1458,FV14095723,M14107442,KZ150201690,KZ170200673,JW190513279} to
describe all the above different emergent generalized symmetries in a unified
way \cite{Ji:2019jhk,KZ200308898,Kong:2020cie,Freed:2022qnc}.  Since 
gravitational anomalies correspond to topological orders in one higher
dimension \cite{KW1458,KZ150201690,KZ170200673}, this leads to
symmetry/topological-order (Symm/TO) correspondence
\cite{Ji:2019jhk,KZ200308898,Kong:2020cie,Freed:2022qnc}.

\begin{figure}[t]
\begin{center}
\includegraphics[scale=0.6]{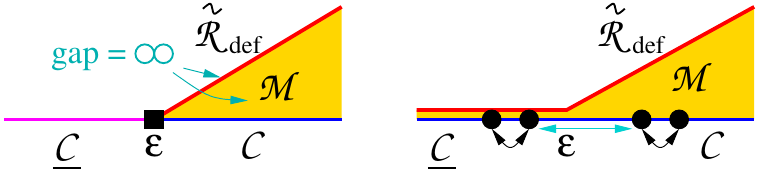}
\end{center}
\caption{To define a homomorphism between quantum field theories that preserves
properties of excitations, \Rf{KZ150201690} introduced the isomorphism $\veps$
(a low energy equivalence that preserves, for example, low energy partition
functions \cite{CW221214432}) between two (gapped or gapless) quantum field theories,
$\underline{\cC}$ and $\cC\boxtimes_{\eM} \t\cR $, where the bulk topological
order $\eM$ and the gapped boundary $\t\cR$ have infinite energy gaps.  The
equivalence $\veps$ exposes the emergent symmetry described by the fusion
higher category $\t\cR$ and/or the symmetry-TO $\eM$ in quantum field theory
$\underline{\cC}$.  } \label{HoloDecom} 
\end{figure}

The relation between topological orders and generalized symmetries becomes
manifest in the isomorphic holographic decomposition described by
Fig.~\ref{HoloDecom}
\cite{KZ150201690,Ji:2019jhk,KZ200308898,Kong:2020cie,Freed:2022qnc,
Freed:2022iao}.  The composite system $\cC\boxtimes_{\eM} \t\cR $ (a slab of
bulk topological order $\eM$ with a gapped boundary $\t\cR$ and a low energy
boundary $\cC$) exactly simulates the low energy system $\underline{\cC}$, for
example $\underline{\cC}$ and $\cC\boxtimes_{\eM} \t\cR $ have the same
partition function, in the limit where the gaps of $\eM$ and $\t\cR$ approach
infinity. If a low energy theory $\underline{\cC}$ has such an isomorphic
holographic decomposition, then the $\underline{\cC}$ has an emergent symmetry
described by $\t\cR$ and $\eM=\eZ(\t\cR)$ (where $\eZ()$ is the generalized
Drinfeld center).  If $\t\cR$ is a local fusion higher category, then the
symmetry (which is usually non-invertible) is an anomaly-free algebraic higher
symmetry \cite{KZ200308898,Kong:2020cie} (or a fusion category symmetry for
1+1D case \cite{Thorngren:2019iar}). If $\t\cR$ is not local, then the symmetry
is anomalous.

If we only look at the symmetry within the symmetric sub-Hilbert space
$\cV_\text{symmetric}$, then some different symmetries, such as $\Z_2\times
\Z_2$ symmetry with mixed anomaly and $\Z_4$ symmetry in 1+1D
\cite{Chatterjee:2022kxb,ZL220601222}, become equivalent, which is called
holo-equivalent \cite{Kong:2020cie}.  
It was proposed that a holo-equivalence
class of symmetries is described fully by symmetry-TO $\eM$ in one higher
dimension \cite{Ji:2019jhk,Kong:2020cie}.\footnote{Mathematically, two fusion
higher categories, $\t\cR$ and $\t\cR'$, describe holo-equivalent symmetries if
they are Morita equivalent $ \eZ(\t\cR) =\eZ(\t\cR')$.  The symmetry-TO was
called ``categorical symmetry'' in \cite{Ji:2019jhk, Kong:2020cie}.} Thus,
symmetry-TO (and its mathematical description -- braided fusion higher category
in the trivial Witt class), replacing group and higher group, becomes a theory for
generalized symmetry (up to holo-equivalence). Such a holographic picture was
discussed for 1+1D in, e.g., \cite{JW190513279, Thorngren:2019iar, Ji:2019jhk,
Lichtman:2020nuw, Gaiotto2021, Aasen:2020jwb, Chatterjee:2022kxb, Moradi:2022lqp, LOST220805495, Lin2023uvm, ZC230401262}. 
The holographic picture was also used to study dualities \cite{Freed:2018cec}. See
Sec.~\ref{sec: Motivation} for more details of the holographic picture of symmetries.  The symmetry-TO $\eM$ is also
called a symmetry topological field theory (symmetry-TFT)
\cite{Apruzzi:2021nmk}. The name ``symmetry-TO'' stresses the existence of
lattice UV completion and the absence of any lattice symmetry.

We know that a (generalized) symmetry is defined by an algebra of local
symmetric operators without involving anything in one higher dimension.
\Rf{Ji:2019jhk, Chatterjee:2022kxb, lan2023quantum} try to calculate the braided
fusion higher category from the local symmetric operators directly.  Usually,
one introduces commutant operators that commute with all the local symmetric
operators, and then uses the algebra of commutant operators to describe and define 
symmetry.  But here, to calculate braided fusion higher category,
\Rf{Chatterjee:2022kxb, lan2023quantum} introduced commutant patch operators. 
A commutant patch operator is formed by local symmetric operators glued together on a patch of any dimension.\footnote{In the examples discussed in \cite{Ji:2019jhk, Chatterjee:2022kxb}, a commutant patch operator reduces to the product of local symmetric operators on a patch.}
It commutes with all the local symmetric operators
far away from the boundary of the patch.
Moreover, it is a symmetric operator by itself because it consists of local symmetric operators.
\Rf{Chatterjee:2022kxb,lan2023quantum} conjecture that the algebra of commutant
patch operators encodes a braided fusion higher category, which in turn
describes a symmetry-TO if the braided fusion higher category is finite.
A similar idea is also presented in
\cite{Lootens:2021tet, Lootens:2022avn}, where it was shown that certain 1+1D
lattice models with Morita equivalent symmetries have isomorphic algebras of
symmetric local operators.  A related problem is also studied in the context of
algebraic quantum field theory \cite{Haag96} under the name of the DHR theory
\cite{Doplicher:1969tk, Doplicher:1969kp, Doplicher:1971wk, Doplicher:1973at},
see, e.g., \cite{SzV1993, Nill:1995nk, jones2023dhr} for the reconstruction of
2+1D topological orders from 1+1D quantum spin chains in this context.

The conjecture in \cite{Chatterjee:2022kxb} implies that the data of
topological orders are encoded in the algebra of commutant patch operators in
one lower dimension.  Indeed, in \cite{Ji:2019jhk, Chatterjee:2022kxb}, the
anyon data of the 2+1D toric code and double-semion topological orders were
explicitly computed from commutant patch operators in 1+1D systems with
non-anomalous and anomalous $\mathbb{Z}_2$ symmetries.  A similar computation
of the anyon data of the toric code was also performed in \cite{Kong:2021equ}
based on topological excitations in 1+1D gapped phases with non-anomalous
$\mathbb{Z}_2$ symmetry.  This computation was later generalized to the case of
an arbitrary non-anomalous abelian group symmetry in \cite{Xu:2022rtj}.  The
above results motivate us to expect that it would be possible to reconstruct
more general topological orders from commutant patch operators in one lower
dimension.  However, thus far, topological orders that have been explicitly
reconstructed from patch operators are limited to several abelian topological
orders mentioned above.

In this paper, we generalize the analysis in \cite{Ji:2019jhk,
Chatterjee:2022kxb} so that we can reconstruct more general 2+1D topological
orders from 1+1D systems with finite symmetries that are generally described
by fusion categories \cite{BT2018, CLSWY2019, Thorngren:2019iar}. In
particular, based on the holographic picture of 1+1D systems with finite
symmetries, we will argue that the anyon data of 2+1D topological orders should
be encoded in commutant patch operators in 1+1 dimensions. 
The commutant patch operators will also be called symmetric transparent connectable patch operators in the subsequent sections due to the properties of these operators.\footnote{The name ``transparent patch operator" was coined in \cite{Chatterjee:2022kxb} to emphasize that a commutant patch operator is transparent to local symmetric operators away from its boundary.}
We will then propose a general method to compute
the anyon data of 2+1D topological orders from these patch operators.  As an
example, we will write down symmetric transparent connectable patch operators
in 1+1D systems with a general non-anomalous finite group symmetry $G$ and
verify our proposal by explicitly computing the anyon data of the corresponding
topological orders in 2+1D.  The anyon data that we will compute include
topological invariants beyond modular data, such as those associated with the
Borromean rings and the Whitehead link.

The rest of the paper is organized as follows.  In Sec.~\ref{sec: Review of
topological orders in 2+1 dimensions}, we will give a brief review of
topological orders in 2+1 dimensions.  In Sec.~\ref{sec: Topological orders
in 2+1D from patch operators in 1+1D}, we will propose a general scheme to
compute the anyon data of 2+1D topological orders by using symmetric
transparent connectable patch operators in 1+1 dimensions. In Sec.~\ref{sec: Reconstruction of Kitaev's quantum double topological order}, we will
apply our computational scheme to the topological order realized by Kitaev's
quantum double model for a general finite group.  More specifically, we will
demonstrate that various topological invariants associated with anyon diagrams
of Kitaev's quantum double model can be computed from patch operators in 1+1D
systems with finite group symmetry.  
In Sec.~\ref{sec: Discussion}, we will summarize the results and argue that symmetric transparent connectable patch operators would serve as order and disorder operators for gapped phases with finite symmetries.
In App.~\ref{sec: Ribbon operators on
the boundary of Kitaev's quantum double model}, we will discuss a relation
between patch operators in 1+1D systems with a finite group symmetry and ribbon
operators on the rough boundary of Kitaev's quantum double model.  Throughout
the paper, we will only consider bosonic systems.

\section{Review of topological orders in 2+1 dimensions}
\label{sec: Review of topological orders in 2+1 dimensions}
In this section, we briefly review the algebraic description of 2+1D topological orders.

\subsection{General case}
It is widely believed that topological orders in 2+1 dimensions (up to invertible ones) are mathematically described by non-degenerate braided fusion categories, which axiomatize the fusion and braiding statistics of anyons.
The basic data of a non-degenerate braided fusion category consist of the following ingredients (see, e.g., \cite{Moore:1988qv, Kitaev:2005hzj, EGNO2015} for more details):
\begin{itemize}
\item A finite set of anyon types $\{1, a, b, c, \cdots \}$.
This set is equipped with an involution $a \mapsto \overline{a}$, where $\overline{a}$ is called the dual of $a$. The distinguished element $1$ corresponds to a trivial anyon.

\item Fusion rules $a \otimes b = \bigoplus_c N_{ab}^c c$, where $N_{ab}^c$ is a non-negative integer called a fusion coefficient. The trivial anyon $1$ behaves as a unit under the fusion, i.e., $1 \otimes a = a \otimes 1 = a$.

\item Finite dimensional vector spaces $\mathop{\mathrm{Hom}}(a \otimes b, c)$ and $\mathop{\mathrm{Hom}}(c, a \otimes b)$ called a fusion space and a splitting space, whose dimensions are equal to the fusion coefficient $N_{ab}^c$. Elements of these vector spaces are called morphisms.

\item $F$-symbols $(F^{abc}_d)_{(e; \mu, \nu), (f; \rho, \sigma)} \in \mathbb{C}$ that describe the crossing relations of worldlines of anyons:
\begin{equation}
\adjincludegraphics[valign = c, width = 1.8cm]{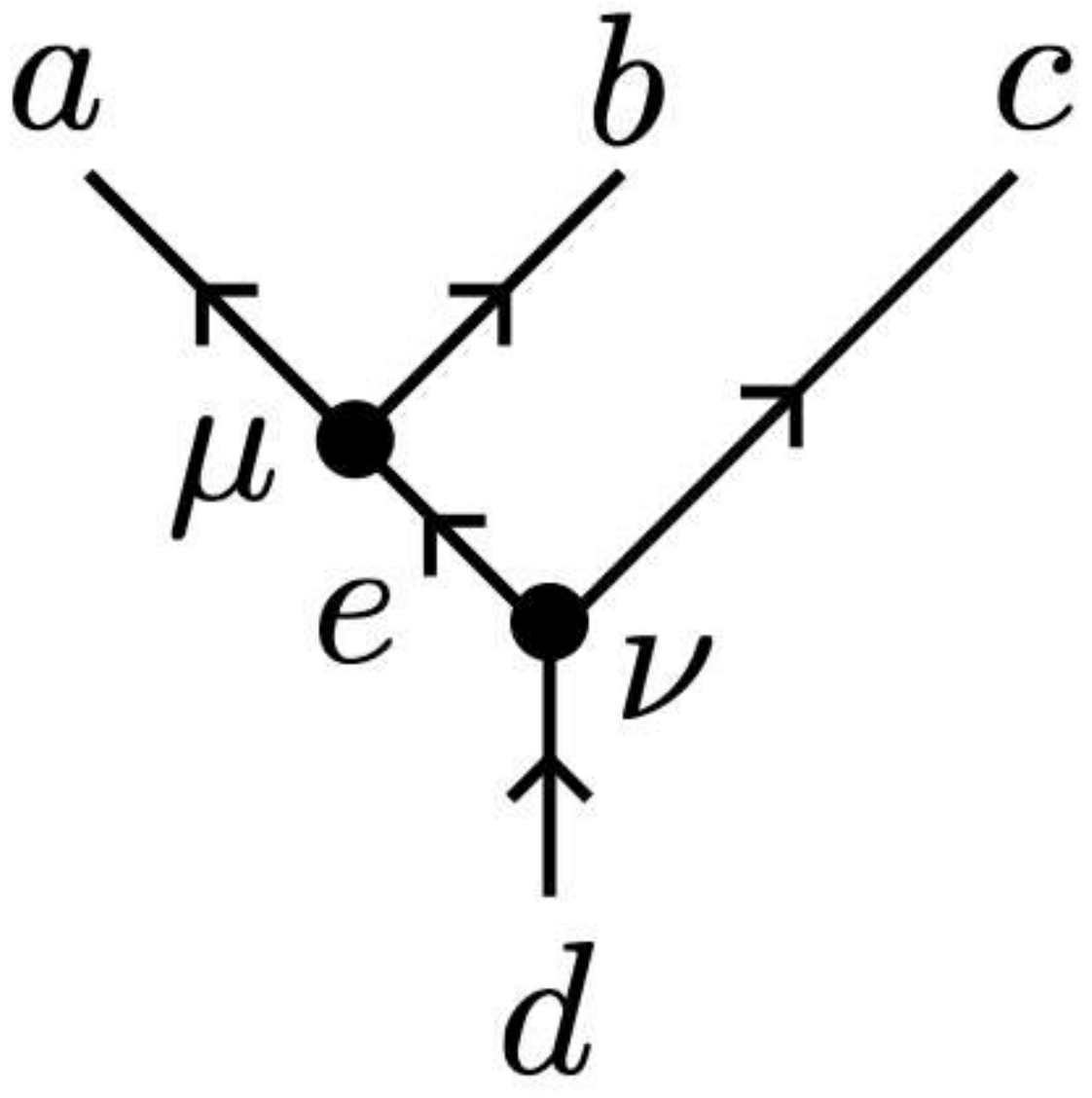} = \sum_{f, \rho, \sigma} (F^{abc}_d)_{(e; \mu, \nu), (f; \rho, \sigma)} \adjincludegraphics[valign = c, width = 1.8cm]{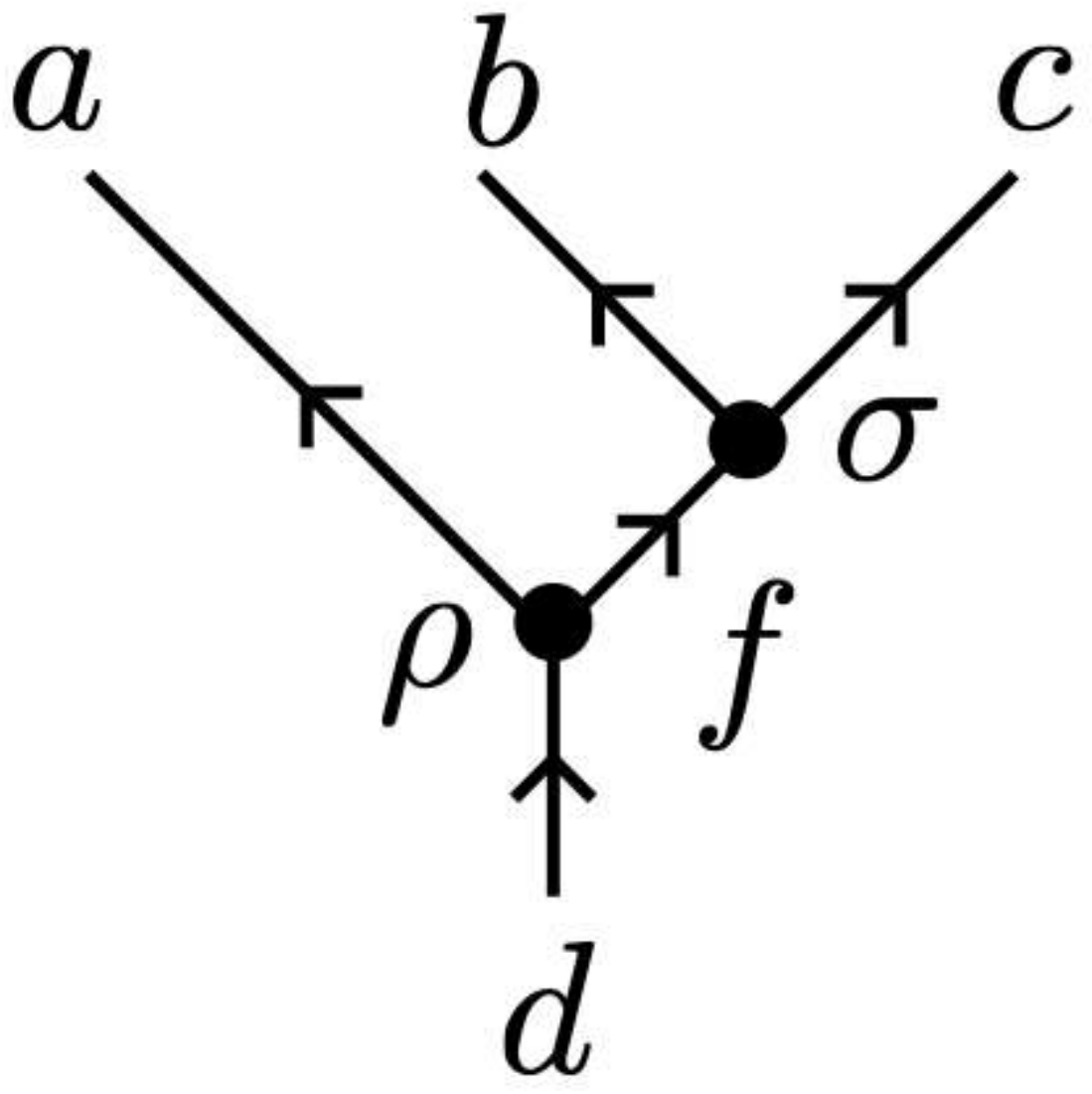}.
\label{eq: F}
\end{equation}
The summation on the right-hand side is taken over fusion channels $f \subset b \otimes c$ and basis morphisms $\rho \in \mathop{\mathrm{Hom}}(d, a \otimes f)$ and $\sigma \in \mathop{\mathrm{Hom}}(f, b \otimes c)$.
The $F$-symbols must satisfy consistency conditions known as the pentagon equation \cite{Moore:1988qv}.
We note that the $F$-symbols are not gauge invariant, i.e., they depend on the choice of bases of the splitting spaces.

\item $R$-symbols $(R^{ab}_c)_{\mu, \nu} \in \mathbb{C}$ that describe the braiding of anyon lines:
\begin{equation}
\adjincludegraphics[valign = c, width = 1cm]{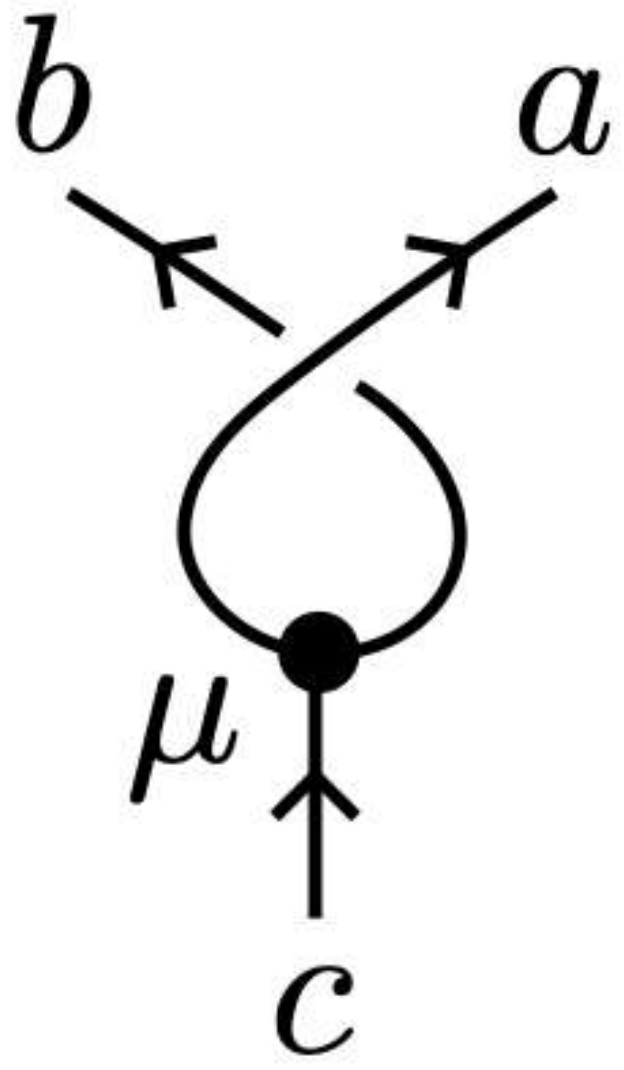} = \sum_{\nu} (R^{ab}_c)_{\mu, \nu} \adjincludegraphics[valign = c, width = 1cm]{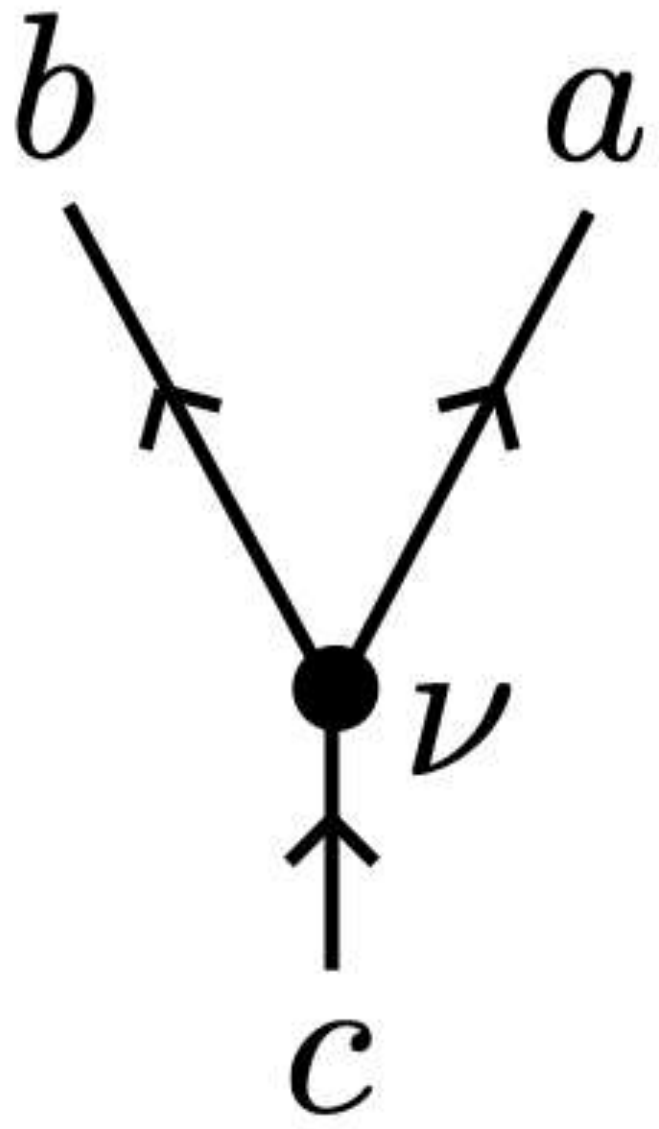}.
\label{eq: R}
\end{equation}
The summation on the right-hand side is taken over basis morphisms $\nu \in \mathop{\mathrm{Hom}}(c, b \otimes a)$.
The $R$-symbols must satisfy consistency conditions known as the hexagon equations \cite{Moore:1988qv}.
We note that the $R$-symbols are also gauge dependent.

\item A left evaluation morphism $\epsilon^L_a \in \mathop{\mathrm{Hom}}(\overline{a} \otimes a, 1)$ and a left coevaluation morphism $\eta^L_a \in \mathop{\mathrm{Hom}}(1, a \otimes \overline{a})$ that describe the annihilation and creation of a pair of anyons:
\begin{equation}
\epsilon^L_a = \adjincludegraphics[valign = c, width = 1.2cm]{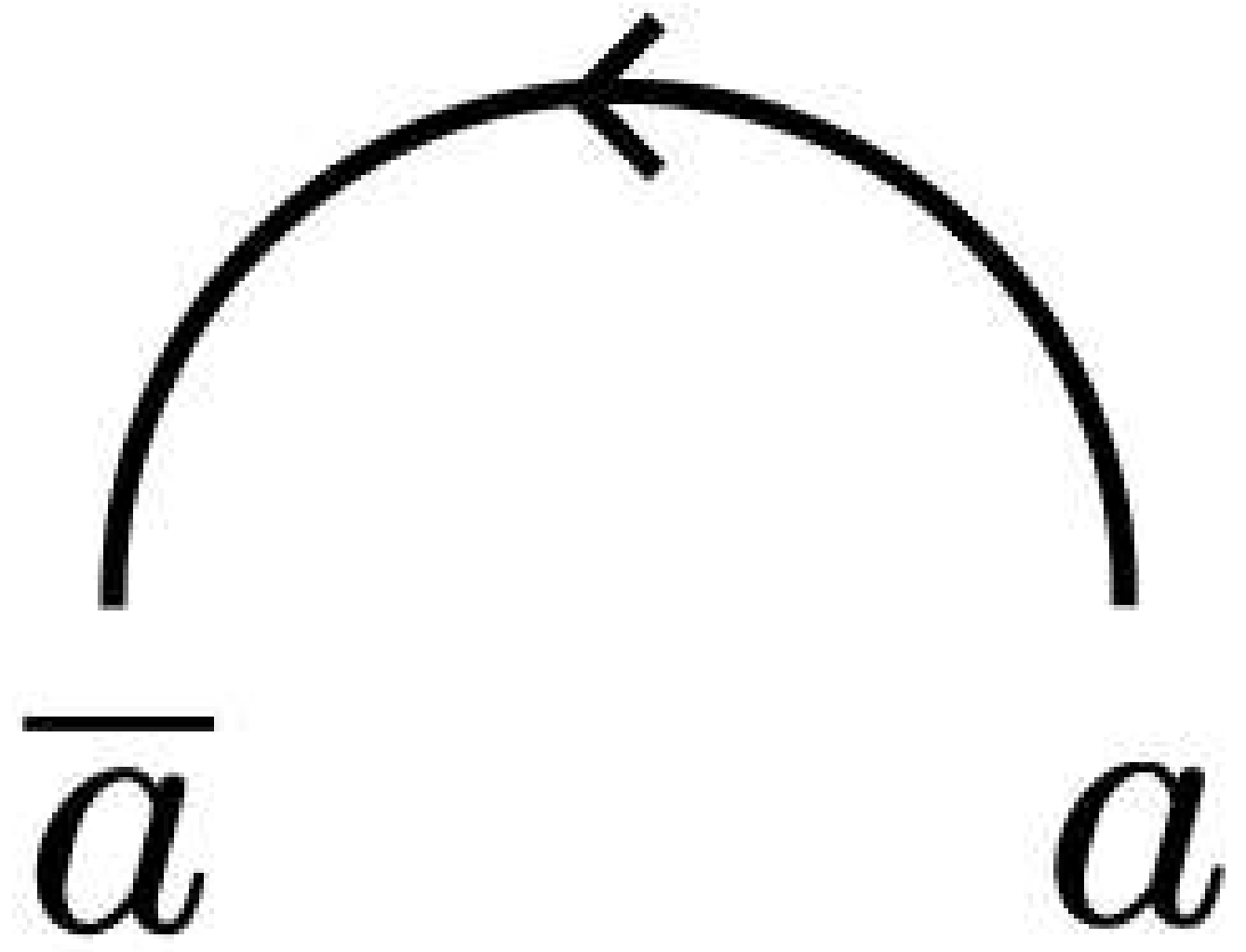}, \quad \eta^L_a = \adjincludegraphics[valign = c, width = 1.2cm]{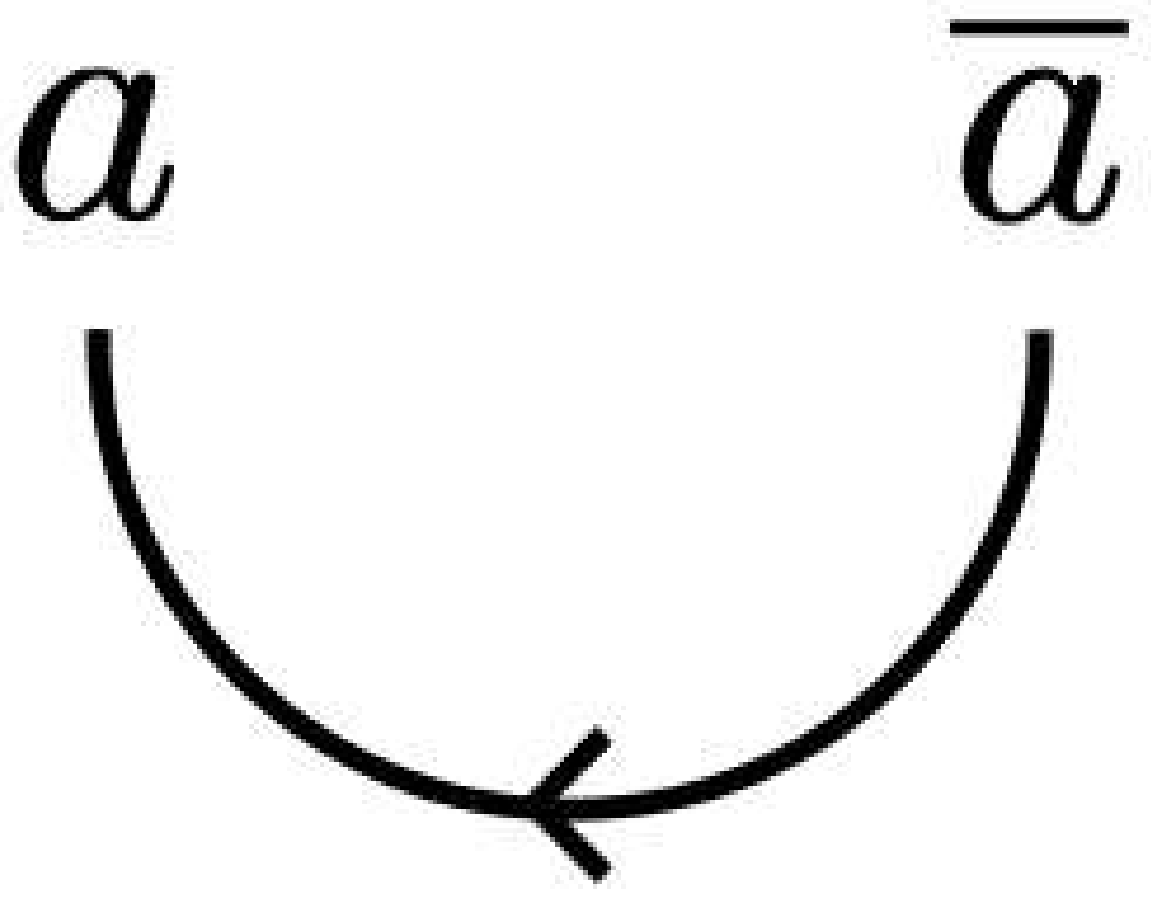},
\end{equation}
where the worldline of $\overline{a}$ is represented by the orientation reversal of that of $a$.
The evaluation and coevaluation morphisms must satisfy the following zigzag identities:
\begin{equation}
\adjincludegraphics[valign = c, width = 1.7cm]{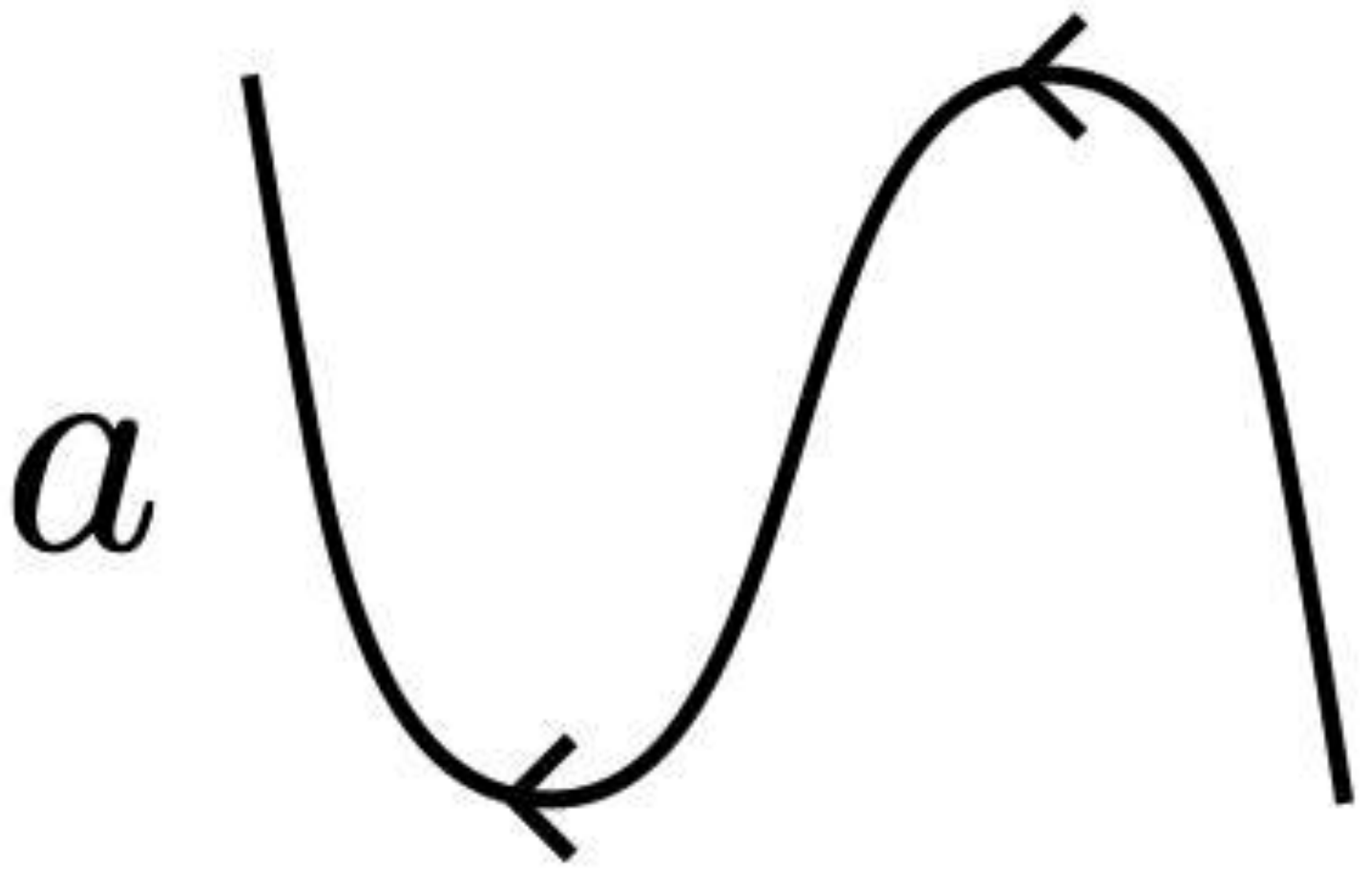} = \adjincludegraphics[valign = c, width = 0.5cm]{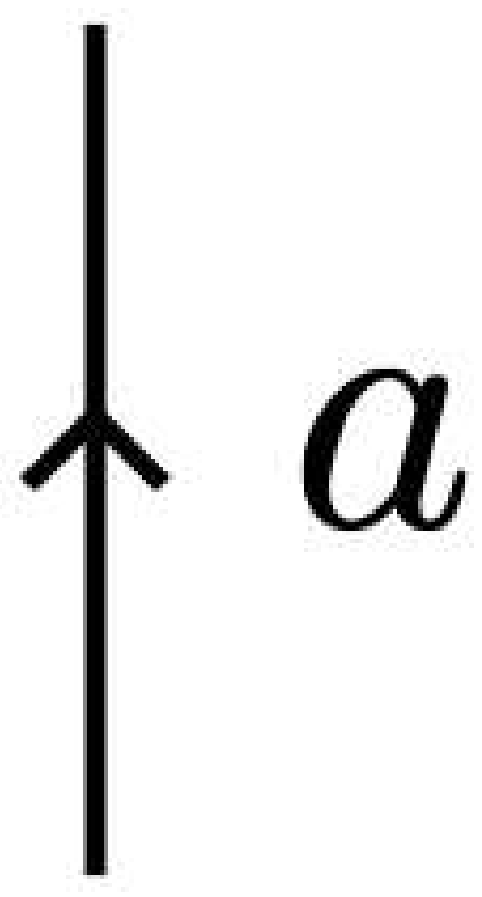}, \quad \adjincludegraphics[valign = c, width = 1.7cm]{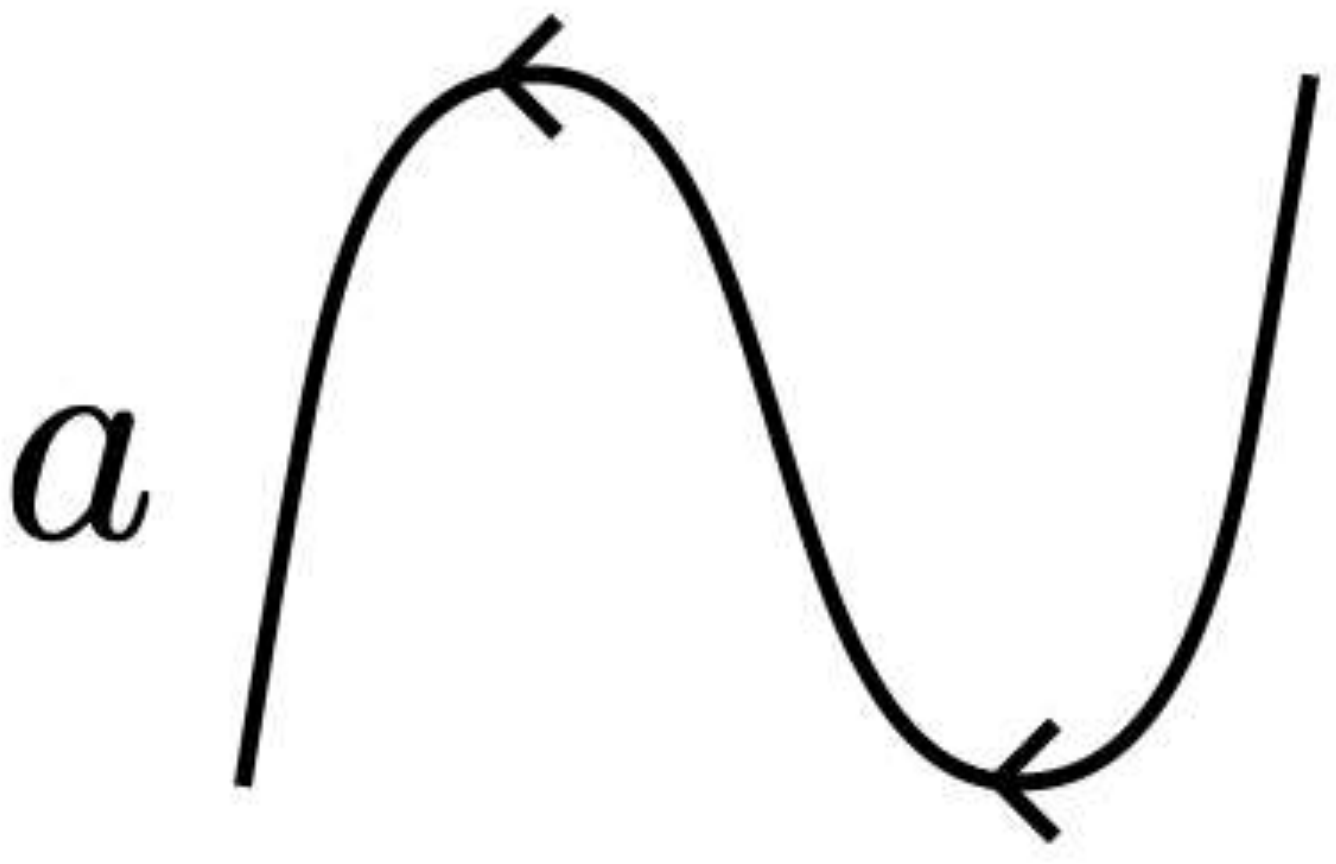} = \adjincludegraphics[valign = c, width = 0.5cm]{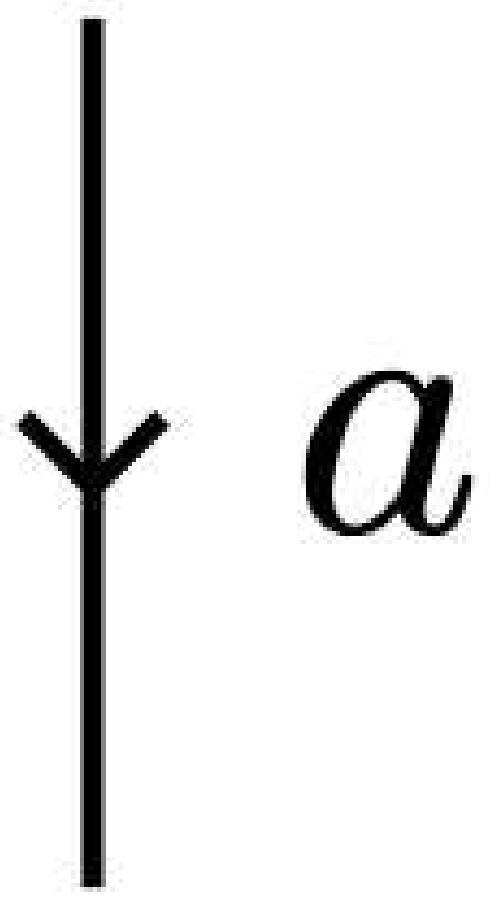}.
\label{eq: zigzag}
\end{equation}

\item A right evaluation morphism $\epsilon^R_a \in \mathop{\mathrm{Hom}}(a \otimes \overline{a}, 1)$ and a right coevaluation morphism $\eta^R_a \in \mathop{\mathrm{Hom}}(1, \overline{a} \otimes a)$ represented by the diagrams
\begin{equation}
\epsilon^R_a = \adjincludegraphics[valign = c, width = 1.2cm]{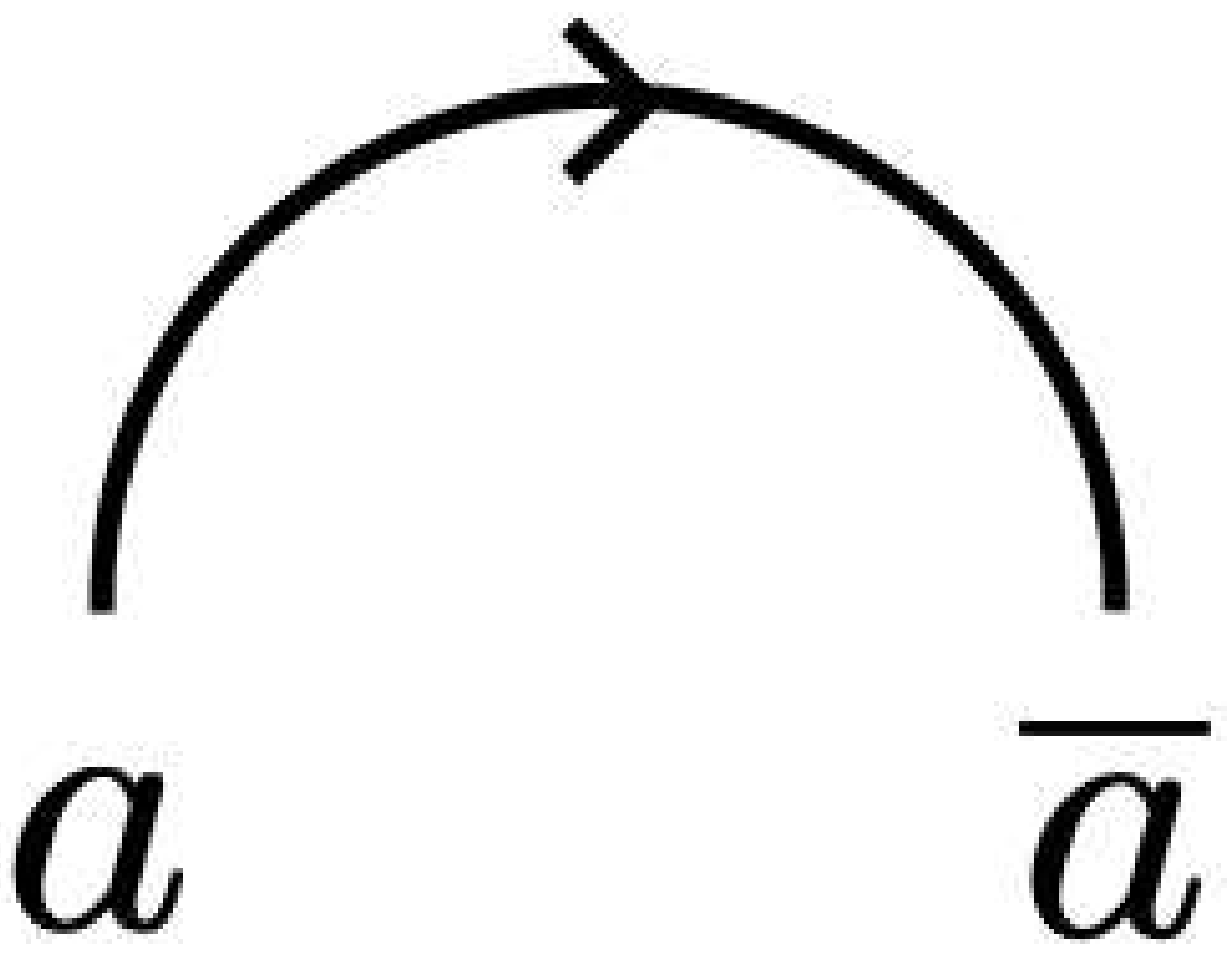}, \quad \eta^R_a = \adjincludegraphics[valign = c, width = 1.2cm]{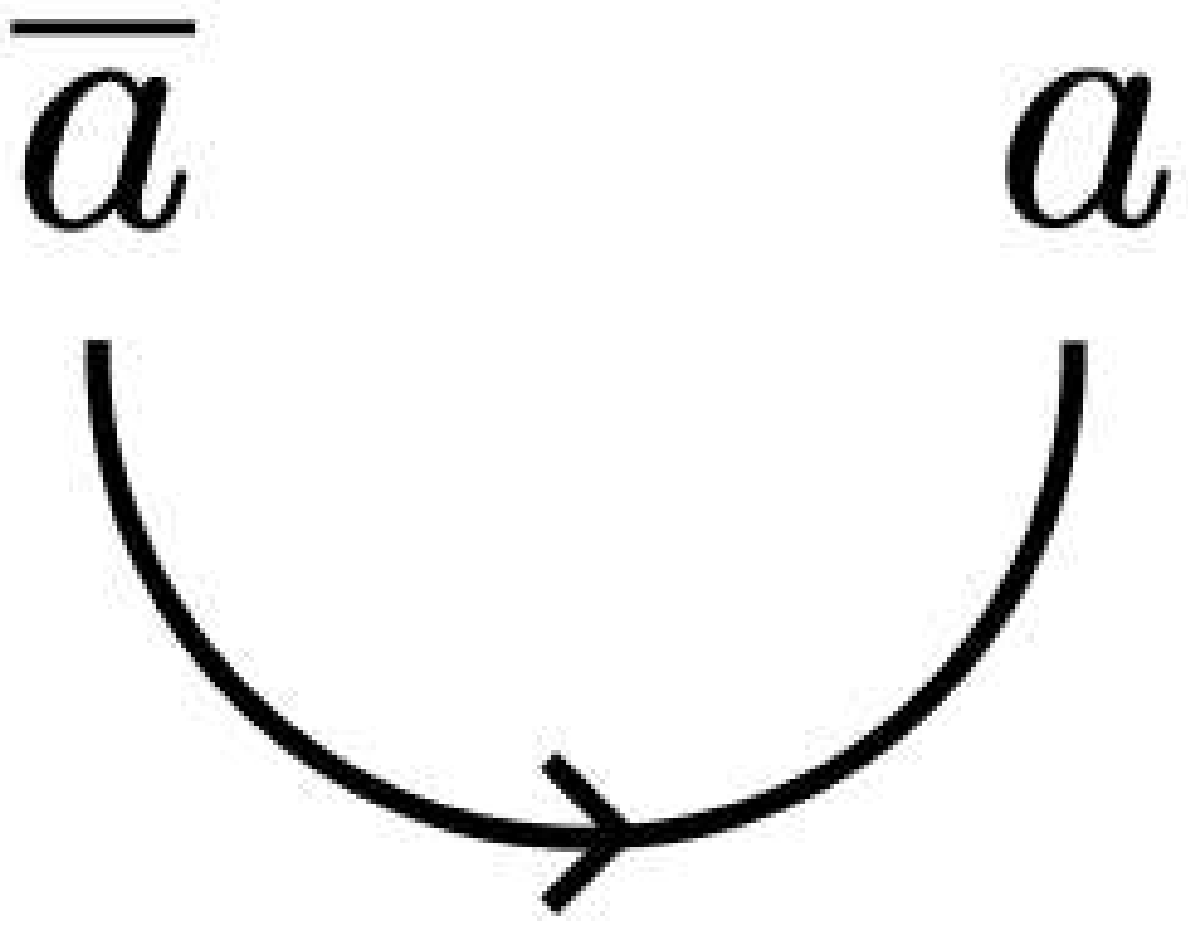},
\end{equation}
which satisfy the zigzag identities analogous to Eq.~\eqref{eq: zigzag}.
In a unitary (braided) fusion category, the left and right evaluation/coevaluation morphisms are related by the Hermitian conjugation.
\end{itemize}

Given a non-degenerate braided fusion category, we can associate a complex number with any closed diagram consisting of anyon lines.
This complex number can be expressed in terms of the $F$-symbols, $R$-symbols, and evaluation/coevaluation morphisms.
In particular, when anyon lines form a framed knot or link, the associated complex number is gauge invariant.
We will call such a gauge invariant quantity simply a topological invariant.
Of particular importance among topological invariants are the following quantities: 
\begin{itemize}
\item The quantum dimension $d_a$ of each anyon $a$. This is a topological invariant associated with a loop of an anyon line:
\begin{equation}
d_a = \adjincludegraphics[valign = c, width = 1.2cm]{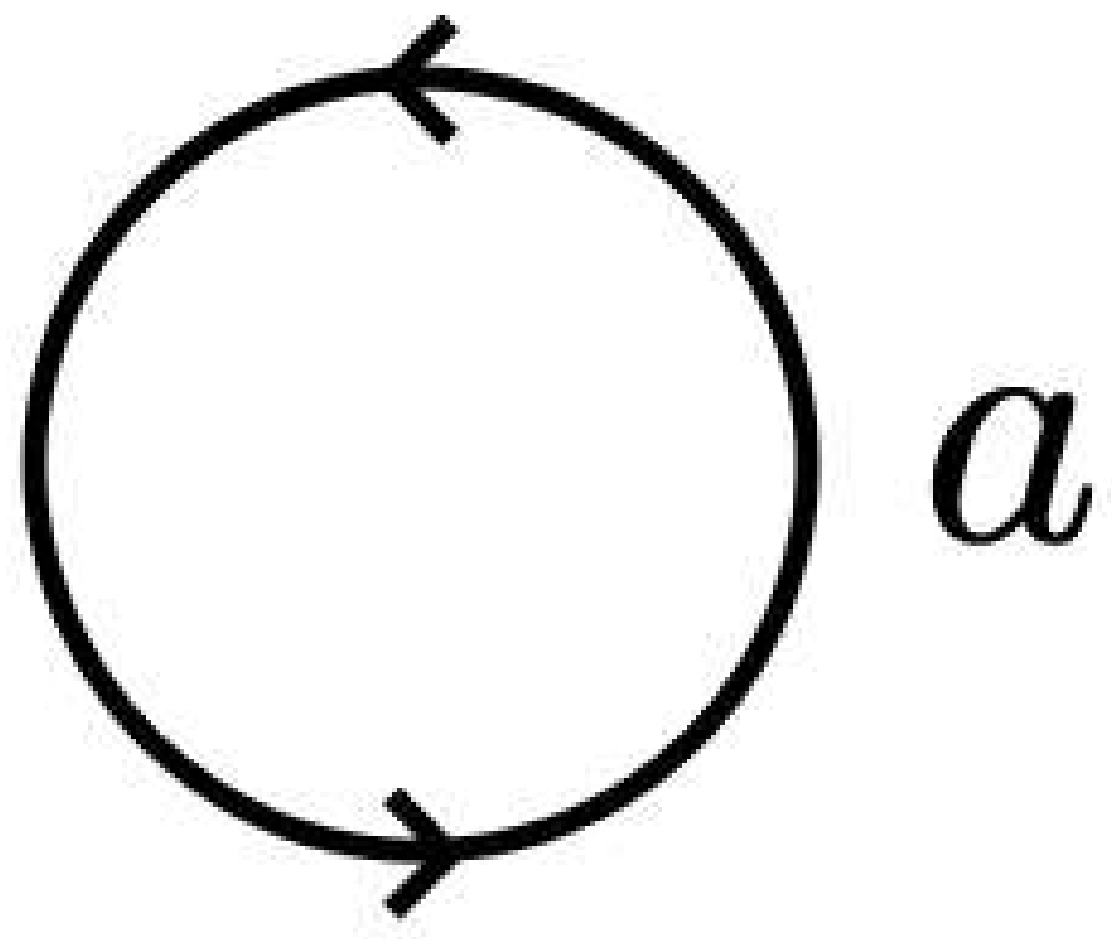} = \adjincludegraphics[valign = c, width = 1.2cm]{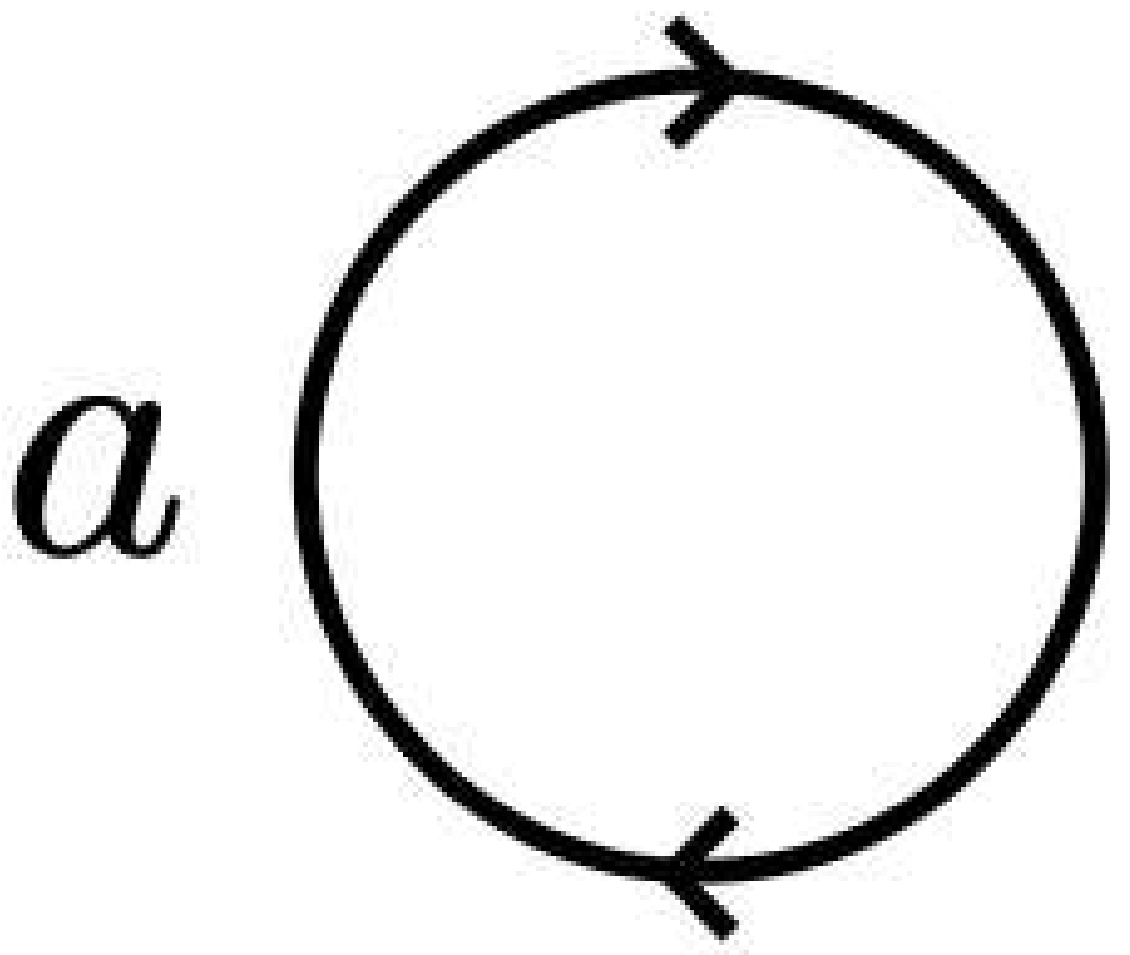}.
\end{equation}
The quantum dimension in a unitary (braided) fusion category is a positive real number that is greater than or equal to one.
When $d_a = 1$, $a$ is called an abelian anyon. 
Otherwise, it is called a non-abelian anyon.
\item The topological spin $\theta_a \in \mathrm{U}(1)$ of each anyon $a$.
This is a topological invariant associated with an anyon line forming a figure of eight:
\begin{equation}
\theta_a = \frac{1}{\mathop{\mathrm{dim}} a} ~ \adjincludegraphics[valign = c, width = 1.7cm]{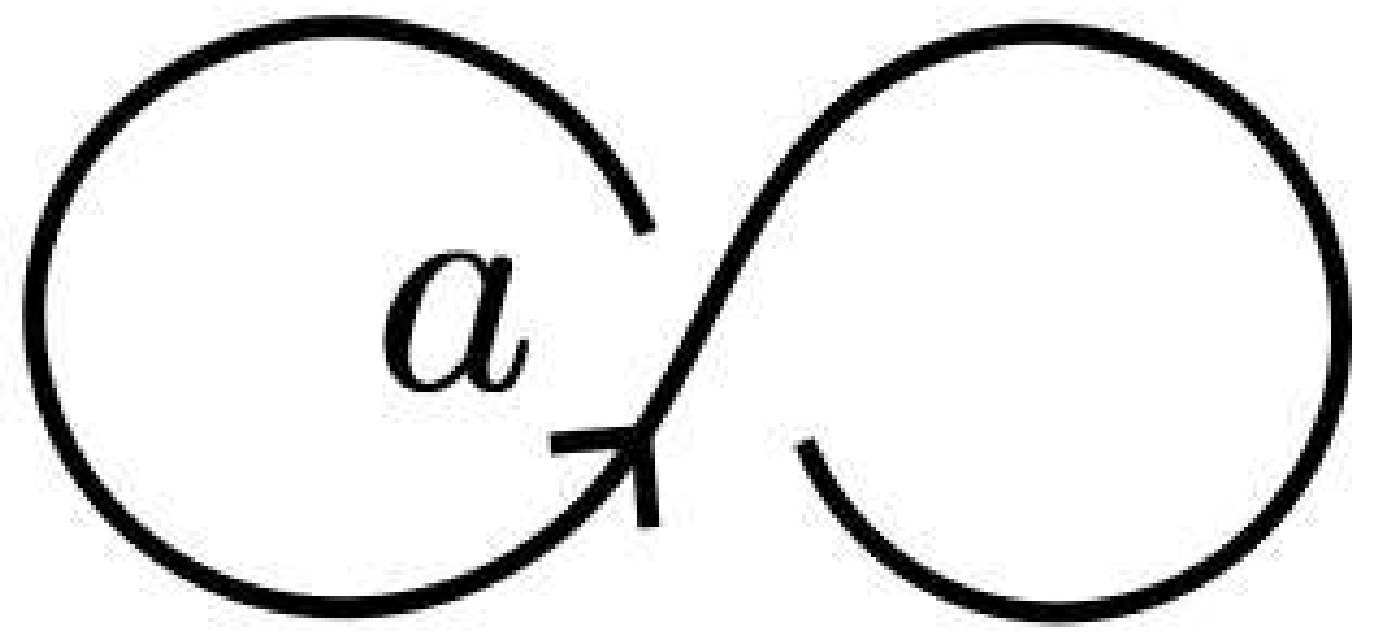}.
\end{equation}
\item The $(a, b)$-component $S_{ab} \in \mathbb{C}$ of the modular $S$-matrix for each pair of anyons $a$ and $b$.
This is a topological invariant associated with the Hopf link:
\begin{equation}
S_{ab} = \frac{1}{\sqrt{\mathcal{D}}} ~ \adjincludegraphics[valign = c, width = 1.8cm]{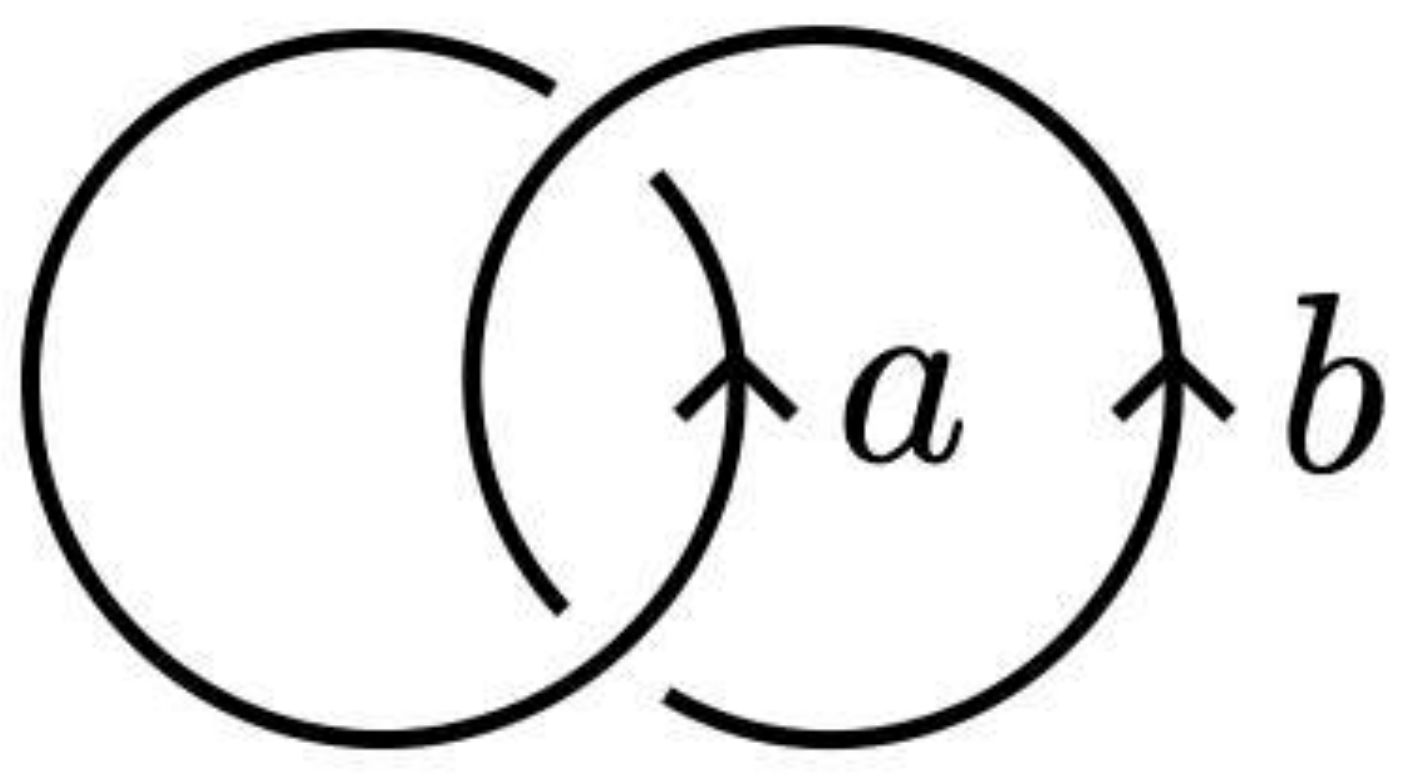}.
\end{equation}
Here, $\mathcal{D} := \sum_a d_a^2$ is the total dimension.
The modular $S$-matrix is non-degenerate in a non-degenerate braided fusion category.
\end{itemize}
The topological spins and modular $S$-matrix are called modular data.

Although the modular data largely characterize a non-degenerate braided fusion category, they are not complete invariants.
Namely, different non-degenerate braided fusion categories can have the same modular data \cite{mignard2021modular}. 
Several topological invariants have been proposed to distinguish topological orders that share the same modular data \cite{Bonderson:2018ryx, Delaney:2018xkw, kulkarni2021topological, Wen:2019ylt, Delaney:2021owx}.
We may expect that the set of all topological invariants associated with framed knots and links of anyon lines uniquely determines a non-degenerate braided fusion category.
However, a simple set of topological invariants that completely characterize a non-degenerate braided fusion category has not been worked out yet.

A 2+1D topological order is said to be non-chiral if it admits a topological (i.e., gapped) boundary condition, while it is chiral otherwise.
Mathematically, a 2+1D topological order is non-chiral if and only if the non-degenerate braided fusion category describing anyons is equivalent to the Drinfeld center of a fusion category \cite{Fuchs:2012dt, Freed:2020qfy}.
In general, 2+1D non-chiral topological orders are realized by the Levin-Wen model \cite{Levin:2004mi}, whose low energy limit is described by the Turaev-Viro-Barrett-Westbury topological field theory \cite{Turaev:1992hq, Barrett:1993ab}.
In the rest of this paper, we will only consider non-chiral topological orders due to their intimate relation to finite symmetries in 1+1D.

\subsection{Example: Kitaev's quantum double topological order}
\label{sec: Example: Kitaev's quantum double topological order}
A simple example of a 2+1D topological order is realized by Kitaev's quantum double model \cite{Kitaev:1997wr}, which is a Hamiltonian formulation of a topological finite gauge theory known as the (untwisted) Dijkgraaf-Witten theory \cite{Dijkgraaf:1989pz}.\footnote{In general, we can twist Kitaev's quantum double model by a third group cohomology $\omega \in H^3(G, \mathrm{U}(1))$ \cite{Hu:2012wx}. The topological order of this model is described by the topological gauge theory with the Dijkgraaf-Witten twist $\omega$ \cite{Dijkgraaf:1989pz}. We will not consider such a twisted version of the quantum double model in this paper.}
We will denote Kitaev's quantum double model based on a finite group $G$ as $\mathrm{QD}(G)$.
In this subsection, we summarize the anyon data of Kitaev's quantum double model $\mathrm{QD}(G)$, or equivalently, a topological $G$-gauge theory (see, e.g., \cite{deWildPropitius:1995cf} for a review).

The anyons of the quantum double model $\mathrm{QD}(G)$ are labeled by pairs $([g], \alpha)$, where $[g] := \{hgh^{-1} \mid h \in G\}$ is the conjugacy class of $g \in G$ and $\alpha$ is a unitary irreducible representation of the centralizer $C(g) := \{h \in G \mid hg = gh\}$ of $g$.
\begin{itemize}
\item The quantum dimension of an anyon labeled by $([g], \alpha)$ is given by
\begin{equation}
d_{[g], \alpha} = |[g]| \mathop{\mathrm{dim}}\alpha,
\label{eq: qdim QD}
\end{equation}
where $|[g]|$ is the number of elements in $[g]$ and $\mathop{\mathrm{dim}} \alpha$ is the dimension of the representation $\alpha$.

\item The topological spin of an anyon $([g], \alpha)$ is given by
\begin{equation}
\theta_{[g], \alpha} = \frac{\mathop{\mathrm{tr}}\alpha(g)}{\mathop{\mathrm{dim}}\alpha},
\label{eq: topological spin QD}
\end{equation}
where, by a slight abuse of notation, $\alpha(g)$ denotes the representation matrix of $g$ and the trace is taken over the representation space of $\alpha$.

\item The $\left(([g], \alpha), ([h], \beta)\right)$-component of the modular $S$-matrix is given by
\begin{equation}
\begin{aligned}
& \quad S_{([g], \alpha), ([h], \beta)} \\
& = \frac{1}{|G|} \sum_{\substack{a \in [g], ~ b \in [h] \\ \text{s.t. } ab = ba}} \mathop{\mathrm{tr}} \overline{\alpha} \left( (x_a^g)^{-1} b x_a^g \right) \mathop{\mathrm{tr}} \overline{\beta} \left( (x_b^h)^{-1} a x_b^h \right),
\end{aligned}
\label{eq: modular S QD}
\end{equation}
where $\overline{\alpha}$ and $\overline{\beta}$ are the complex conjugate representations of $\alpha$ and $\beta$ respectively.
The group element $x_a^g \in G$ is a representative of a coset in $G/C(g)$, i.e., an arbitrary element that satisfies $a = x_a^g g (x_a^g)^{-1}$ for $a \in [g]$.
\end{itemize}
The complete anyon data of Kitaev's quantum double model $\mathrm{QD}(G)$ is described by the representation category of the quantum double of group algebra $\mathbb{C}[G]$ \cite{Kitaev:1997wr}.
This category is equivalent to the Drinfeld center of the category of $G$-graded vector spaces as a braided fusion category, see, e.g., \cite{EGNO2015}.

When $G$ is abelian, anyons of Kitaev's quantum double model are labeled by pairs $(g, \alpha)$ of a group element $g \in G$ and a unitary irreducible representation $\alpha$ of $G$ because the conjugacy class of $g$ consists only of $g$ and the centralizer of $g \in G$ is the whole group $G$.
Since any irreducible representation $\alpha$ of a finite abelian group is one-dimensional, Eq.~\eqref{eq: qdim QD} implies that all anyons of Kitaev's quantum double model $\mathrm{QD}(G)$ for an abelian group $G$ have quantum dimension one, i.e., they are abelian anyons.
Furthermore, the topological spins and modular $S$-matrix in this case reduce to $\theta_{g, \alpha} = \alpha(g)$ and $S_{(g, \alpha), (h, \beta)} = \overline{\alpha}(h) \overline{\beta}(g)/|G|$ respectively.

\section{Topological orders in 2+1D from patch operators in 1+1D}
\label{sec: Topological orders in 2+1D from patch operators in 1+1D}
A patch operator in 1+1 dimensions is an extended operator that acts only on a finite interval (i.e., a patch) of a one-dimensional space. 
In this section, we propose a general method to compute topological invariants of 2+1D non-chiral topological orders by using patch operators in 1+1D systems with finite symmetries.
The basic idea of computing topological invariants from patch operators was already presented in \cite{Ji:2019jhk, Chatterjee:2022kxb}, where it was demonstrated that the modular data of several abelian topological orders, such as the toric code and double-semion topological orders, can be computed from patch operators in one lower dimension.
Here, we will slightly extend the formulation in \cite{Ji:2019jhk, Chatterjee:2022kxb} so that we can also deal with non-abelian topological orders.
In what follows, topological orders are assumed to be non-chiral and have an infinitely large energy gap unless otherwise stated.

\subsection{Motivation}
\label{sec: Motivation}
\begin{figure*}
\includegraphics[width = 12cm]{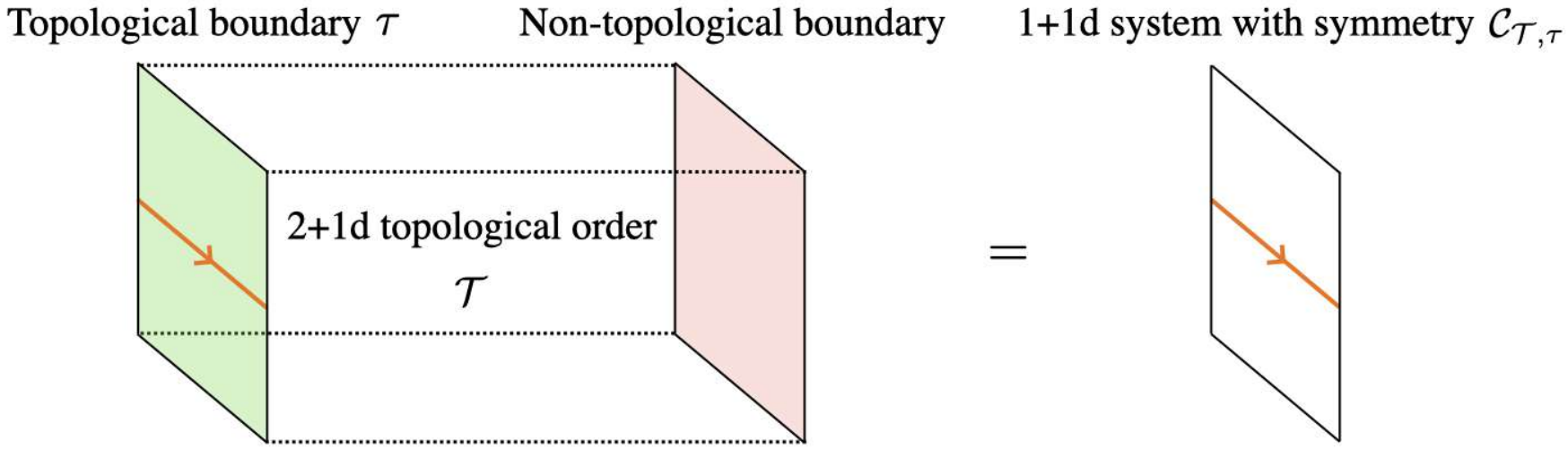}
\caption{A 2+1D topological order on a slab gives rise to a 1+1D system with finite symmetry. The symmetry action in 1+1D is implemented by inserting a topological line on the topological boundary of the slab. In the figure, time is supposed to go up.} 
\label{fig: SymTFT}
\end{figure*}

We first provide motivation behind our expectation that the anyon data of a 2+1D topological order should be encoded in patch operators in 1+1D systems with finite symmetry.
To this end, we first recall the holographic picture that we mentioned in Sec.~\ref{sec: Introduction}.
As illustrated in Fig.~\ref{fig: SymTFT}, a 1+1D system with finite symmetry can be obtained by putting a 2+1D topological order on a slab $[0, 1] \times \Sigma$, where $[0, 1]$ is a finite interval and $\Sigma$ is a two-dimensional oriented surface \cite{Freed:2018cec, Thorngren:2019iar, Lichtman:2020nuw, Gaiotto2021, Aasen:2020jwb, Freed:2022iao, Freed:2022qnc}.
On the left boundary of the slab, we impose a topological boundary condition, while on the right boundary, we impose a non-topological physical boundary condition.
The symmetry of the 1+1D system obtained in this way is described by a fusion category formed by topological lines on the left boundary $\{0\} \times \Sigma$.
We denote this fusion category by $\mathcal{C}_{\eM, \tau}$, where $\eM$ is the 2+1D topological order in the bulk and $\tau$ is the topological boundary condition on the left.
To be more precise, the 1+1D system has symmetry $\mathcal{C}_{\eM, \tau}$ if we view the system from the left side of the 2+1D bulk.\footnote{If we look at the 1+1D system from the other side of the bulk, the symmetry is described by the opposite category $\mathcal{C}_{\eM, \tau}^{\mathrm{op}}$.}
The fusion category $\mathcal{C}_{\eM, \tau}$ and the bulk topological order $\eM$ are related by the boundary-bulk relation, i.e., the bulk topological order $\eM$ is described by the Drinfeld center of $\mathcal{C}_{\eM, \tau}$ \cite{Kapustin:2010hk, Kitaev:2011dxc, Fuchs:2012dt, Lan:2013wia}.\footnote{A similar boundary-bulk relation also holds in higher dimensions \cite{kong2015boundarybulk, KONG201762}.}

\begin{figure*}
\includegraphics[width = 9cm]{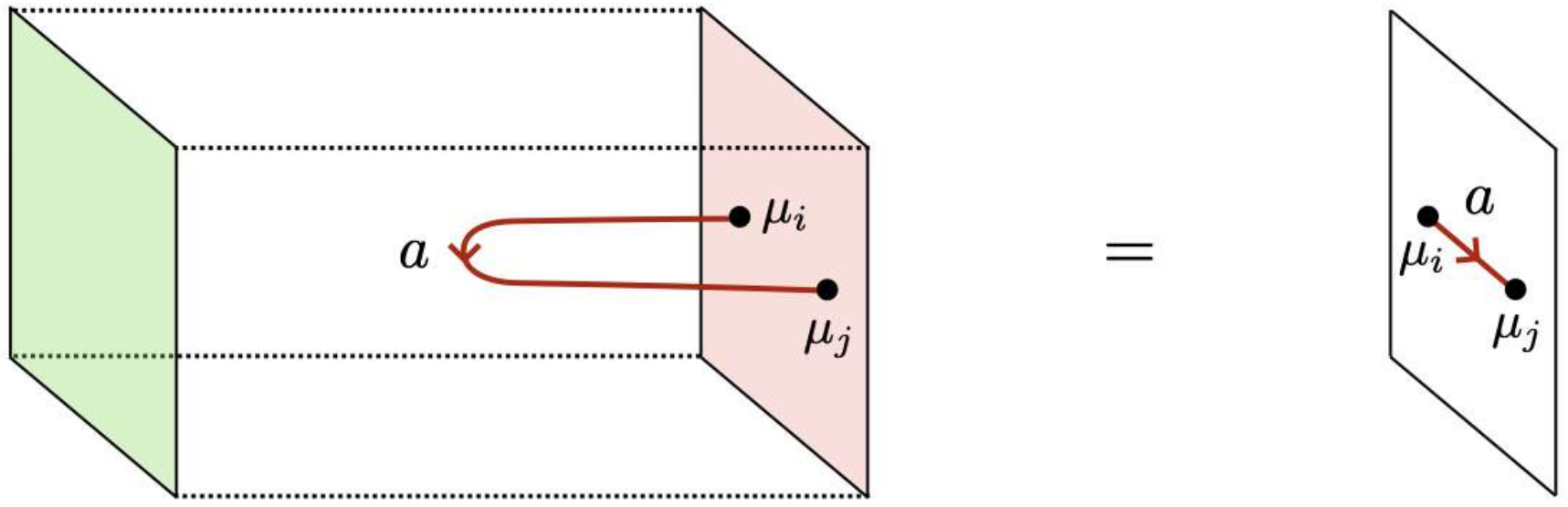}
\caption{The relation between anyons in a 2+1D topological order and patch operators in 1+1D.}
\label{fig: SymTFT patch}
\end{figure*}

The relation between anyons in the bulk topological order and patch operators in 1+1D is now clear: a bulk anyon line terminating on the right boundary gives rise to a patch operator of the 1+1D system, see Fig.~\ref{fig: SymTFT patch}.
A similar construction of symmetric operators in 1+1D can be found in \cite{Freed:2018cec, Moradi:2022lqp, Albert:2021vts}.
We note that the endpoints of an anyon line on the right boundary are not necessarily topological.
The patch operator obtained in this way should be labeled by an anyon $a$ of the bulk topological order.
In addition, the endpoints of the patch operator would carry extra indices corresponding to the internal degrees of freedom of an anyon.
For example, as we will see in Sec.~\ref{sec: Reconstruction of Kitaev's quantum double topological order}, the extra indices at the endpoints are associated with the charge of an anyon when the bulk topological order is Kitaev's quantum double model $\mathrm{QD}(G)$ (i.e., a $G$-gauge theory) and the symmetry in 1+1D is described by a finite group $G$.
Extra indices at the endpoints of an anyon line are also observed in more general non-chiral topological orders \cite{Levin:2004mi}.
Thus, the patch operator obtained as in Fig.~\ref{fig: SymTFT patch} should be denoted by $(P_a^{ij})_{\mu_i, \mu_j}$, where $[ij]$ is a finite interval on which the patch operator acts, $a$ is the label of an anyon in the bulk topological order, and $\mu_i$ and $\mu_j$ are extra indices at the two ends $i$ and $j$ respectively.
For later convenience, we express a patch operator $(P_a^{ij})_{\mu_i, \mu_j}$ diagrammatically as follows:
\begin{equation}
(P_a^{ij})_{\mu_i, \mu_j} = ~ \adjincludegraphics[valign = c, width = 2.2cm]{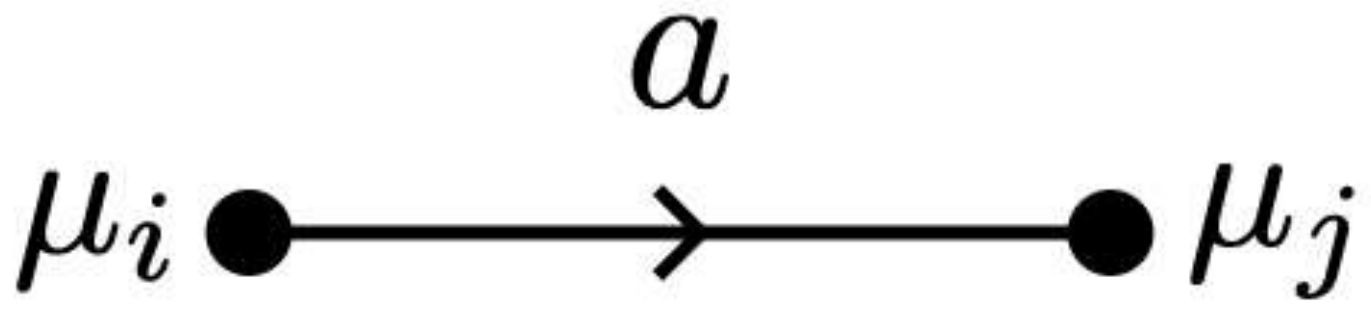}.
\end{equation}
The above arguments suggest that the anyon data of a 2+1D topological order should be encoded in patch operators in 1+1D systems with finite symmetry.

\subsection{Symmetric transparent connectable patch operators}
Although we argued that anyons in 2+1D topological orders give rise to patch operators in 1+1D, the converse is not true in general: there are many patch operators that have nothing to do with anyons in the bulk.
Therefore, in order to reconstruct the anyon data of 2+1D topological orders from 1+1D systems with finite symmetry, we need to identify patch operators that originate from anyons in the bulk.
To this end, in this subsection, we study some properties of the patch operators obtained as in Fig.~\ref{fig: SymTFT patch} and derive necessary conditions for patch operators to be related to anyons.
We will see that the sandwich construction shown in Fig.~\ref{fig: SymTFT patch} naturally leads us to the notion of symmetric transparent patch operators that were introduced in \cite{Ji:2019jhk, Chatterjee:2022kxb}.

\begin{figure*}
\includegraphics[width = 10cm]{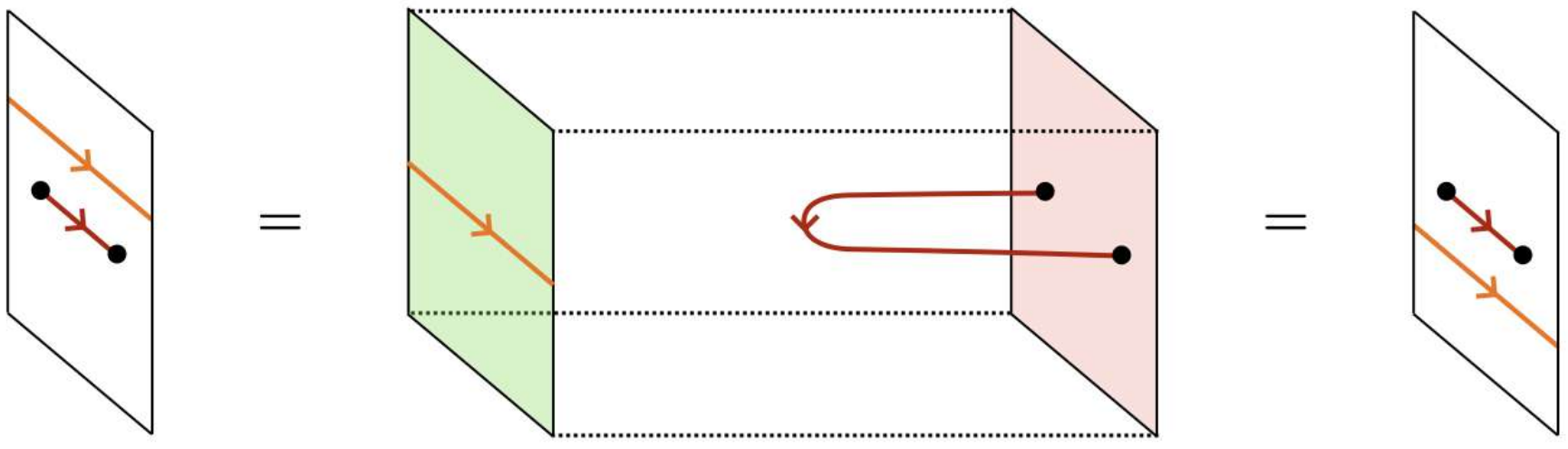}
\caption{The patch operators commute with the symmetry operators.}
\label{fig: symmetricity}
\end{figure*}

By construction, the patch operator $(P_a^{ij})_{\mu_i, \mu_j}$ should satisfy the following properties:
\paragraph{Symmetricity.} 
The patch operator $(P_a^{ij})_{\mu_i, \mu_j}$ is symmetric, i.e., it commutes with the action of fusion category symmetry $\mathcal{C}$ in 1+1D.
This is an immediate consequence of the fact that the symmetry action in 1+1D is implemented by topological lines on the left (i.e., topological) boundary of the 2+1D bulk.
Indeed, this definition of the symmetry action guarantees that symmetry operators in 1+1D can freely go through the patch operators obtained as anyon lines attached to the right boundary, see Fig.~\ref{fig: symmetricity}.
Thus, the symmetry operators and patch operators commute with each other.

\begin{figure*}
\includegraphics[width = 11cm]{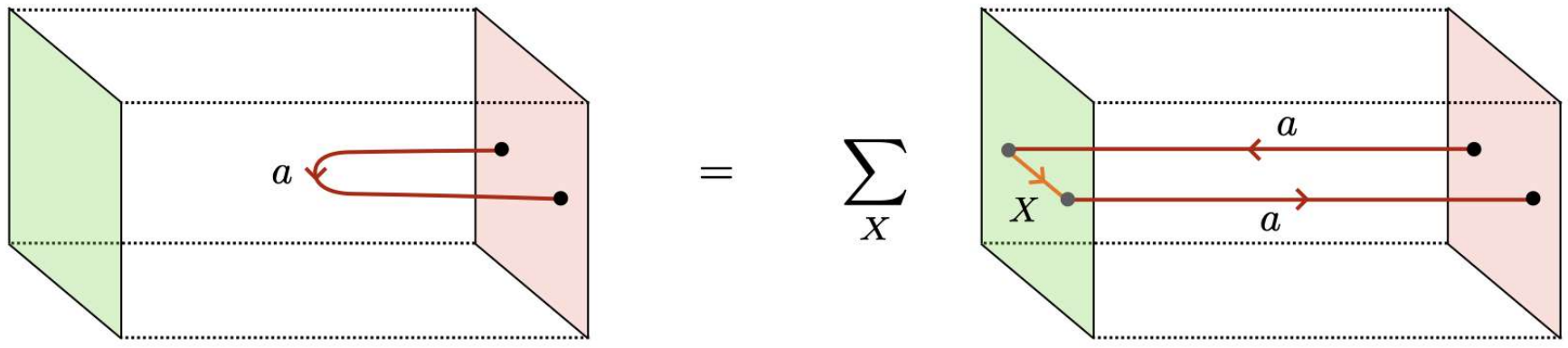}
\caption{A bulk anyon line becomes a sum of topological lines on the topological boundary.}
\label{fig: decomposition}
\end{figure*}

\paragraph{Transparency.}
The patch operator $(P_a^{ij})_{\mu_i, \mu_j}$ is transparent with respect to symmetric operators, i.e., it commutes with any symmetric operators that act non-trivially only in the middle of interval $[ij]$.
More specifically, the patch operator $(P_a^{ij})_{\mu_i, \mu_j}$ can be written as a sum
\begin{equation}
(P_a^{ij})_{\mu_i, \mu_j} = \sum_{X \in \mathcal{C}} (P_a^{ij})^X_{\mu_i, \mu_j},
\label{eq: decomposition}
\end{equation}
where $(P_a^{ij})^X_{\mu_i, \mu_j}$ on the right-hand side is a (generically non-symmetric) patch operator whose action in the middle of interval $[ij]$ is indistinguishable from the symmetry action of an object $X \in \mathcal{C}$. 
The above equation follows from the fact that a bulk anyon line is decomposed into a sum of topological lines when pushed onto a topological boundary as shown in Fig.~\ref{fig: decomposition}.
When $X \in \mathcal{C}$ is not contained in the decomposition of $a$, we define $(P_a^{ij})^X_{\mu_i, \mu_j} = 0$.
On the other hand, when $X \in \mathcal{C}$ is contained in the decomposition of $a$ multiple times, we have a summation over the multiplicity of $X$ on the right-hand side, which is implicit in the above equation.
We note that Eq.~\eqref{eq: decomposition} implies that the patch operator $(P_a^{ij})_{\mu_i, \mu_j}$ commutes with any symmetric operators whose supports are contained in the middle of interval $[ij]$ because the summand $(P_a^{ij})^X_{\mu_i, \mu_j}$ does commute with such operators by definition.

When the anyon $a$ is condensed on the topological boundary,\footnote{An anyon is said to be condensed if its worldline can end topologically on the topological boundary, see \cite{Kong:2013aya} for a general theory of anyon condensation.} the corresponding patch operator $(P_a^{ij})_{\mu_i, \mu_j}$ has an empty bulk because condensed anyons become trivial on the boundary.
Such a patch operator is called a patch charge operator in \cite{Ji:2019jhk, Chatterjee:2022kxb} because it carries point-like charges at the two ends.
The meaning of a charge should be generalized appropriately as in \cite{Bhardwaj:2023ayw, Bartsch:2023wvv} when the symmetry is non-invertible.
On the other hand, when the anyon $a$ is not condensed on the topological boundary, the corresponding patch operator has a non-empty bulk.
Such a patch operator is called a patch symmetry operator in \cite{Ji:2019jhk, Chatterjee:2022kxb} because it looks like a symmetry operator in the middle of the patch.

\paragraph{Connectability.}
Patch operators with opposite orientations should be connectable to each other via a ``left evaluation" tensor $\epsilon_a^L$ and a ``left coevaluation" tensor $\eta_a^L$,\footnote{By abuse of terminology, we will simply call a quantity with indices a tensor in the following.} which are represented by the gray ovals in the following equations:
\begin{equation}
\begin{aligned}
\adjincludegraphics[valign = c, width = 2cm]{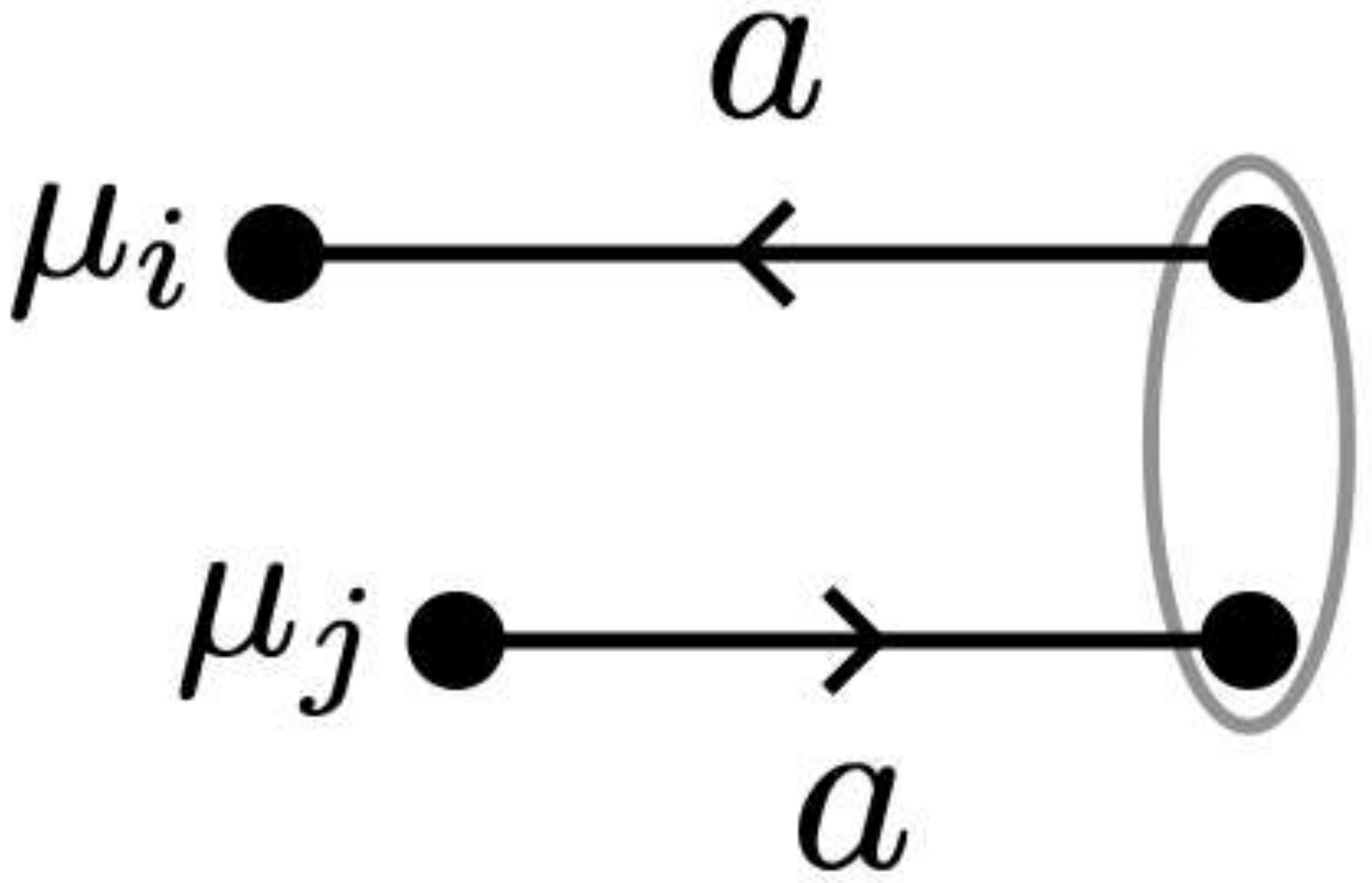} = \sum_{X, X^{\prime}} \sum_{\nu, \nu^{\prime}} (P_a^{ki})^X_{\nu, \mu_i} (\epsilon_a^L)^{\nu, \nu^{\prime}}_{X, X^{\prime}} (P_a^{jk})^{X^{\prime}}_{\mu_j, \nu^{\prime}}, \\[6pt]
\adjincludegraphics[valign = c, width = 2cm]{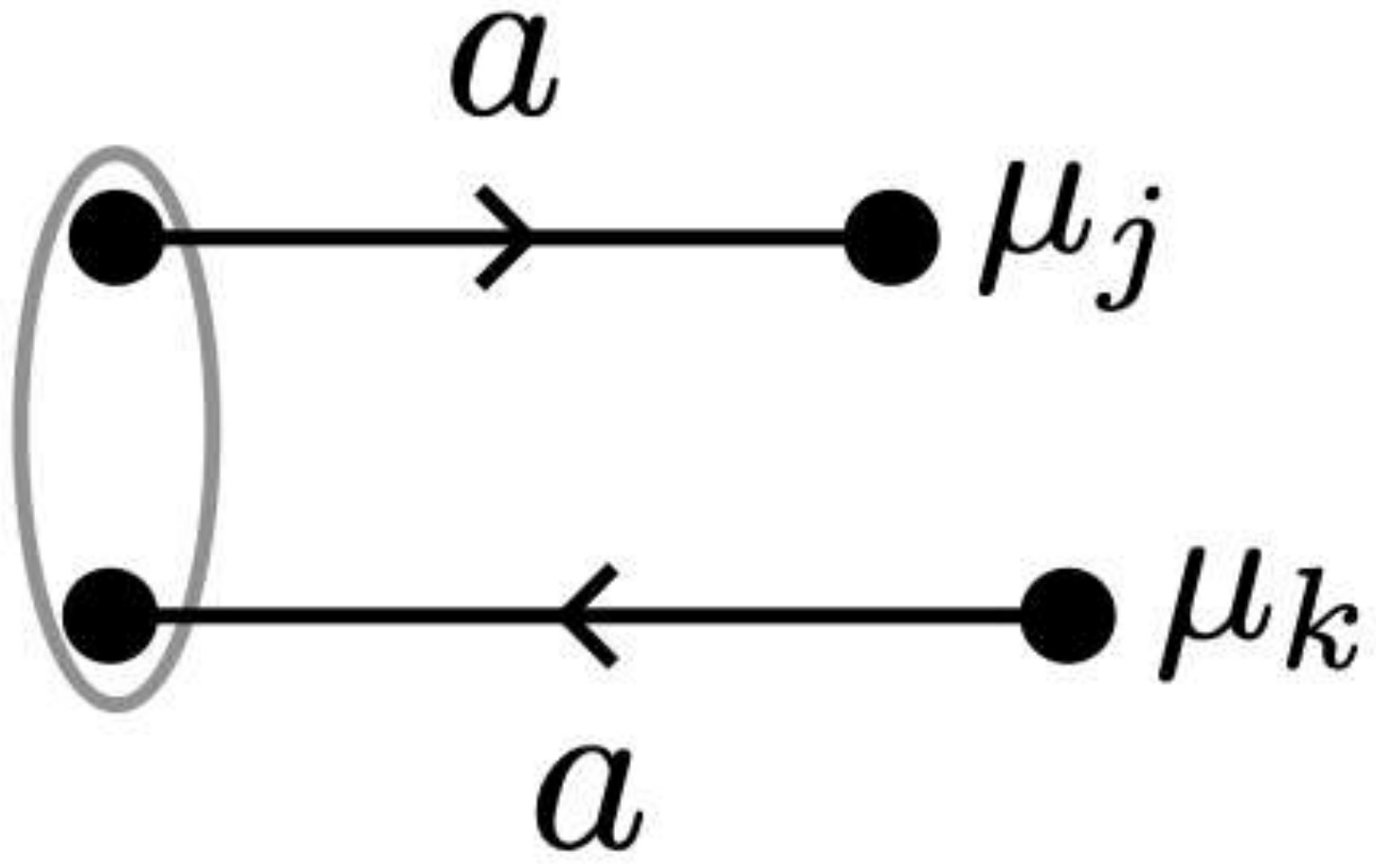} = \sum_{X, X^{\prime}} \sum_{\nu, \nu^{\prime}} (P_a^{ij})^X_{\nu, \mu_j} (\eta_a^L)^{\nu, \nu^{\prime}}_{X, X^{\prime}} (P_a^{ki})^{X^{\prime}}_{\mu_k, \nu^{\prime}},
\end{aligned}
\label{eq: ev/coev tensor}
\end{equation}
where $i < j < k$.
Here, the orientation-reversal of a patch operator is defined by its Hermitian conjugate:
\begin{equation}
(P_a^{ji})_{\mu_j, \mu_i} := (P_a^{ij})_{\mu_i, \mu_j}^{\dagger} = \sum_{X \in \mathcal{C}} \left((P_a^{ij})^X_{\mu_i, \mu_j}\right)^{\dagger}.
\label{eq: orientation reversal}
\end{equation}
The summand on the right-hand side of Eq.~\eqref{eq: orientation reversal} is denoted by $(P_a^{ji})^X_{\mu_j, \mu_i}$ in Eq.~\eqref{eq: ev/coev tensor}, i.e., we have 
\begin{equation}
(P_a^{ji})^X_{\mu_j, \mu_i} := \left((P_a^{ij})^X_{\mu_i, \mu_j}\right)^{\dagger}.
\end{equation}
We expect that the orientation-reversal \eqref{eq: orientation reversal} of the patch operator labeled by anyon $a$ is equivalent to the patch operator labeled by the dual anyon $\overline{a}$.
Namely, we expect that $(P_a^{ji})_{\mu_j, \mu_i}$ and $(P_{\overline{a}}^{ij})_{\mu_i, \mu_j}$ differ only by symmetric local operators around the endpoints.
For the equivalence of patch operators, see the last paragraph of this subsection.
The components $(\epsilon_a^L)^{\nu, \nu^{\prime}}_{X, X^{\prime}}$ and $(\eta_a^L)^{\nu, \nu^{\prime}}_{X, X^{\prime}}$ in Eq.~\eqref{eq: ev/coev tensor} may be non-trivial symmetric local operators around the gray ovals, although they are complex numbers in the examples that we will discuss in Sec.~\ref{sec: Reconstruction of Kitaev's quantum double topological order}.
For consistency of the diagrammatic representations, we impose the condition that the evaluation and coevaluation tensors are related to their Hermitian conjugates as follows:
\begin{equation}
\left((\epsilon_a^L)_{X, X^{\prime}}^{\nu,\nu^{\prime}}\right)^{\dagger} = (\epsilon_a^L)_{X^{\prime}, X}^{\nu^{\prime}, \nu}, \quad
\left((\eta_a^L)_{X, X^{\prime}}^{\nu, \nu^{\prime}}\right)^{\dagger} = (\eta_a^L)_{X^{\prime}, X}^{\nu^{\prime}, \nu}.
\label{eq: unitarity}
\end{equation}
This condition guarantees that the Hermitian conjugates of the right-hand sides of Eq.~\eqref{eq: ev/coev tensor} are represented by the diagrams turned upside down and oriented in the opposite direction.
The evaluation and coevaluation tensors should satisfy the following zigzag identities:
\begin{equation}
\begin{aligned}
\adjincludegraphics[valign = c, width = 2.8cm]{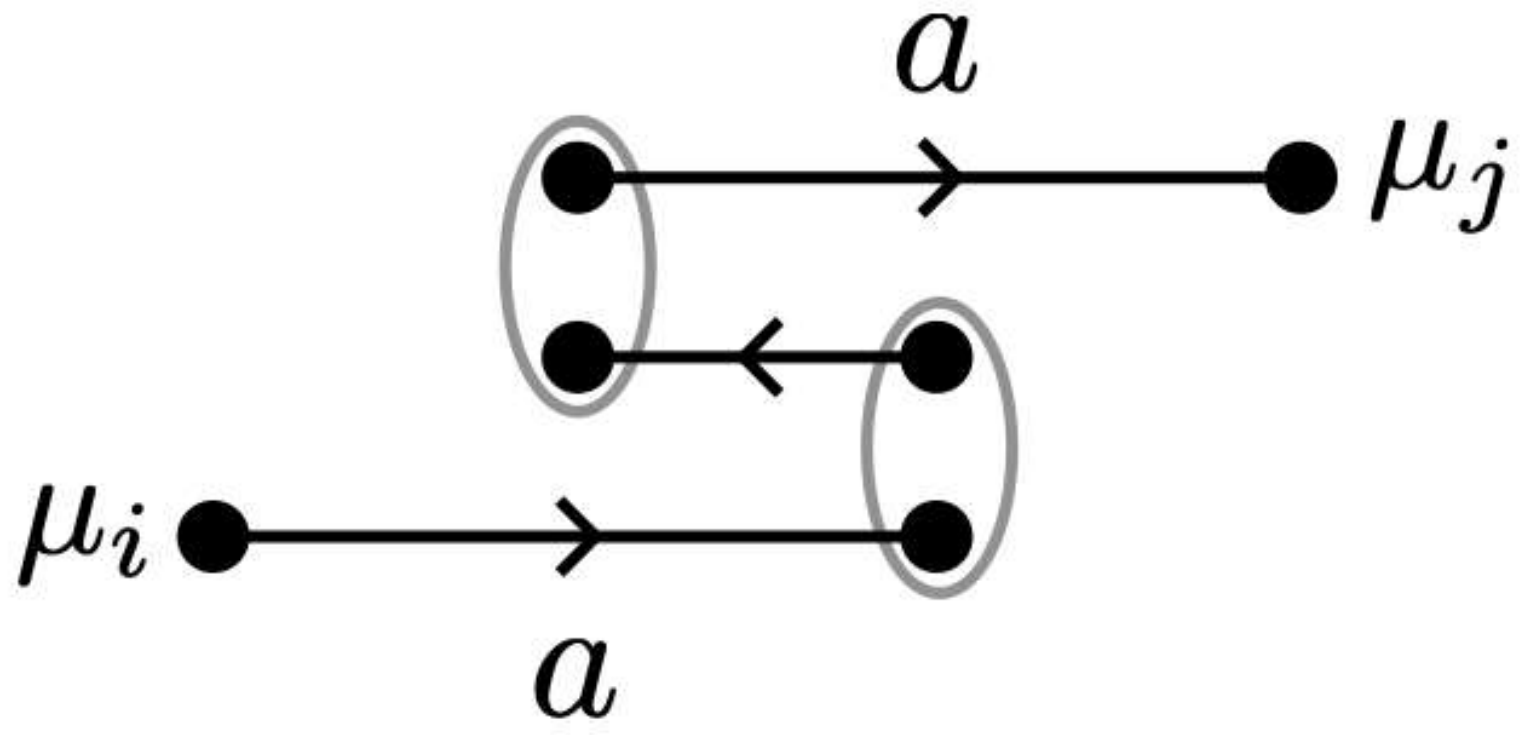} &= \adjincludegraphics[valign = c, width = 2.2cm]{patch1.pdf},\\[6pt]
\adjincludegraphics[valign = c, width = 2.8cm]{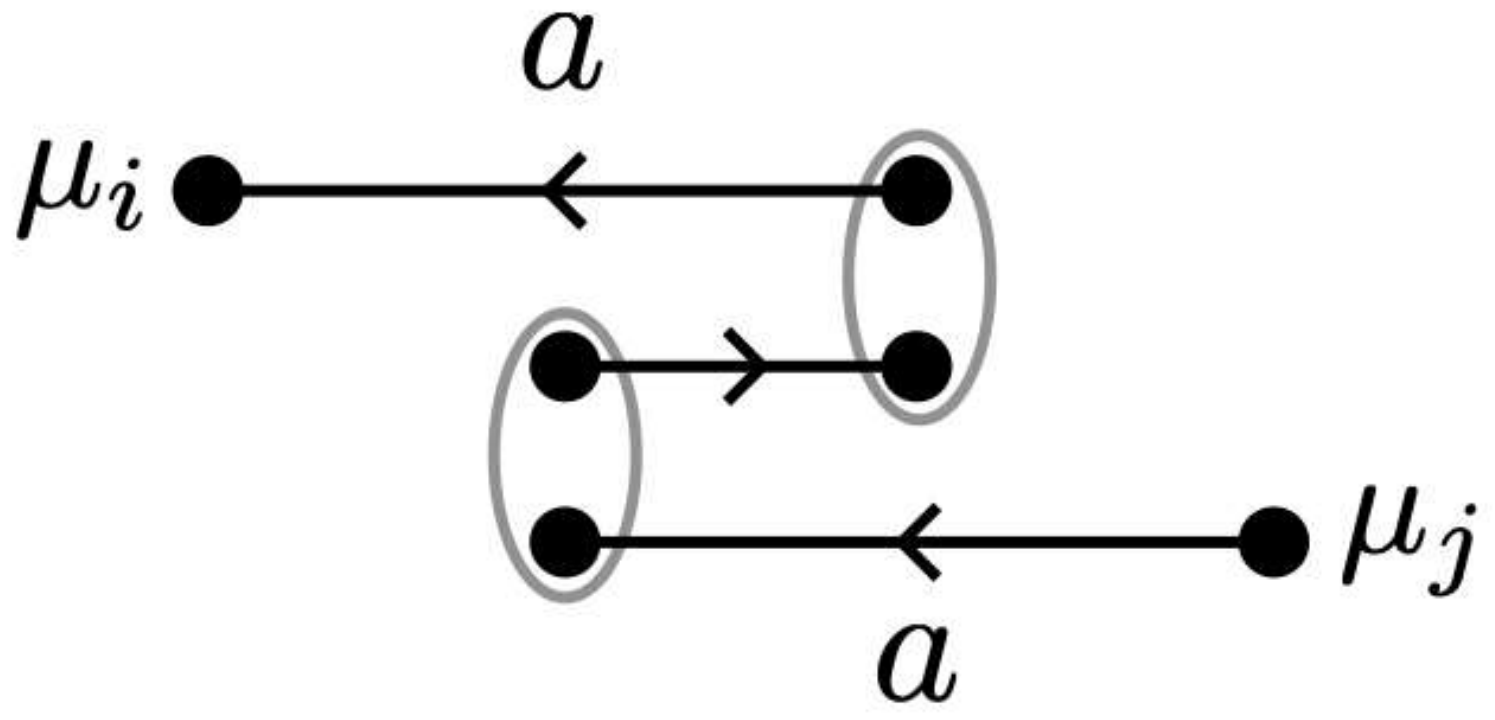} &= \adjincludegraphics[valign = c, width = 2.2cm]{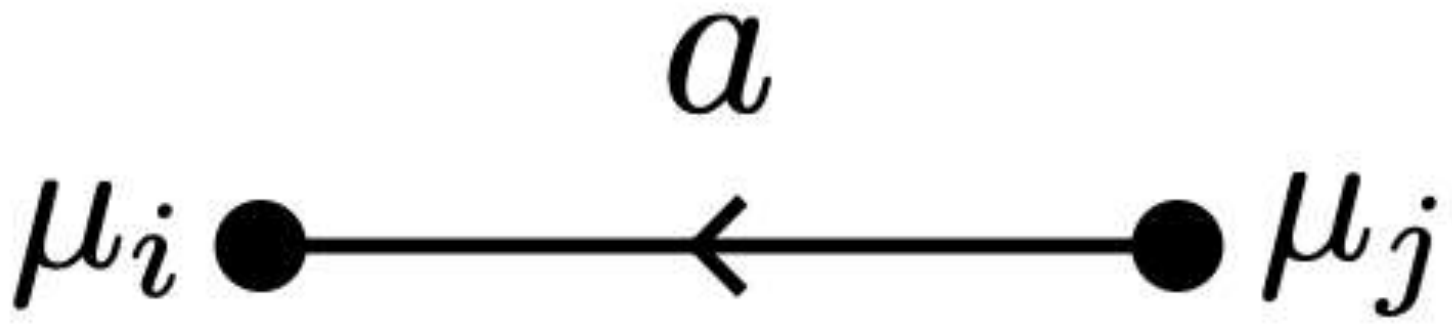}.
\end{aligned}
\label{eq: zigzag contraction}
\end{equation}
We note that the second equality automatically follows from the first equality and Eq.~\eqref{eq: unitarity}.
There are also a ``right evaluation" tensor $\epsilon_a^R$ and a ``right coevaluation" tensor $\eta_a^R$, which are represented by the gray ovals in the following equations:
\begin{equation*}
\begin{aligned}
\adjincludegraphics[valign = c, width = 2cm]{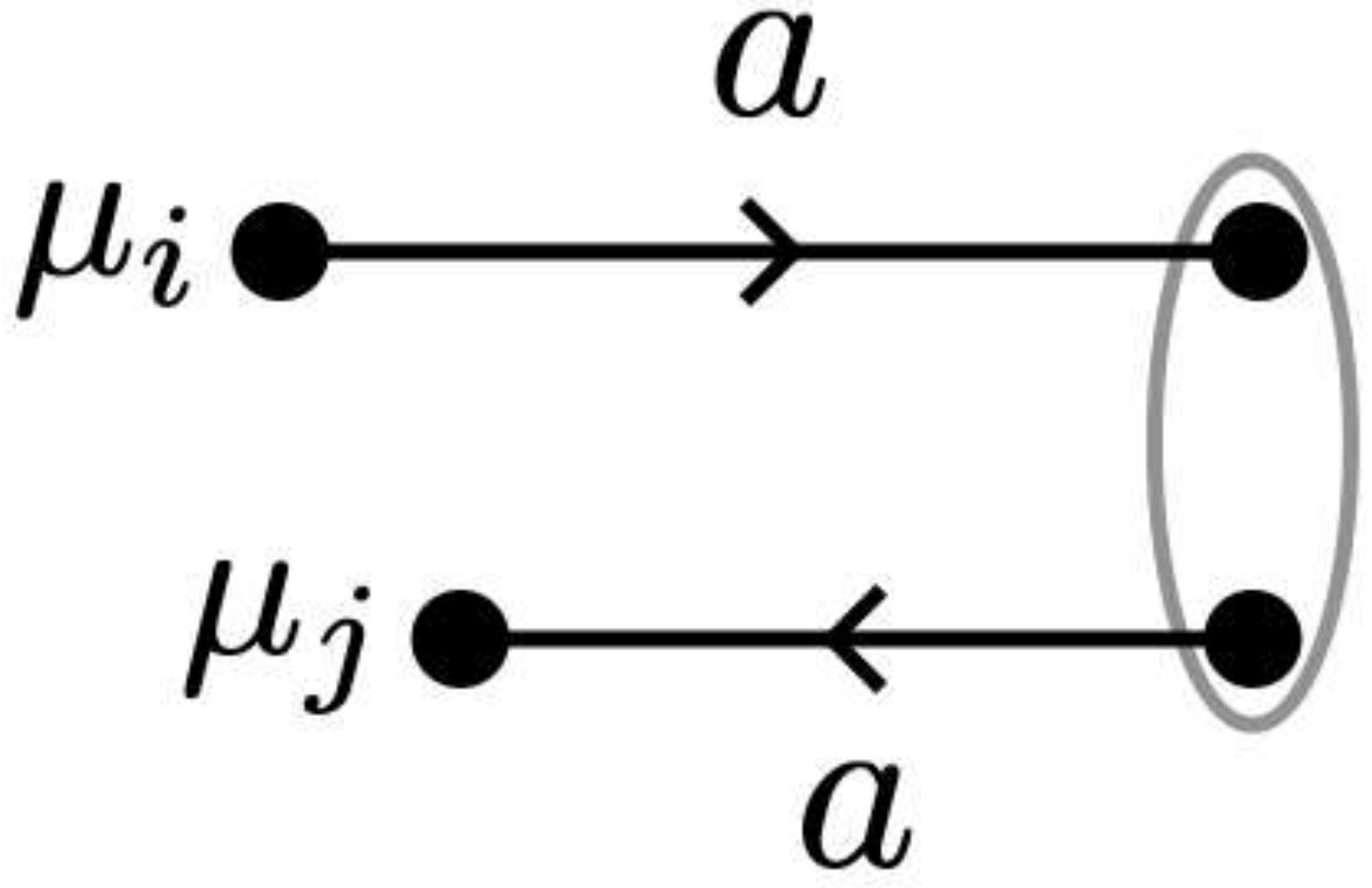} & = \sum_{X, X^{\prime}} \sum_{\nu, \nu^{\prime}} (P_a^{ik})^X_{\mu_i, \nu} (\epsilon_a^{R})^{\nu, \nu^{\prime}}_{X, X^{\prime}} (P_a^{kj})^{X^{\prime}}_{\nu^{\prime}, \mu_j}, \\[6pt]
\adjincludegraphics[valign = c, width = 2cm]{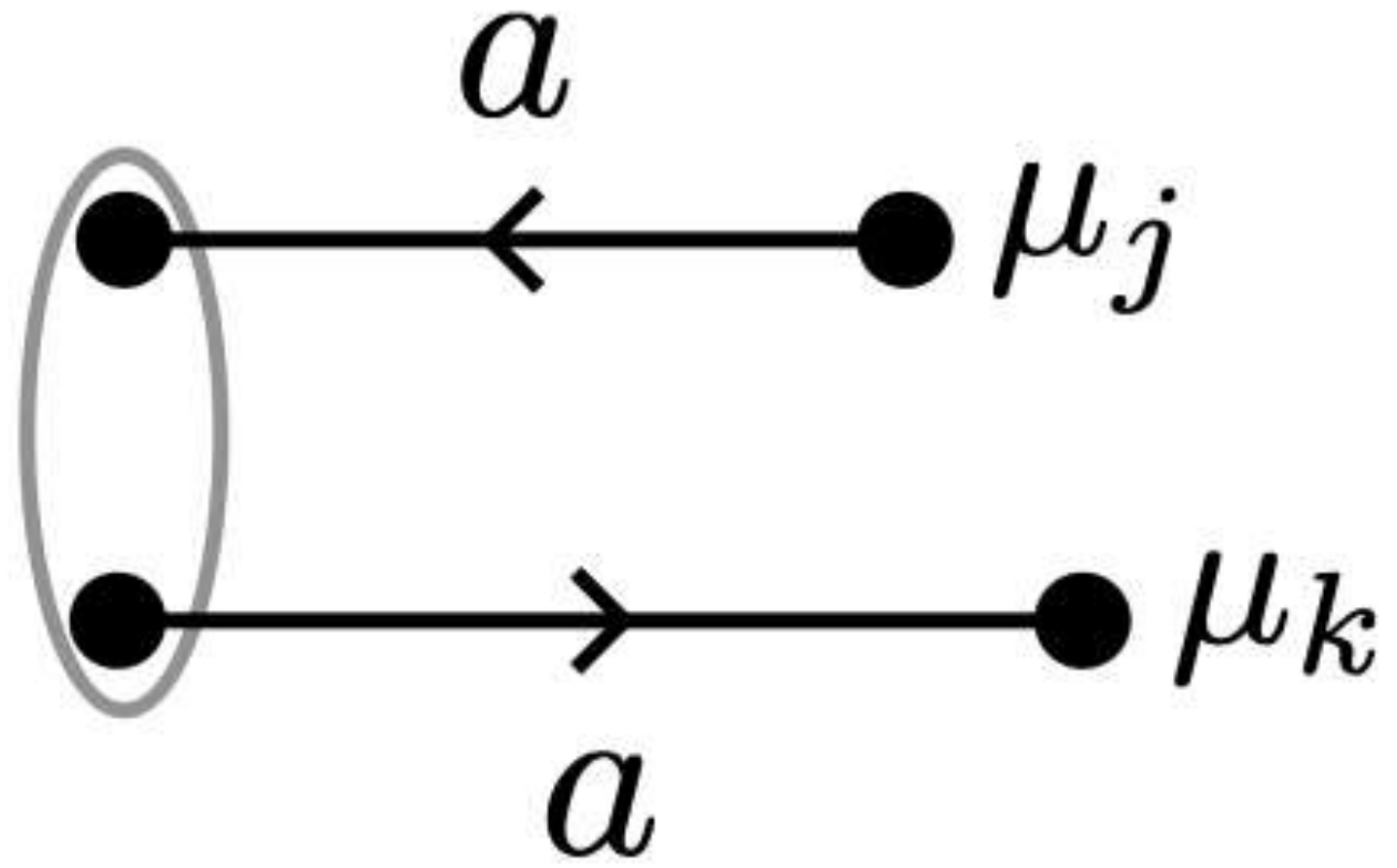} & = \sum_{X, X^{\prime}} \sum_{\nu, \nu^{\prime}} (P_a^{ji})^X_{\mu_j, \nu} (\eta_a^R)^{\nu, \nu^{\prime}}_{X, X^{\prime}} (P_a^{ik})^{X^{\prime}}_{\nu^{\prime}, \mu_k}.
\end{aligned}
\end{equation*}
These tensors should also satisfy the zigzag identities analogous to Eq.~\eqref{eq: zigzag contraction}.

The three properties listed above would automatically hold if the patch operator originates from an anyon line of the bulk topological order as shown in Fig.~\ref{fig: SymTFT patch}.
Therefore, these properties are necessary conditions for patch operators to be related to anyons.
The importance of these properties was already noticed in \cite{Ji:2019jhk, Chatterjee:2022kxb}, where the patch operators with these properties were utilized to reconstruct the anyon data of several topological orders only with abelian anyons.
Here, we slightly extended the formulation in \cite{Ji:2019jhk, Chatterjee:2022kxb} so that we can handle more general 2+1D topological orders that may have non-abelian anyons.
We emphasize that the explicit form of patch operators with the above properties does not depend on the Hamiltonian of the 1+1D system: it depends only on the symmetry and its representation on the state space.

Based on the above discussion, we conjecture that the anyon data of a 2+1D non-chiral topological order is encoded in the set of symmetric transparent connectable patch operators in 1+1D.
More specifically, we conjecture that any topological invariants of a 2+1D topological order described by the Drinfeld center of a fusion category $\mathcal{C}$ can be computed from symmetric transparent connectable patch operators in 1+1D systems with symmetry $\mathcal{C}$.
A general computational scheme will be presented in the next subsection.

Before proceeding, we notice that symmetric transparent connectable patch operators have ambiguities around the endpoints because we can multiply symmetric local operators around the endpoints without violating the symmetricity, transparency, and connectability.
However, the multiplication by symmetric local operators would not affect the anyon data contained in the patch operators as long as they remain to be symmetric, transparent, and connectable.
Therefore, two patch operators should be considered to be equivalent to each other if they differ only by local symmetric operators around the endpoints.
We expect that equivalence classes of symmetric transparent connectable patch operators are sufficient to reconstruct the anyon data of 2+1D topological orders.

\subsection{General Scheme to compute topological invariants}
\label{sec: General Scheme for computing topological invariants}
In this subsection, we propose a general scheme to compute topological invariants of 2+1D topological orders by using symmetric transparent connectable patch operators in 1+1D.
To this end, we first introduce twisted evaluation and coevaluation tensors, which allow us to connect two patch operators in the presence of another patch operator in between.
Pictorially, the twisted evaluation tensor $\epsilon_{a; b}^L$ and the twisted coevaluation tensor $\eta_{a; b}^L$ are expressed as follows:
\begin{equation*}
\begin{aligned}
&\quad \adjincludegraphics[valign = c, width = 3.75cm]{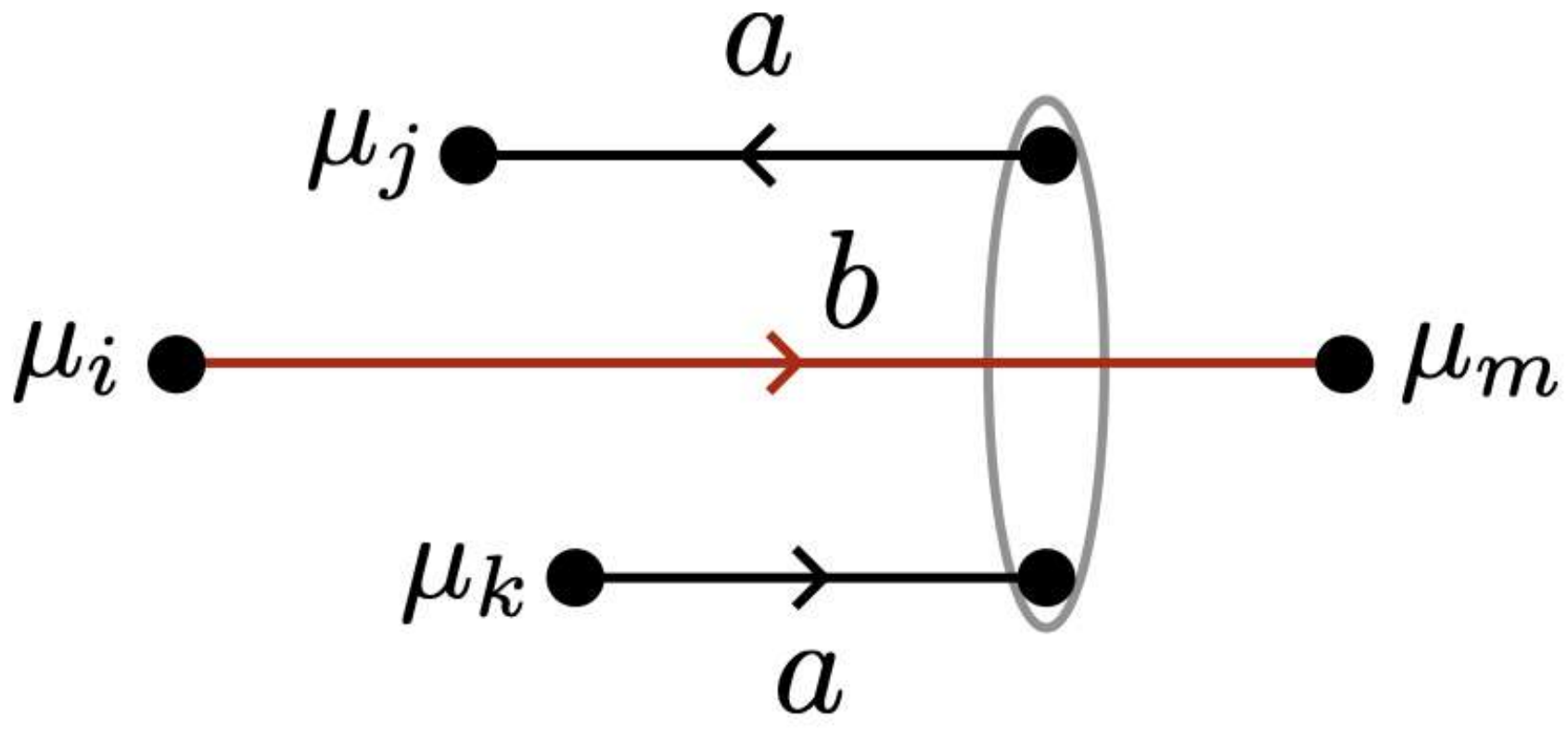}\\
&= \sum_{X, X^{\prime}, Y} \sum_{\nu, \nu^{\prime}} (P_a^{lj})_{\nu, \mu_j}^X (P_b^{im})_{\mu_i, \mu_m}^Y (\epsilon_{a; b}^L)_{X, X^{\prime}; Y}^{\nu, \nu^{\prime}} (P_a^{kl})_{\mu_k, \nu^{\prime}}^{X^{\prime}},
\end{aligned}
\end{equation*}
\begin{equation*}
\begin{aligned}
&\quad \adjincludegraphics[valign = c, width = 3.75cm]{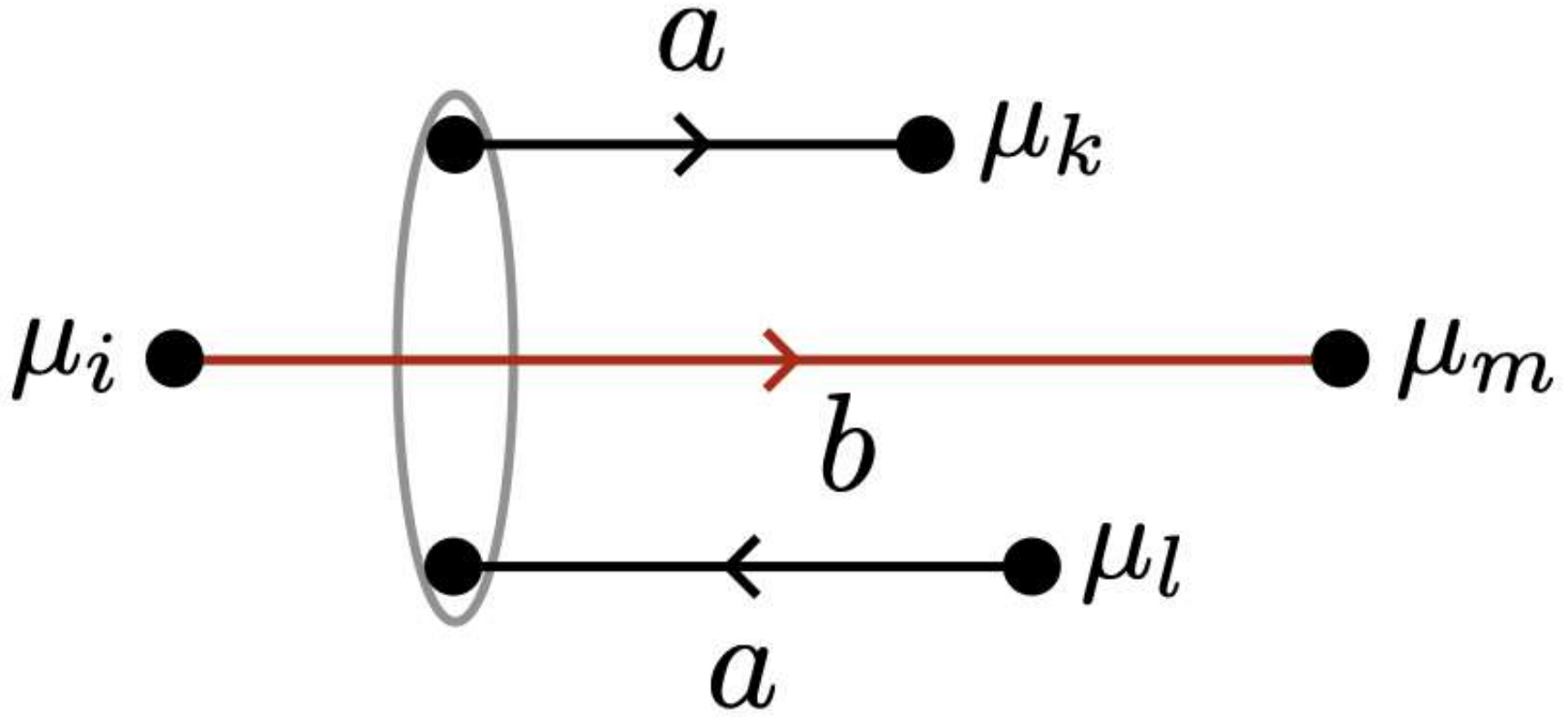}\\
&= \sum_{X, X^{\prime}, Y} \sum_{\nu, \nu^{\prime}} (P_a^{jk})_{\nu, \mu_k}^X (P_b^{im})_{\mu_i, \mu_m}^Y (\eta_{a; b}^L)^{\nu, \nu^{\prime}}_{X, X^{\prime}; Y} (P_a^{lj})_{\mu_l, \nu^{\prime}}^{X^{\prime}},
\end{aligned}
\end{equation*}
where $i < j < k < l < m$.
The components $(\epsilon_{a; b}^L)_{X, X^{\prime}; Y}^{\nu, \nu^{\prime}}$ and $(\eta_{a; b}^L)^{\nu, \nu^{\prime}}_{X, X^{\prime}; Y}$ are supposed to be symmetric local operators.
In particular, they commute with $(P_b^{im})_{\mu_i, \mu_m}^Y$ in the above equations.
When the patch operator inserted in between is oriented in the opposite direction, we use the Hermitian conjugates of $\epsilon_{a; b}^L$ and $\eta_{a; b}^L$ to connect two patch operators.
Similarly, there also exist twisted versions of the right evaluation and coevaluation tensors, which are expressed diagrammatically as follows:
\begin{equation*}
\begin{aligned}
&\quad \adjincludegraphics[valign = c, width = 3.75cm]{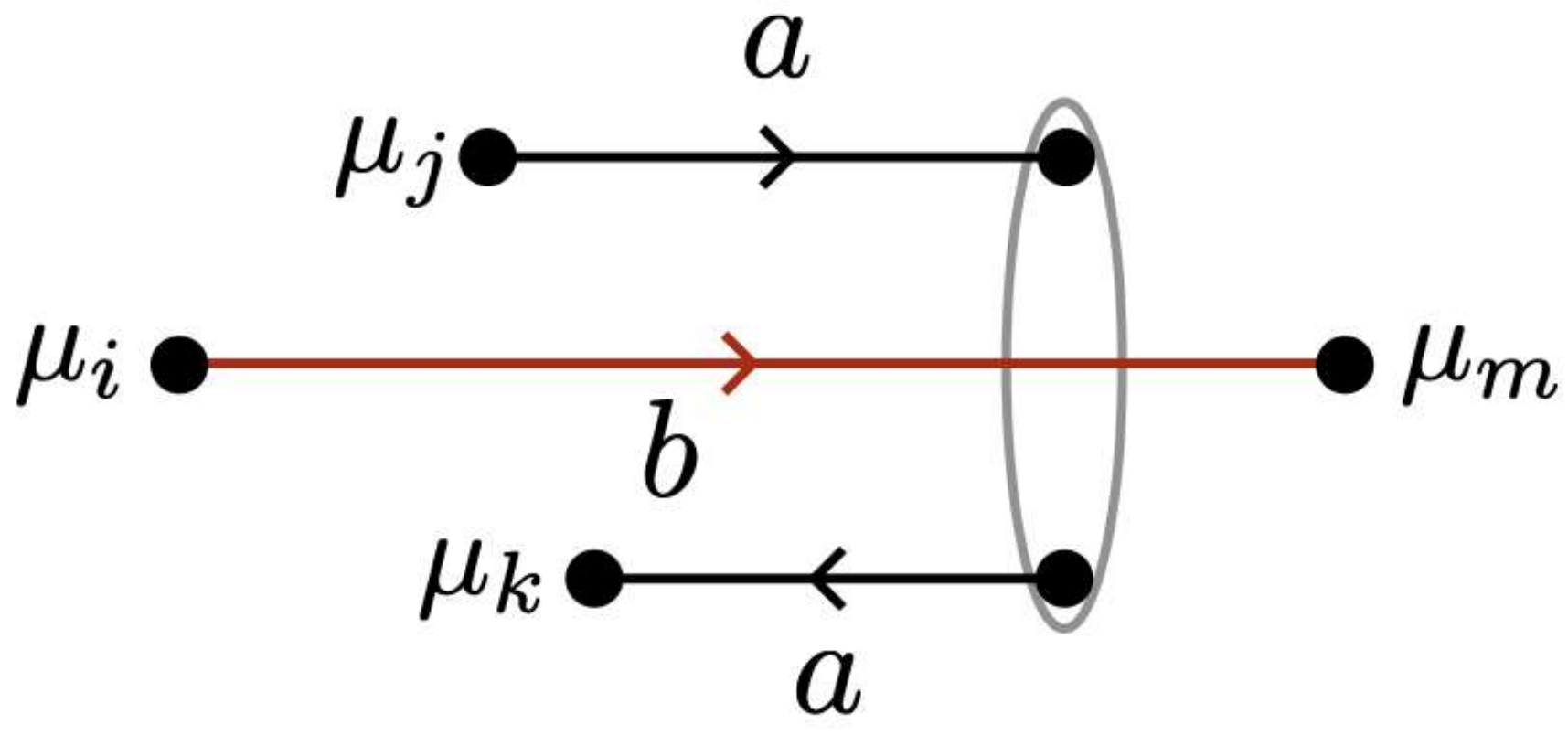}\\
&= \sum_{X, X^{\prime}, Y} \sum_{\nu, \nu^{\prime}} (P_a^{jl})^X_{\mu_j, \nu} (P_b^{im})_{\mu_i, \mu_m}^Y (\epsilon_{a; b}^{R})^{\nu, \nu^{\prime}}_{X, X^{\prime}; Y} (P_a^{lk})^{X^{\prime}}_{\nu^{\prime}, \mu_k},
\end{aligned}
\end{equation*}
\begin{equation*}
\begin{aligned}
&\quad \adjincludegraphics[valign = c, width = 3.75cm]{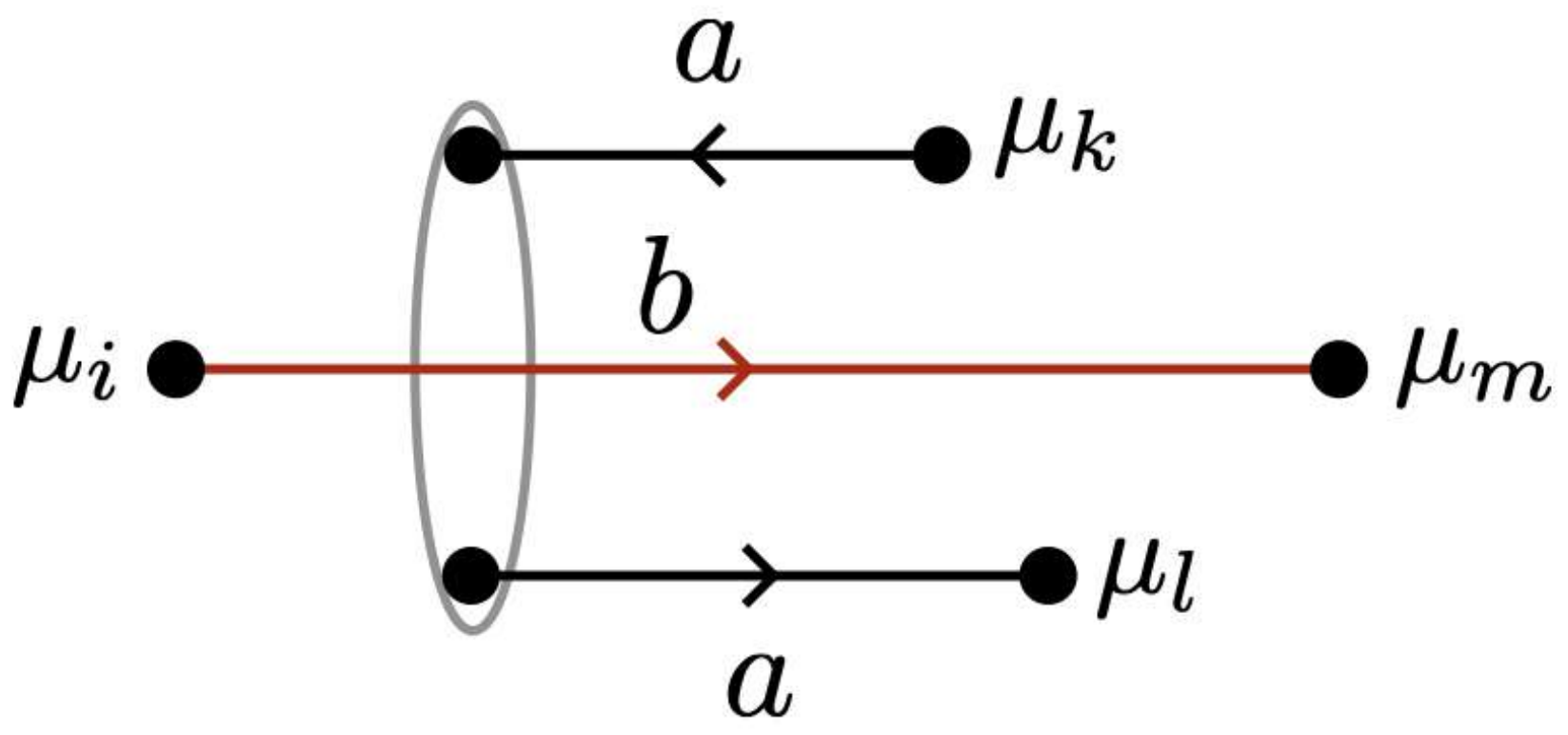}\\
&= \sum_{X, X^{\prime}, Y} \sum_{\nu, \nu^{\prime}} (P_a^{kj})^X_{\mu_k, \nu} (P_b^{im})_{\mu_i, \mu_m}^Y (\eta_{a; b}^R)^{\nu, \nu^{\prime}}_{X, X^{\prime}; Y} (P_a^{jl})^{X^{\prime}}_{\nu^{\prime}, \mu_l}.
\end{aligned}
\end{equation*}
In the above equations, we should be able to freely move the patch operators inserted in the middle to the top or bottom because these patch operators correspond to anyon lines in the bulk. 
More specifically, we should impose the following consistency conditions on the twisted evaluation and coevaluation tensors:
\begin{equation}
\begin{aligned}
\adjincludegraphics[valign = c, width = 1.8cm]{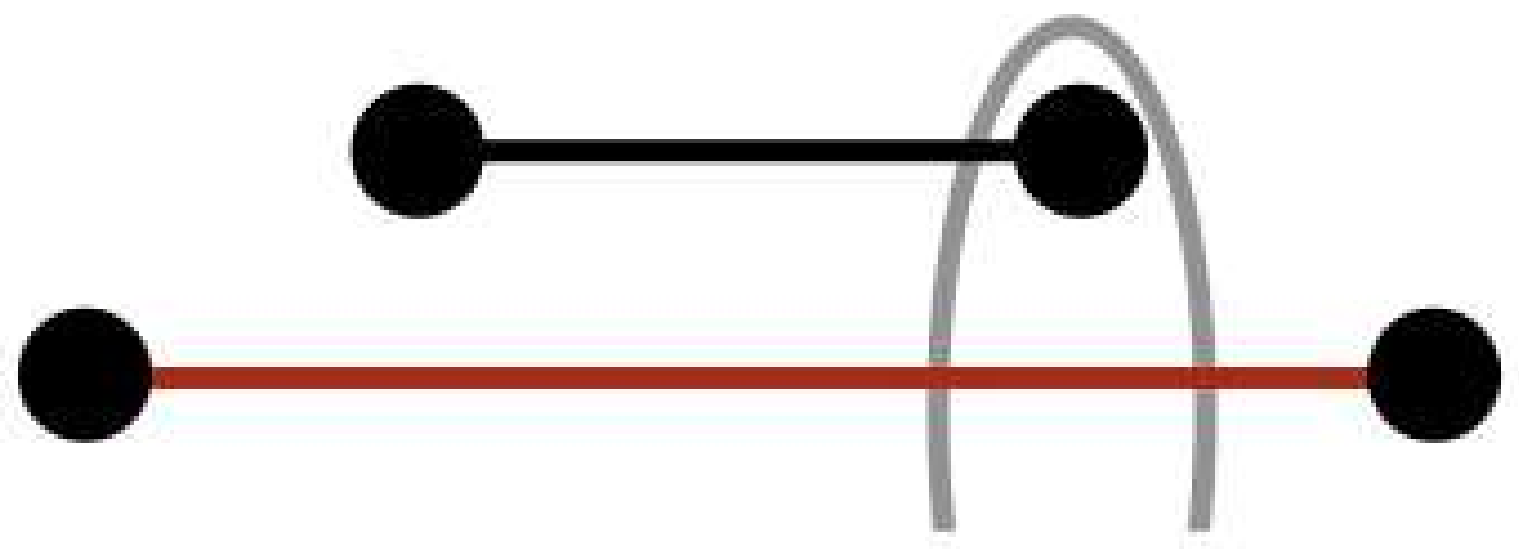} ~ &= ~ \adjincludegraphics[valign = c, width = 1.8cm]{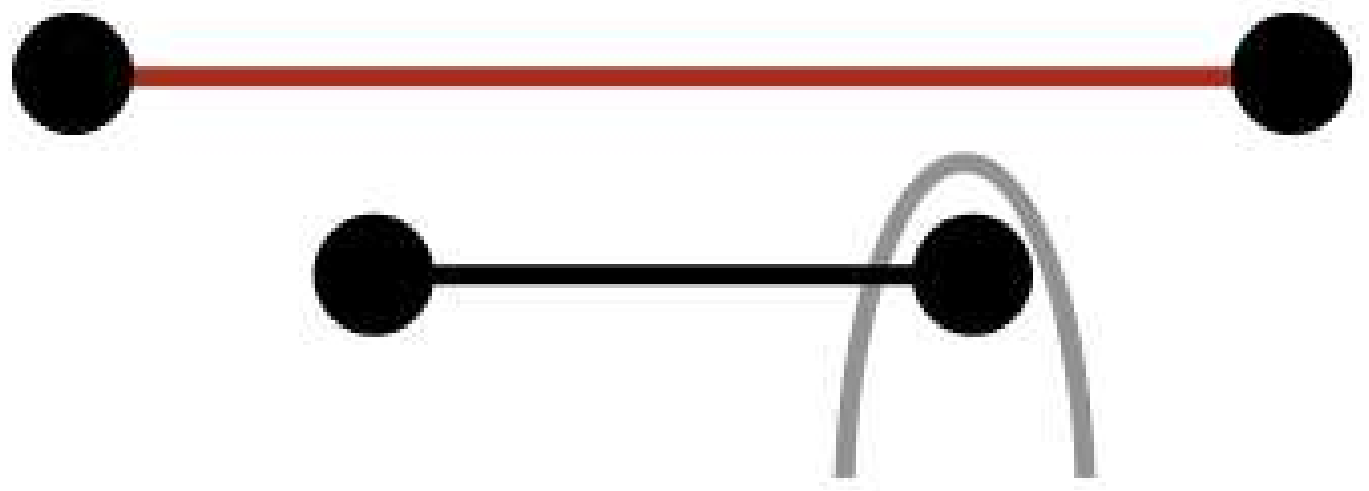}, \\[10pt]
\adjincludegraphics[valign = c, width = 1.8cm]{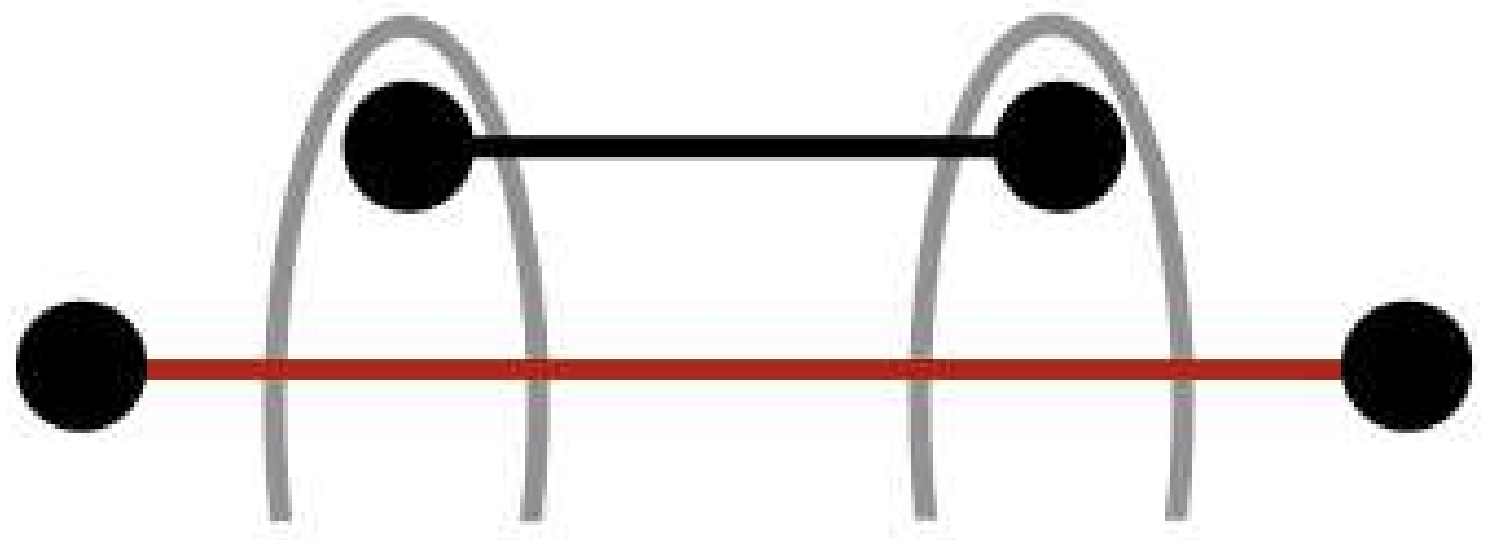} ~ &= ~ \adjincludegraphics[valign = c, width = 1.8cm]{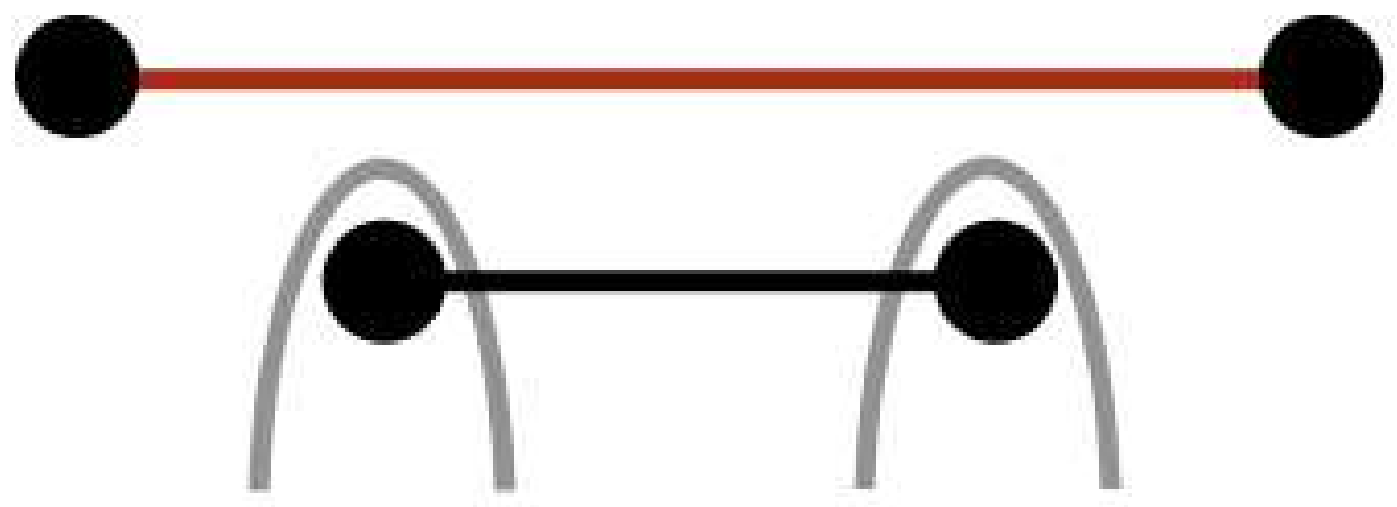}, \\[10pt]
\adjincludegraphics[valign = c, width = 1.8cm]{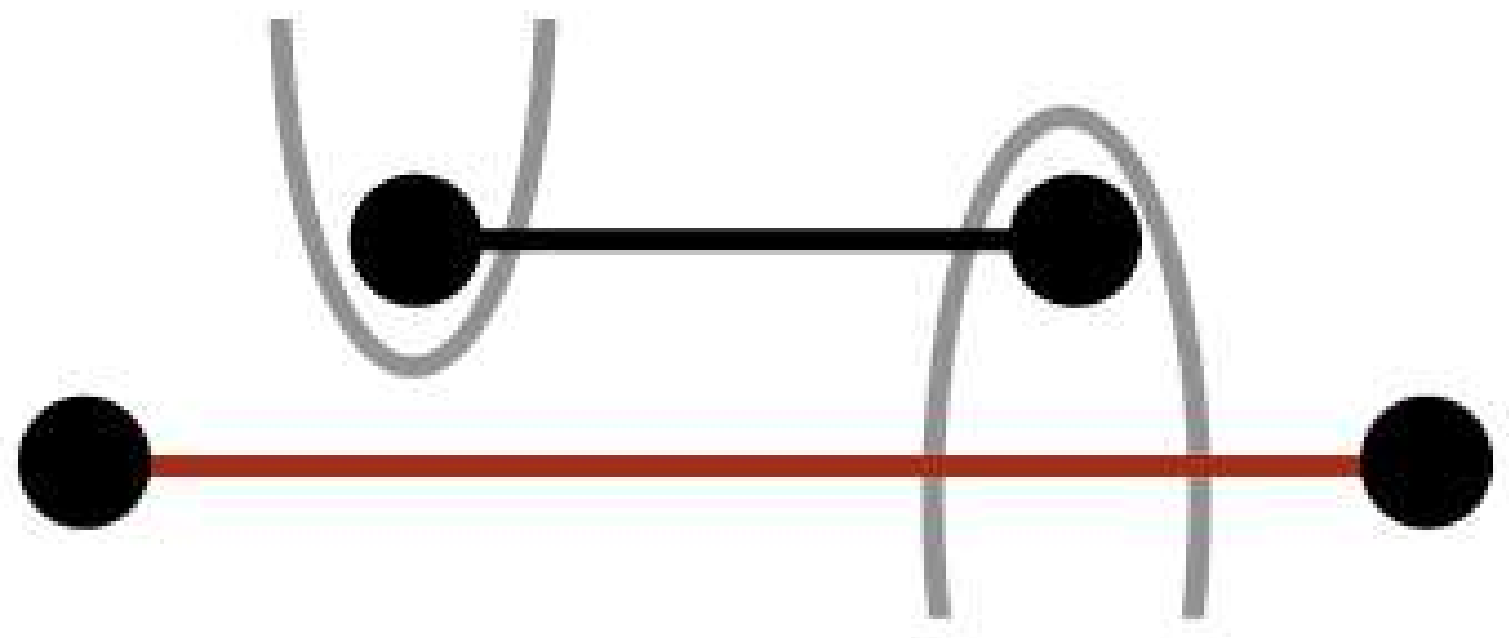} ~ &= ~ \adjincludegraphics[valign = c, width = 1.8cm]{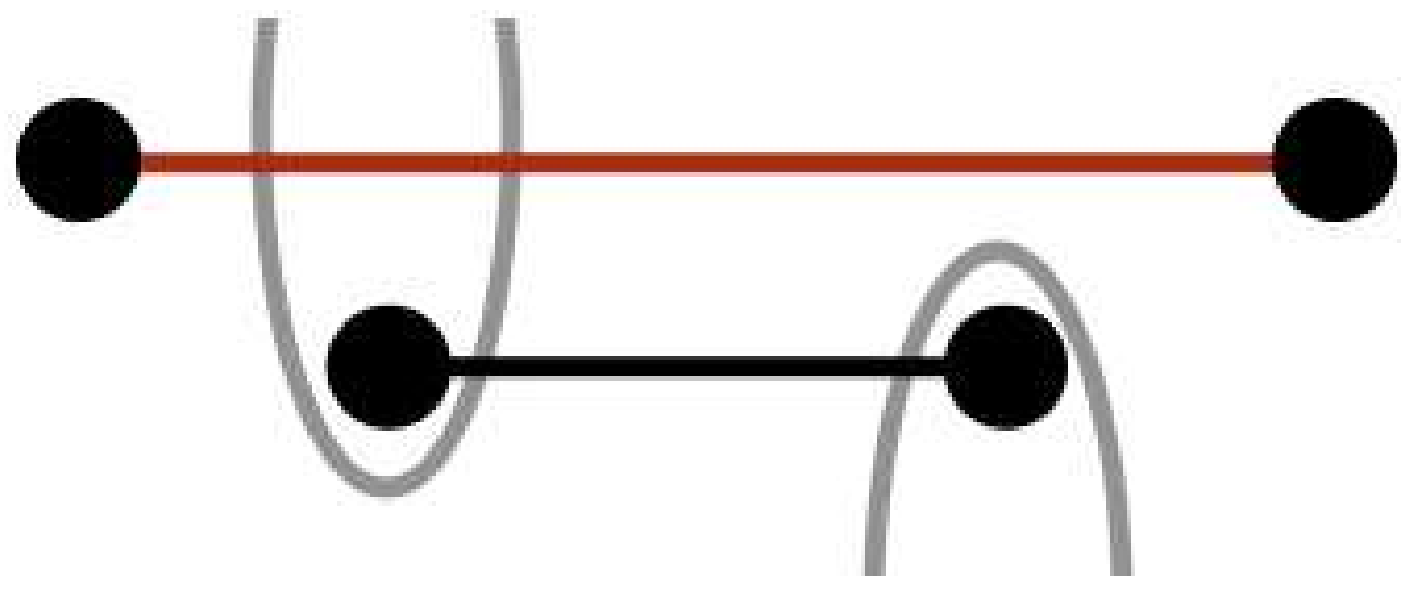}.
\end{aligned}
\label{eq: twist consistency}
\end{equation}
We also have the consistency conditions obtained by flipping the diagrams horizontally and/or vertically in the above equations.
These consistency conditions are supposed to be satisfied regardless of the orientations of the patch operators.
Equation \eqref{eq: twist consistency} can be expressed more explicitly in terms of $P_a^{ij}$, $\epsilon_{a; b}$, and $\eta_{a; b}$.
For example, the first equality in Eq.~\eqref{eq: twist consistency} can be written as
\begin{equation*}
\begin{aligned}
&\quad \sum_{X} \sum_{\nu} (P_a^{kj})^X_{\nu, \mu_j} (P_b^{il})^Y_{\mu_i, \mu_l} (\epsilon_{a; b}^L)_{X, X^{\prime}; Y}^{\nu, \nu^{\prime}}\\
&= \sum_{X} \sum_{\nu} (P_b^{il})^Y_{\mu_i, \mu_l} (P_a^{kj})^X_{\nu, \mu_j} (\epsilon_a^L)_{X, X^{\prime}}^{\nu, \nu^{\prime}}, \quad \forall Y, X^{\prime}, \nu^{\prime},
\end{aligned}
\end{equation*}
where we chose specific orientations of the patch operators for concreteness.
The other two conditions in Eq.~\eqref{eq: twist consistency} can also be expressed in a similar manner.

Now, we propose a general method to compute topological invariants of 2+1D topological orders by using symmetric transparent connectable patch operators and twisted evaluation/coevaluation tensors.
Our computational scheme goes as follows (see also Fig.~\ref{fig: scheme} for an illustrative example):
\begin{figure}
\subfloat[\label{fig: scheme1}]{
\includegraphics[width = 2cm]{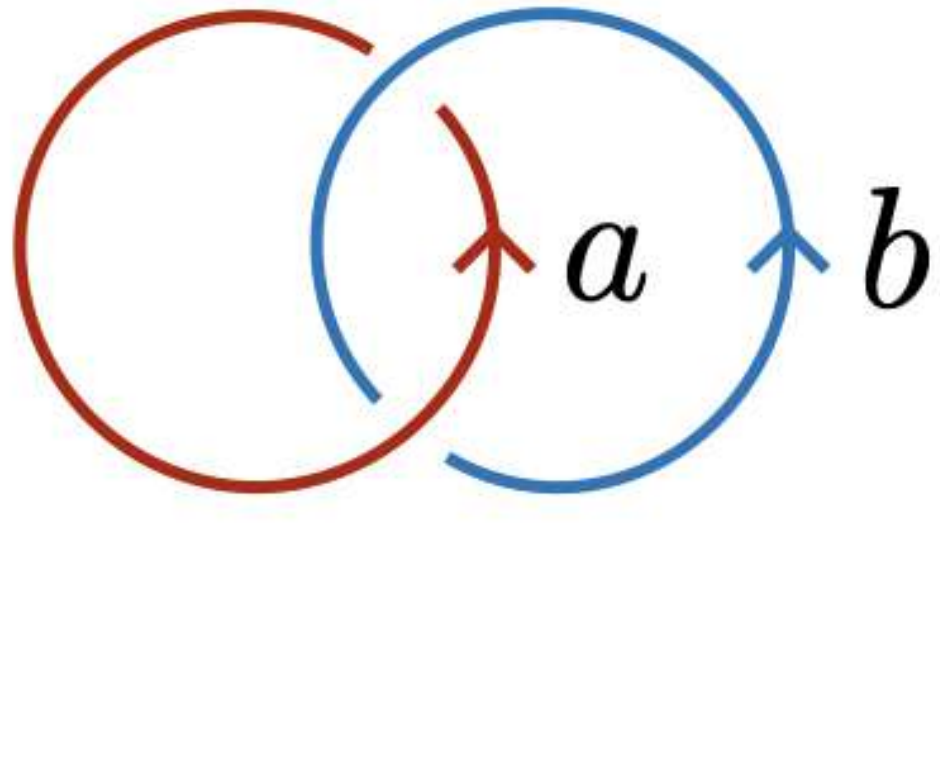}
}\hfill
\subfloat[\label{fig: scheme2}]{
\includegraphics[width = 2.25cm]{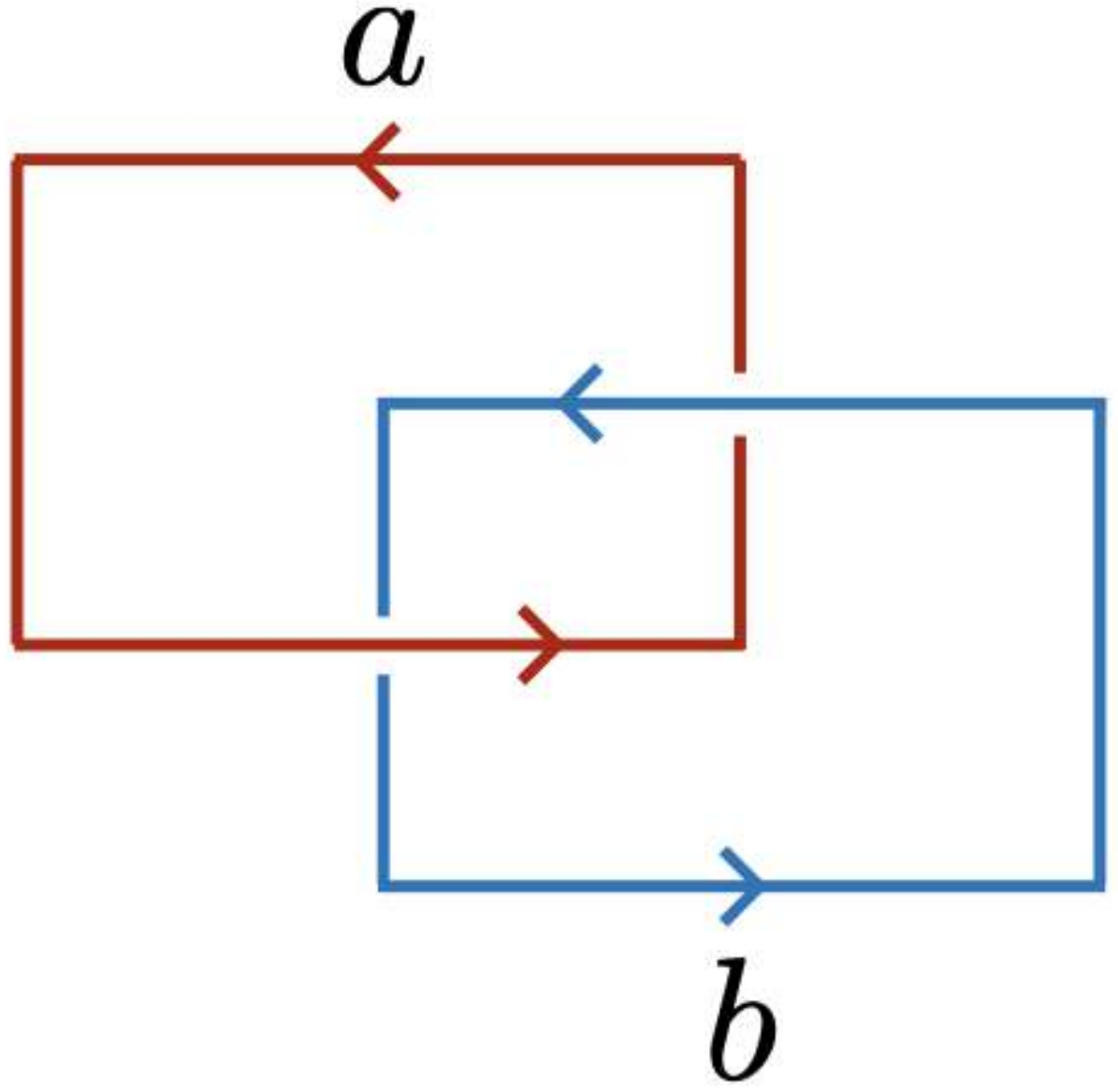}
}\hfill
\subfloat[\label{fig: scheme3}]{
\includegraphics[width = 2.5cm]{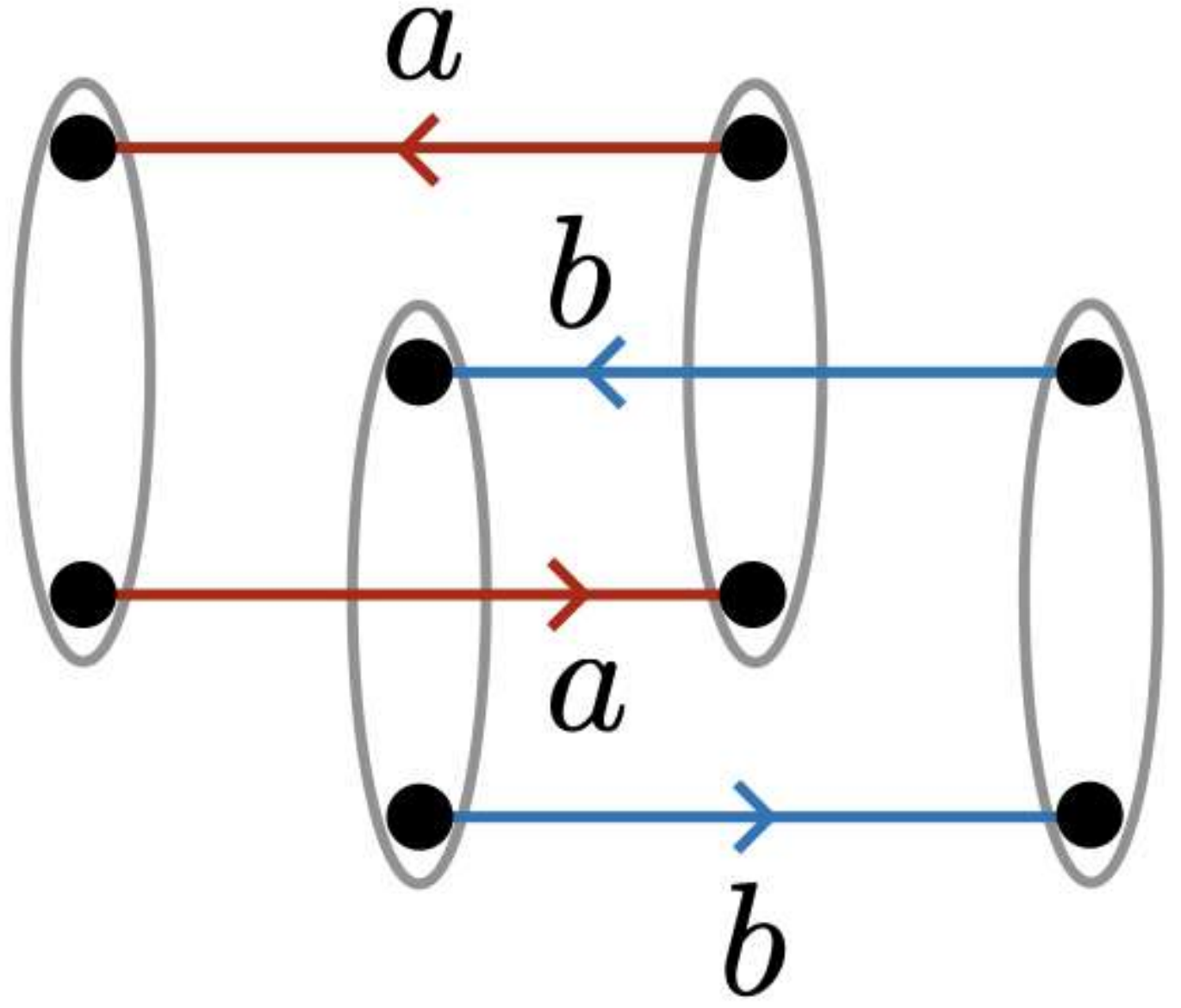}
}
\caption{A general scheme to compute topological invariants of 2+1D topological orders by using patch operators in 1+1D. (a) An anyon diagram drawn on a 2d plane. (b) A deformed anyon diagram consisting only of horizontal and vertical lines. A vertical line is above a horizontal line when they intersect. (c) A contracted product of patch operators translated from the corresponding anyon diagram.}
\label{fig: scheme}
\end{figure}
\begin{enumerate}
\item First, we draw a closed anyon diagram on a two-dimensional plane.
\item We then continuously deform the anyon diagram so that it consists only of horizontal lines and vertical lines. As a convention, we require that a horizontal line is always above a vertical line when they intersect each other. 
The reason for adopting this convention will be explained shortly.
\item Finally, we translate the anyon diagram into a product of patch operators whose indices at the endpoints are contracted by using the twisted evaluation and coevaluation tensors.
Here, horizontal lines of an anyon diagram correspond to patch operators, while vertical lines correspond to twisted evaluation and coevaluation tensors.
\end{enumerate}
The idea of identifying an anyon diagram with a contracted product of patch operators would be justified by the relation between anyon lines and patch operators illustrated in Fig.~\ref{fig: SymTFT patch}.
The convention adopted in the second step is due to the fact that the 1+1D system obtained as in Fig.~\ref{fig: SymTFT patch} is viewed from the left side of the 2+1D bulk as we mentioned in Sec.~\ref{sec: Motivation}.
Indeed, if we view the system from the left side of the bulk, anyon lines are always in front of twisted evaluation and coevaluation tensors living on the right boundary.
Thus, the third step of the above prescription makes sense only when the horizontal lines of an anyon diagram are above vertical lines when they intersect.

The above computational scheme would be valid for any topological invariants associated with framed knots and links of anyon lines in 2+1D.
In particular, the result of a computation should be invariant under continuous deformations of the anyon diagram due to the defining properties of the patch operators and the consistency conditions on the twisted evaluation/coevaluation tensors.
We conjecture that the above prescription enables us to reconstruct the Drinfeld center $Z(\mathcal{C})$ of a fusion category $\mathcal{C}$ from symmetric transparent connectable patch operators in 1+1D systems with symmetry $\mathcal{C}$.\footnote{If we adopt the convention that horizontal lines are below vertical lines in the second step of the prescription, we end up with the reverse category $Z(\mathcal{C})^{\mathrm{rev}}$, which is equivalent to the Drinfeld center of the opposite category $\mathcal{C}^{\mathrm{op}}$ \cite{EGNO2015}. This is consistent with the fact that the symmetry of the 1+1D system becomes $\mathcal{C}^{\mathrm{op}}$ if we view the system from the right side of the 2+1D bulk.}
In the subsequent section, we will see that the above method does work in the case of Kitaev's quantum double topological order $\mathrm{QD}(G)$ for general finite group $G$.

Let us write down explicit formulae for several topological invariants in terms of patch operators.
\paragraph{Quantum dimension.}
Since the quantum dimension is the topological invariant associated with a loop of an anyon line, it can be computed as the contracted product of two patch operators oriented in opposite directions:
\begin{equation}
d_a = \adjincludegraphics[valign = c, width = 1.6cm]{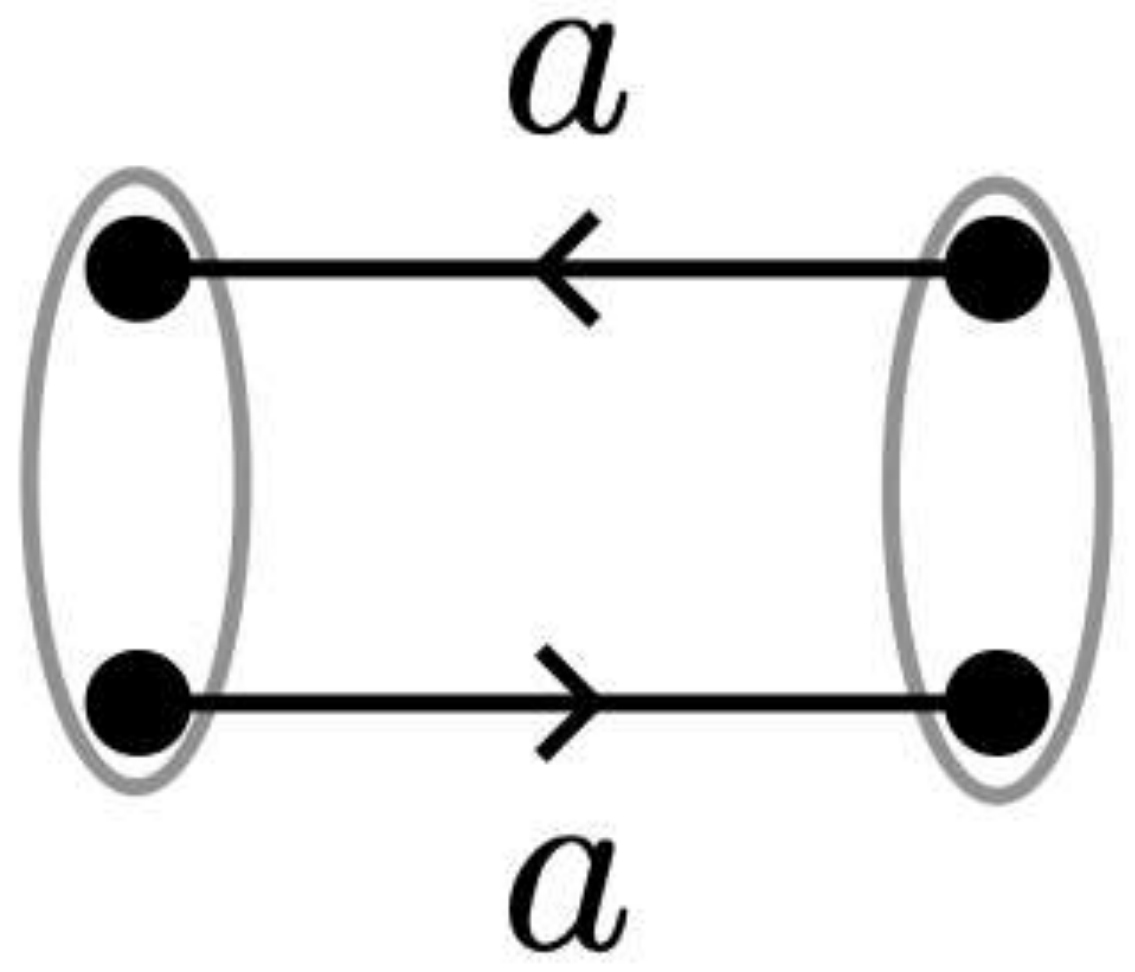}.
\label{eq: qdim}
\end{equation}
More concretely, the quantum dimension $d_a$ can be expressed as
\begin{equation*}
d_a = \sum_{X, X^{\prime}} \sum_{\mu_i, \mu_i^{\prime}} \sum_{\mu_j, \mu_j^{\prime}} (P_a^{ji})^X_{\mu_j, \mu_i} (\eta_a^R)^{\mu_i, \mu_i^{\prime}}_{X, X^{\prime}} (\epsilon_a^L)^{\mu_j, \mu_j^{\prime}}_{X, X^{\prime}} (P_a^{ij})^{X^{\prime}}_{\mu_i^{\prime}, \mu_j^{\prime}}.
\end{equation*}

\paragraph{Topological spin.}
The topological spin is the topological invariant associated with an anyon line forming a figure of eight.
This can be computed as a contracted product of four patch operators labeled by the same anyon:
\begin{equation}
\theta_a = \frac{1}{d_a} ~ \adjincludegraphics[valign = c, width = 2.5cm]{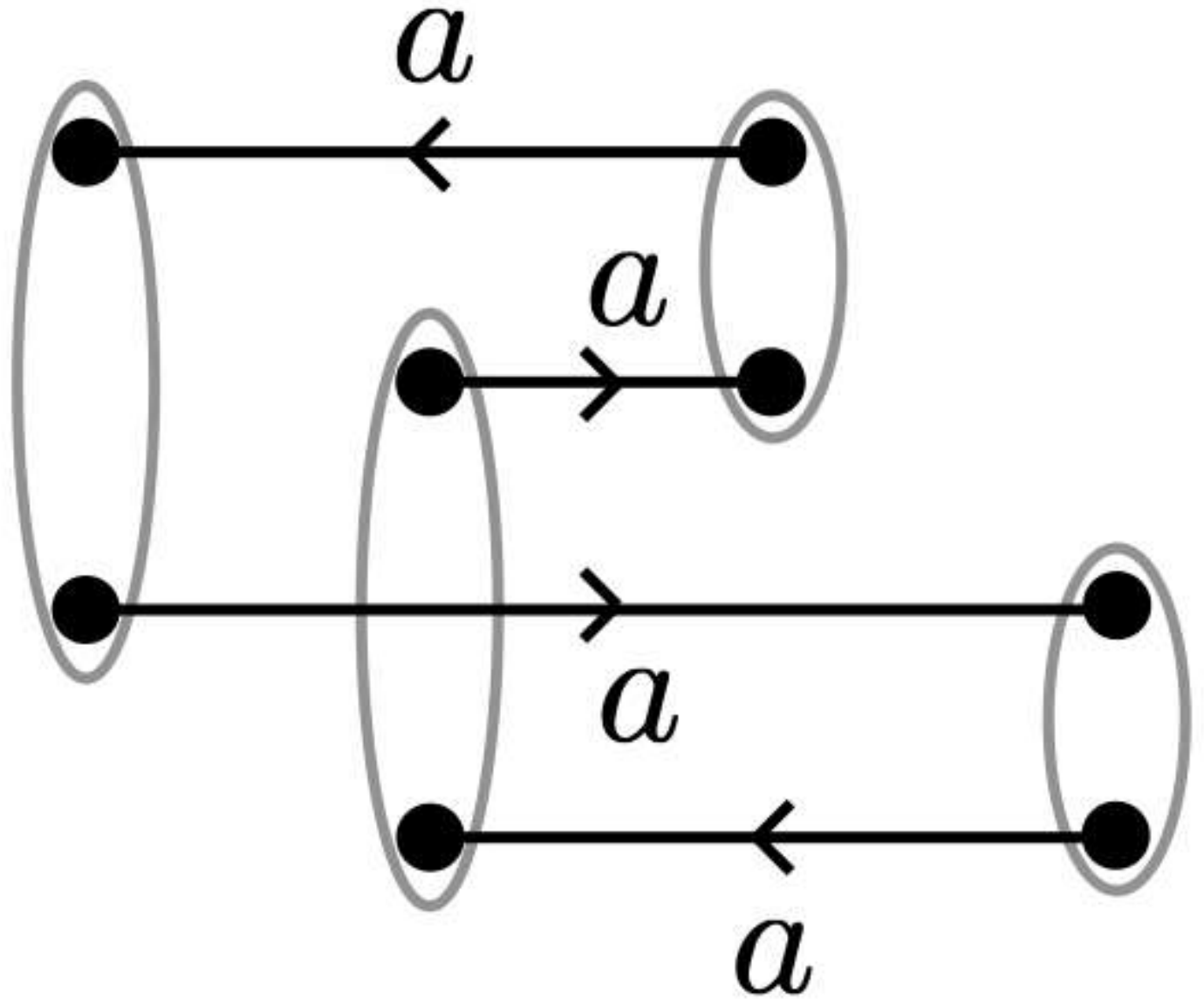}
\label{eq: topological spin}
\end{equation}

\paragraph{Modular $S$-matrix.}
The modular $S$-matrix is the topological invariant associated with the Hopf link.
As shown in Fig.~\ref{fig: scheme}, it can be computed as a contracted product of four patch operators:
\begin{equation}
S_{ab} = \frac{1}{\sqrt{\mathcal{D}}} ~ \adjincludegraphics[valign = c, width = 2.5cm]{S_patch.pdf}.
\label{eq: modular S}
\end{equation}

\begin{figure}
\subfloat[\label{fig: trefoil}]{
\includegraphics[width = 2.1cm]{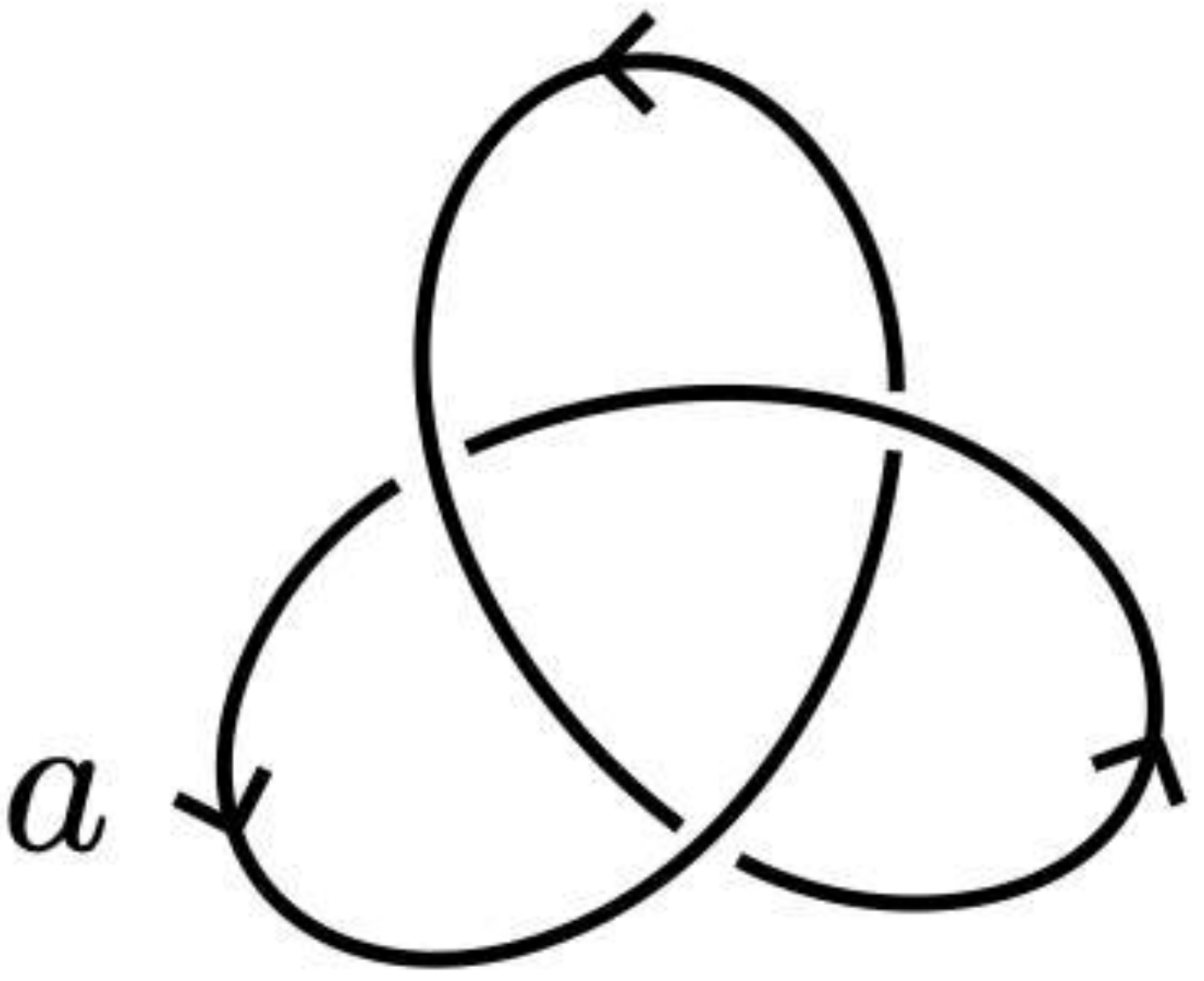}
}\qquad
\subfloat[\label{fig: Borromean}]{
\includegraphics[width = 2.1cm]{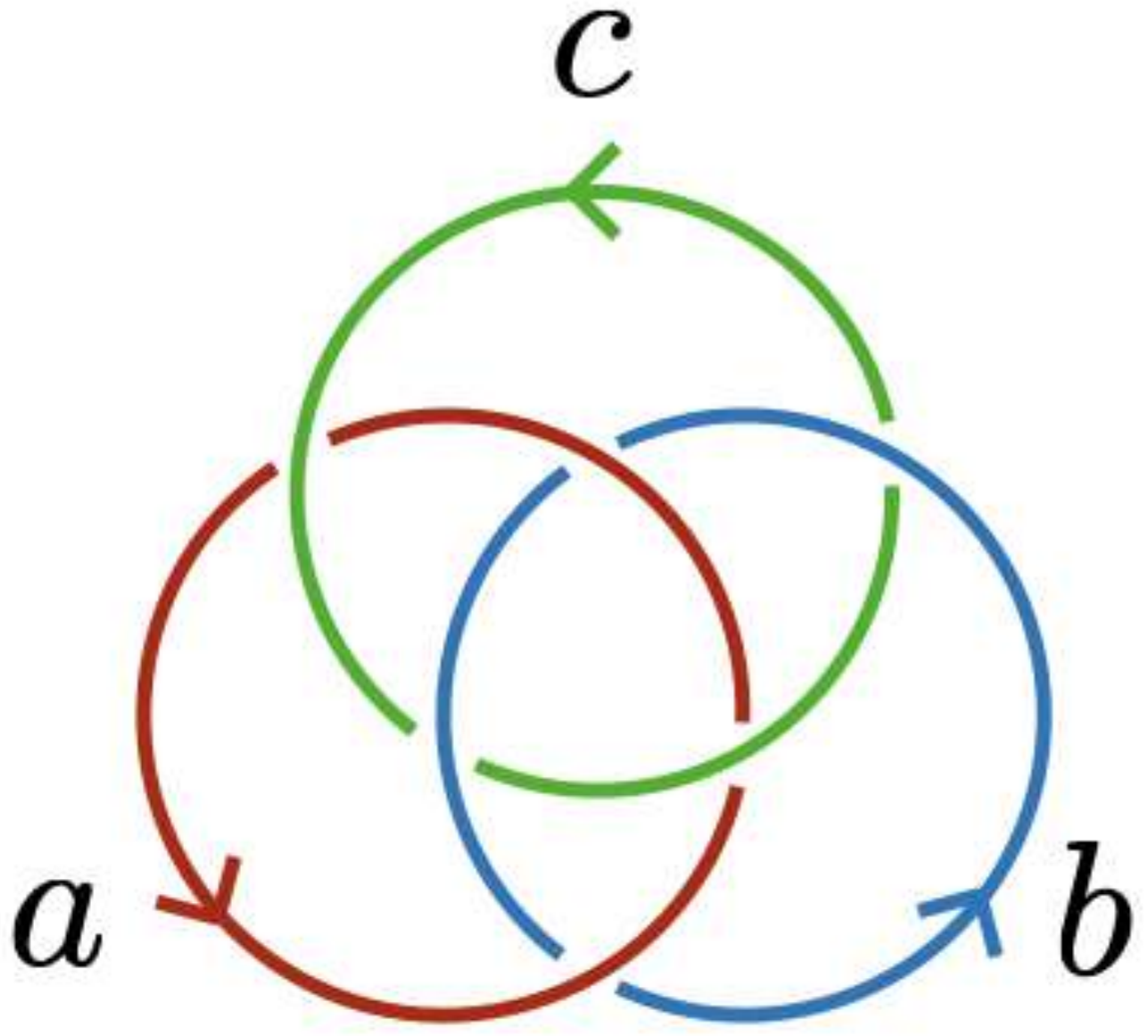}
}\qquad
\subfloat[\label{fig: Whitehead}]{
\includegraphics[width = 2.1cm]{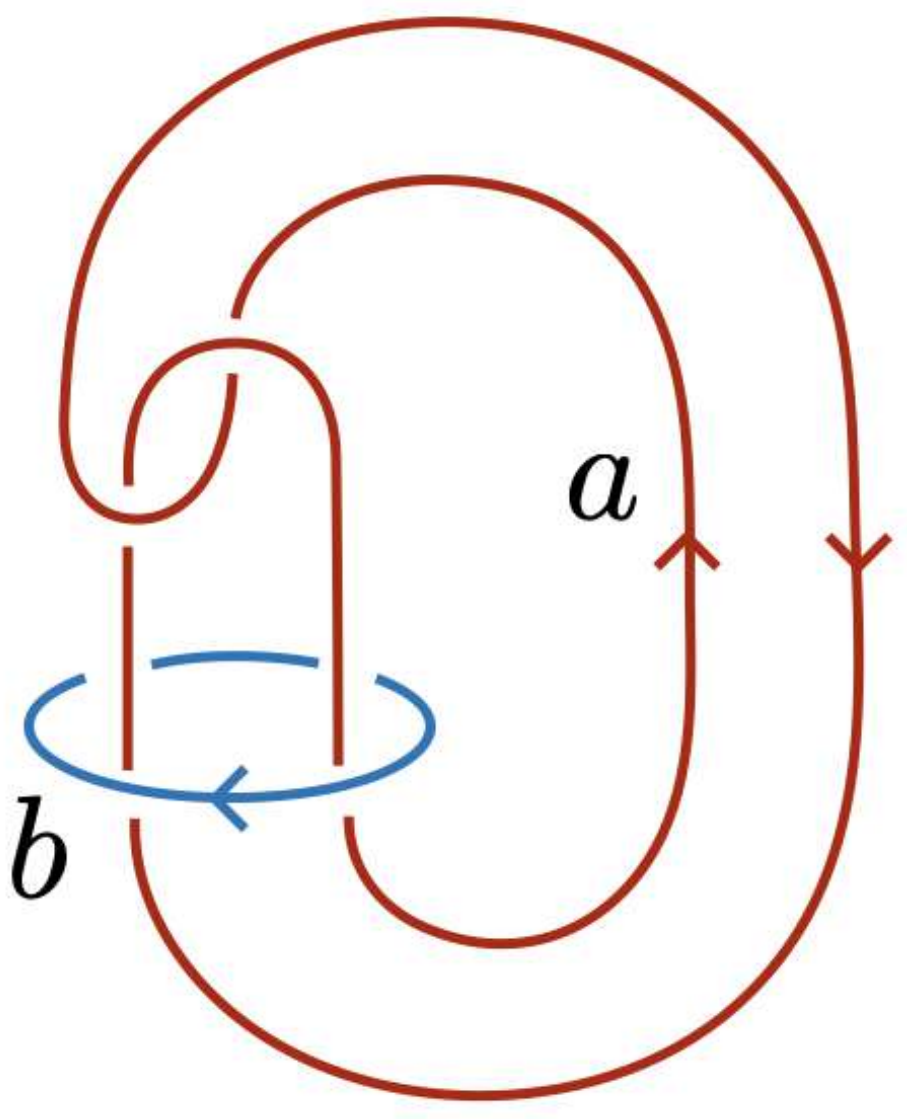}
}
\caption{(a) The trefoil knot. (b) The Borromean rings. (c) The Whitehead link.}
\end{figure}

\paragraph{Trefoil knot.}
The trefoil knot is a non-trivial knot shown in Fig.~\ref{fig: trefoil}.
The associated topological invariant, which we denote by $T_a$, can be computed as a contracted product of six patch operators as follows:\footnote{The authors thank Arkya Chatterjee and Nathanan Tantivasadakarn for discussions on the derivation of equations \eqref{eq: trefoil} and \eqref{eq: Borromean}.}
\begin{equation}
T_a = \adjincludegraphics[valign = c, width = 3.5cm]{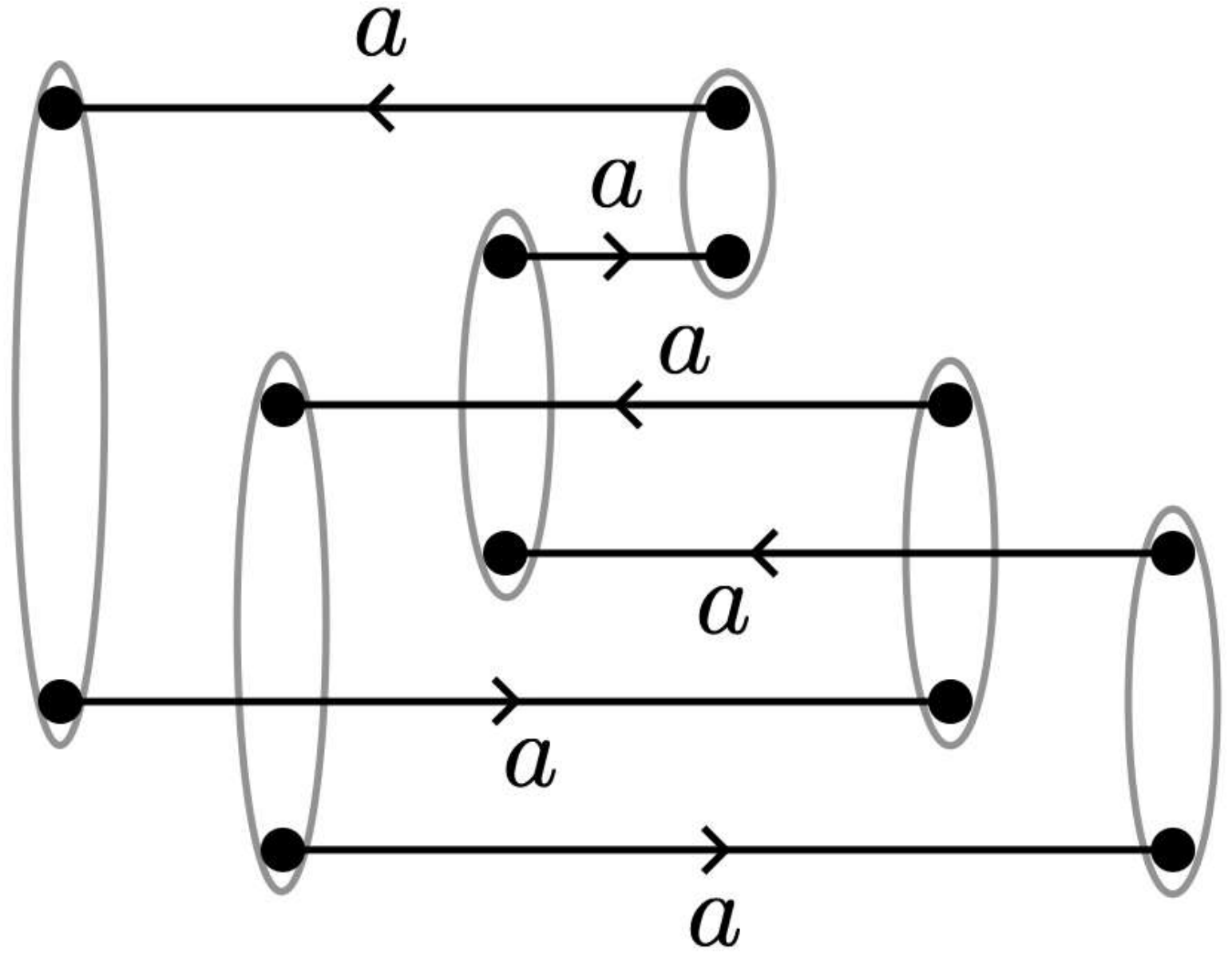}.
\label{eq: trefoil}
\end{equation}

\paragraph{Borromean rings.}
The Borromean rings shown in Fig.~\ref{fig: Borromean} consist of three linking loops, any two of which are not linked together.
The associated topological invariant $B_{abc}$ can be computed as a contracted product of 10 patch operators as follows:
\begin{equation}
B_{abc} = \adjincludegraphics[valign = c, width = 5.5cm]{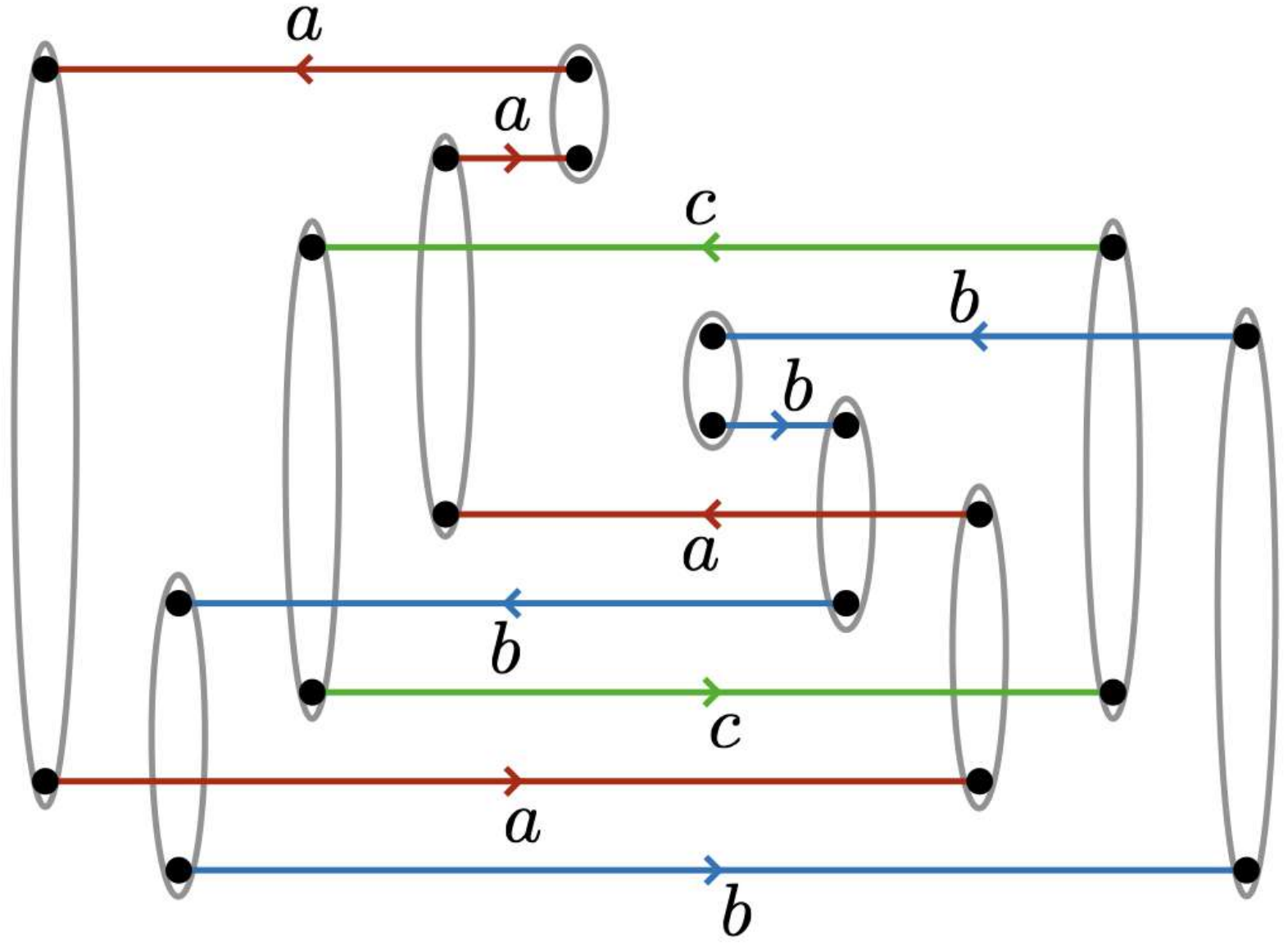}.
\label{eq: Borromean}
\end{equation}
We note that $B_{abc}$ is an example of a topological invariant beyond modular data \cite{Delaney:2018xkw}.

\paragraph{Whitehead link}
The Whitehead link is a framed link consisting of two components as shown in Fig.~\ref{fig: Whitehead}.
The associated topological invariant $\widetilde{W}_{ab}$ can be computed as a contracted product of 10 patch operators as follows:
\begin{equation}
\widetilde{W}_{ab} = \adjincludegraphics[valign = c, width = 5.5cm]{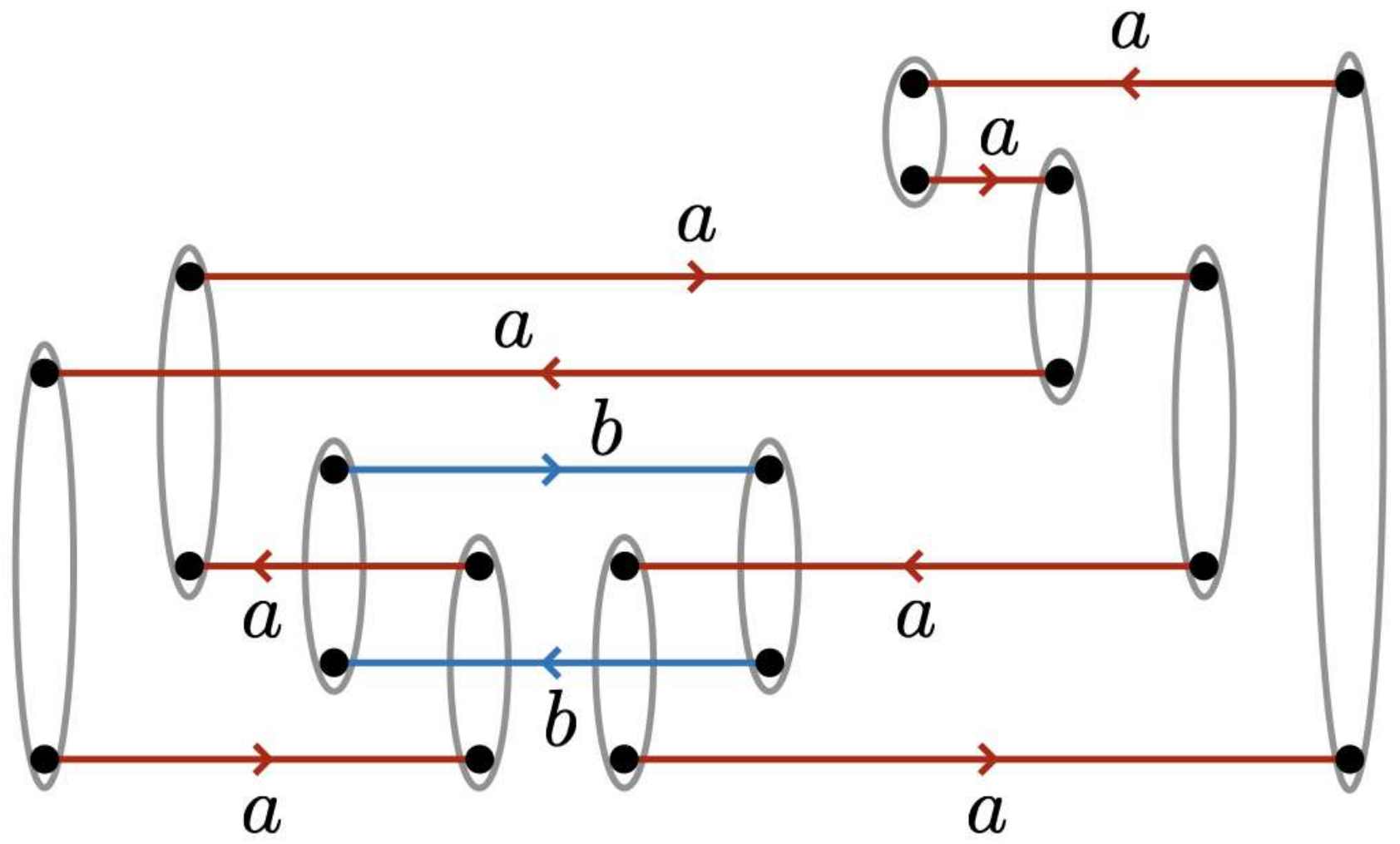}.
\label{eq: W}
\end{equation}
The matrix whose $(a, b)$-component is given by $W_{a, b} = \frac{\theta_a}{\theta_b} \widetilde{W}_{ab}$ is called the $W$-matrix and is shown to be beyond modular data \cite{Bonderson:2018ryx}

The product of patch operators with boundary indices being contracted as in the above equations gives rise to a complex number on the entire state space of the 1+1D system.
This is because an operator defined by the contracted product of patch operators can be shifted freely to the left and right by using the zigzag identities like Eq.~\eqref{eq: zigzag contraction}, which implies that this operator is proportional to the identity operator.
This is in contrast to the fact that a similar computation based on ribbon operators in 2+1D results in the correct topological invariants only on the ground state subspace of the topological order.

\section{Reconstruction of Kitaev's quantum double topological order}
\label{sec: Reconstruction of Kitaev's quantum double topological order}
In this section, based on the general scheme presented in Sec.~\ref{sec: General Scheme for computing topological invariants}, we compute the anyon data of Kitaev's quantum double model $\mathrm{QD}(G)$ from symmetric transparent connectable patch operators in 1+1D systems with non-anomalous finite group symmetry $G$.

\subsection{Patch operators for $\mathbb{Z}_2$ symmetry and the toric code model}
As the simplest example, we begin with the case where $G = \mathbb{Z}_2$.
In this case, Kitaev's quantum double model $\mathrm{QD}(G)$ reduces to the toric code model \cite{Kitaev:1997wr}.
The computation of the anyon data of the toric code was discussed in detail in \cite{Ji:2019jhk, Chatterjee:2022kxb} from the point of view of patch operators in 1+1 dimensions.
Here, we review this computation with an eye toward a generalization to the case of a general finite group $G$.

We first write down patch operators of 1+1D systems with $\mathbb{Z}_2$ symmetry.
To this end, we need to specify the state space on which the patch operators act.
For simplicity, we suppose that the total Hilbert space $\mathcal{H}$ is given by the tensor product of two-dimensional on-site Hilbert spaces, i.e., we have $\mathcal{H} = \bigotimes_i \mathcal{H}_i$ where $i$ denotes a site of a one-dimensional lattice and $\mathcal{H}_i \cong \mathbb{C}^2$ is the state space on site $i$.
In other words, we consider a one-dimensional chain of qubits.
The symmetry operator $U$ for the $\mathbb{Z}_2$ symmetry is defined by the product of the Pauli $X$ operators: $U = \bigotimes_i X_i$.
In this situation, we can find the following four patch operators that satisfy the symmetricity, transparency, and connectability conditions \cite{Ji:2019jhk, Chatterjee:2022kxb}:
\begin{equation}
\begin{aligned}
P_1^{ij} &= \mathrm{id}, \quad P_e^{ij} = Z_i Z_j, \\
P_m^{ij} &= \prod_{i \leq k \leq j} X_k, \quad P_f^{ij} = P_e^{ij} P_m^{ij},
\end{aligned}
\label{eq: patch Z2}
\end{equation}
where $Z_i$ is the Pauli $Z$ operator acting on site $i$.
The existence of the above four patch operators is consistent with the fact that the toric code model has four anyon types $1, e, m$, and $f$.\footnote{In the notation in Sec.~\ref{sec: Example: Kitaev's quantum double topological order}, we can write $1 = (+, \mathsf{triv}), e = (+, \mathsf{sign}), m = (-, \mathsf{triv}), f = (-, \mathsf{sign})$, where $+$ and $-$ are the unit element and the generator of $\mathbb{Z}_2$, $\mathsf{triv}$ is the trivial representation of $\mathbb{Z}_2$, and $\mathsf{sign}$ is the sign representation of $\mathbb{Z}_2$.}
We note that the above patch operators are written in the form of Eq.~\eqref{eq: decomposition} where the summand on the right-hand side is unique and does not have extra indices at the endpoints of a patch.
The twisted evaluation and coevaluation tensors for these patch operators are trivial, i.e., we have
\begin{equation}
\epsilon_{a; b}^L = \epsilon_{a; b}^R = \eta_{a;b}^L = \eta_{a; b}^R = 1,
\label{eq: contraction Z2}
\end{equation}
for all anyon types $a, b = 1, e, m, f$.

Following the prescription given in Sec.~\ref{sec: General Scheme for computing topological invariants}, we can compute the anyon data of the toric code model from the above patch operators.
For example, the quantum dimension of an anyon $a$ is computed as the contracted product of two patch operators $P_a^{ij}$ and $P_a^{ji}$ as in Eq.~\eqref{eq: qdim}. 
Since the evaluation and coevaluation tensors \eqref{eq: contraction Z2} are trivial, the contracted product reduces to the ordinary product of $P_a^{ij}$ and $P_a^{ji}$, which is equal to one because $P_a^{ij}$ in Eq.~\eqref{eq: patch Z2} is unitary for all $a$:
\begin{equation}
d_a = P_a^{ji} P_a^{ij} = 1.
\end{equation}
This result agrees with the fact that the anyons of the toric code model are all abelian.
Similarly, the topological spins \eqref{eq: topological spin} and modular $S$-matrix \eqref{eq: modular S} can be computed as
\begin{equation*}
\theta_a = \frac{1}{d_a} P_a^{ki} P_a^{jk} P_a^{il} P_a^{lj}=
\begin{cases}
1 &\text{for } a = 1, e, m, \\
-1 &\text{for } a = f,
\end{cases}
\end{equation*}
\begin{equation*}
S_{ab} = \frac{1}{2}P_a^{ki} P_b^{lj} P_a^{ik} P_b^{jl} = 
\begin{cases}
\frac{1}{2} &\text{for } a = 1, b = 1, \text{or } a = b,\\
-\frac{1}{2} &\text{otherwise},
\end{cases}
\end{equation*}
where $i < j < k < l$.
These agree with the correct values of the topological spins and modular $S$-matrix of the toric code anyons \cite{Kitaev:1997wr}.

\subsection{Patch operators for $G$ symmetry and the quantum double model}
We generalize the above computation to the case of a general finite group $G$.
To this end, we first write down patch operators in 1+1D systems with finite group symmetry $G$.
For simplicity, we suppose that the state space $\mathcal{H}$ of the system is given by the tensor product $\mathcal{H} = \bigotimes_i \mathcal{H}_i$, where the on-site Hilbert space $\mathcal{H}_i \cong \mathbb{C}^{|G|}$ is the regular representation of $G$.
Namely, the basis states on every site are labeled by group elements of $G$, and the symmetry $G$ acts on them by the left multiplication.
Specifically, the symmetry operator for $g \in G$ is given by the tensor product of on-site operators $U_g^i$ defined by $U_g^i \ket{h}_i = \ket{gh}_i$ for all $\ket{h}_i \in \mathcal{H}_i$.
In this situation, we can find symmetric transparent connectable patch operators labeled by pairs $([g], \alpha)$, where $[g]$ is the conjugacy class of $g$ and $\alpha$ is a unitary irreducible representation of the centralizer $C(g)$.
The patch operator labeled by $([g], \alpha)$ can be explicitly written as
\begin{equation}
(P_{[g], \alpha}^{ij})_{\mu_i, \mu_j} = \sum_{h \in [g]} U_h^{ij} (Z_{\alpha; h}^{ij})_{\mu_i, \mu_j}.
\label{eq: patch G}
\end{equation}
Here, $U_h^{ij}$ is the symmetry operator acting only on interval $[ij]$, i.e., $U_h^{ij} = \bigotimes_{i \leq k \leq j} U_h^k$.
On the other hand, the operator $(Z_{\alpha; h}^{ij})_{\mu_i, \mu_j}$ is a charge operator that acts only on the endpoints of interval $[ij]$.
More specifically, the action of $(Z_{\alpha; h}^{ij})_{\mu_i, \mu_j}$ on a basis state $\ket{g_i}_i \otimes \ket{g_j}_j \in \mathcal{H}_i \otimes \mathcal{H}_j$ is given by
\begin{equation*}
\begin{aligned}
&\quad (Z_{\alpha; h}^{ij})_{\mu_i, \mu_j} \ket{g_i}_i \otimes \ket{g_j}_j\\
&= \overline{\alpha} \left((x^g_{g_i^{-1}hg_i})^{-1} g_i^{-1}g_j x^g_{g_j^{-1}hg_j}\right)_{\mu_i, \mu_j} \ket{g_i}_i \otimes \ket{g_j}_j,
\end{aligned}
\end{equation*}
where $x^g_l$ for $l \in [g]$ is an arbitrary element of $G$ that satisfies $l = x^g_l g (x^g_l)^{-1}$.
The right-hand side of the above equation makes sense because the product $(x^g_{g_i^{-1}hg_i})^{-1} g_i^{-1}g_j x^g_{g_j^{-1}hg_j}$ is always in the centralizer of $g \in G$.
We note that $U_h^{ij}$ and $(Z_{\alpha; h}^{ij})_{\mu_i, \mu_j}$ commute with each other.
The twisted evaluation and coevaluation tensors for the patch operators \eqref{eq: patch G} are given by 
\begin{equation}
\begin{aligned}
&\quad (\epsilon_{([g], \alpha); ([h], \beta)}^{L})_{s, s^{\prime}; t}^{\mu, \mu^{\prime}} = (\eta_{([g], \alpha); ([h], \beta)}^{L})_{s, s^{\prime}; t}^{\mu, \mu^{\prime}}\\
&= (\epsilon_{([g], \alpha); ([h], \beta)}^{R})_{s, s^{\prime}; t}^{\mu, \mu^{\prime}} = (\eta_{([g], \alpha); ([h], \beta)}^{R})_{s, s^{\prime}; t}^{\mu, \mu^{\prime}}\\
&= \delta_{st, ts^{\prime}} \delta_{\mu, \mu^{\prime}},
\end{aligned}
\end{equation}
where $s, s^{\prime} \in [g]$, $t \in [h]$, and $\delta$ denotes the Kronecker delta.
As we will see in Appendix \ref{sec: Ribbon operators on the boundary of Kitaev's quantum double model}, the patch operators \eqref{eq: patch G} can also be obtained as ribbon operators on the rough boundary of Kitaev's quantum double model.

When $G$ is abelian, the patch operator \eqref{eq: patch G} can be written as $P_{g, \alpha}^{ij} = U_g^{ij} Z^i_{\alpha} Z^j_{\overline{\alpha}}$, where $Z^i_{\alpha}$ is the on-site charge operator defined by $Z^i_{\alpha} \ket{h}_i = \alpha(h) \ket{h}_i$ for all $\ket{h}_i \in \mathcal{H}_i$.
In particular, when $G = \mathbb{Z}_2$, equation \eqref{eq: patch G} reduces to the patch operators \eqref{eq: patch Z2} that we wrote down in the previous subsection.

The above expression of the patch operators allows us to explicitly compute various topological invariants following the general scheme presented in Sec.~\ref{sec: General Scheme for computing topological invariants}.
First of all, the quantum dimension of an anyon labeled by a pair $([g], \alpha)$ can be computed as
\begin{equation}
d_{[g], \alpha} = \sum_{h \in [g]} (P_{[g], \alpha}^{ji})_{\mu_j, \mu_i}^h (P_{[g], \alpha}^{ij})_{\mu_i, \mu_j}^h = |[g]| \mathop{\mathrm{dim}}\alpha,
\label{eq: qdim patch QD}
\end{equation}
where $(P_{[g], \alpha}^{ij})_{\mu_i, \mu_j}^h := U_h^{ij} (Z_{\alpha; h}^{ij})_{\mu_i, \mu_j}$ denotes the summand on the right-hand side of Eq.~\eqref{eq: patch G}, and $(P_{[g], \alpha}^{ji})^h_{\mu_j, \mu_i}$ is its complex conjugate.
We note that Eq.~\eqref{eq: qdim patch QD} agrees with the correct quantum dimension \eqref{eq: qdim QD} of an anyon labeled by $([g], \alpha)$.
Similarly, we can compute the topological spins and modular $S$-matrix by taking the contracted products of four patch operators as in Eqs.~\eqref{eq: topological spin} and \eqref{eq: modular S}:
\begin{widetext}
\begin{equation}
\theta_{[g], \alpha} = \frac{1}{d_{[g], \alpha}} \sum_{h \in [g]} (P_{[g], \alpha}^{ki})^h_{\mu_k, \mu_i} (P_{[g], \alpha}^{jk})^h_{\mu_j, \mu_k} (P_{[g], \alpha}^{il})^h_{\mu_i, \mu_l} (P_{[g], \alpha}^{lj})^h_{\mu_l, \mu_j} = \frac{\mathop{\mathrm{tr}}\alpha(g)}{\mathop{\mathrm{dim}}\alpha},
\label{eq: T patch QD}
\end{equation}
\begin{equation}
\begin{aligned}
S_{([g], \alpha), ([h], \beta)} &= \frac{1}{|G|} \sum_{a \in [g]} \sum_{b \in [h]} (P_{[g], \alpha}^{ki})_{\mu_k, \mu_i}^a (P_{[h], \beta}^{lj})_{\mu_l, \mu_j}^b (P_{[g], \alpha}^{ik})_{\mu_i, \mu_k}^a (P_{[h], \beta}^{jl})_{\mu_j, \mu_l}^b \delta_{ab, ba} \\
&= \frac{1}{|G|} \sum_{\substack{a \in [g], ~ b \in [h] \\ \text{s.t. } ab = ba}} \mathop{\mathrm{tr}} \overline{\alpha} \left( (x_a^g)^{-1} b x_a^g \right) \mathop{\mathrm{tr}} \overline{\beta} \left( (x_b^h)^{-1} a x_b^h \right),
\end{aligned}
\label{eq: S patch QD}
\end{equation}
\end{widetext}
where $i < j < k < l$.
These agree with the correct topological spins \eqref{eq: topological spin QD} and modular $S$-matrix \eqref{eq: modular S QD} of Kitaev's quantum double topological order that we reviewed in Sec.~\ref{sec: Example: Kitaev's quantum double topological order}.
Equations \eqref{eq: T patch QD} and \eqref{eq: S patch QD} can be derived by directly computing the action of the contracted product of the patch operators on an arbitrary basis state $\ket{\cdots, g_i, \cdots, g_l, \cdots}$ of the state space $\mathcal{H}$.
In the derivation of Eq.~\eqref{eq: S patch QD}, we used the fact that $x^g_{g_i^{-1}ag_i}$ and $x^g_a$ are related by the equality
\begin{equation}
x^g_{g_i^{-1}ag_i} = g_i^{-1} x_a y, \quad y \in C(g).
\label{eq: representative}
\end{equation}
We can also compute other topological invariants such as those associated with the trefoil knot $T_{[g], \alpha}$, the Borromean rings $B_{([g], \alpha), ([h], \beta), ([k], \gamma)}$, and the Whitehead link $\widetilde{W}_{([g], \alpha), ([h], \beta)}$.
Direct computation based on the formulae \eqref{eq: trefoil}, \eqref{eq: Borromean}, and \eqref{eq: W} shows
\begin{widetext}
\begin{equation}
T_{[g], \alpha} = |[g]| \mathop{\mathrm{tr}}\overline{\alpha}(g^3) = d_{[g], \alpha} \theta_{[g], \alpha}^{-3},
\label{eq: trefoil QD}
\end{equation}
\begin{equation}
B_{([g], \alpha), ([h], \beta), ([k], \gamma)} = \sum_{\substack{a \in [g], ~ b \in [h], ~ c \in [k] \\ \text{s.t. eq.}\eqref{eq: triple commutativity} \text{ holds.}}} \mathop{\mathrm{tr}} \alpha\left( x_a^{-1} [b, c] x_a \right) \mathop{\mathrm{tr}} \beta \left( x_b^{-1} [c, a] x_b \right) \mathop{\mathrm{tr}} \gamma \left( x_c^{-1} [a, b] x_c \right),
\label{eq: Borromean QD}
\end{equation}
\begin{equation}
\widetilde{W}_{([g], \alpha), ([h], \beta)} = \sum_{\substack{a \in [g], ~ b \in [h] \\ \text{s.t. eq.} \eqref{eq: commutation product} \text{ holds.}}} \mathop{\mathrm{tr}} \overline{\alpha}\left( x_a^{-1} [b^{-1}, a] [b, a] a^2 x_a \right) \mathop{\mathrm{tr}} \overline{\beta}\left( x_b^{-1} [b, a] [b, a^{-1}] x_b \right),
\label{eq: W QD}
\end{equation}
\end{widetext}
where $[a, b] := b^{-1}aba^{-1}$ is the commutation relation between $a$ and $b$, which is equal to the unit element $1 \in G$ if and only if $a$ and $b$ commute with each other.
The summation on the right-hand side of Eq.~\eqref{eq: Borromean QD} is taken over group elements $a \in [g]$, $b \in [h]$, and $c \in [k]$ that satisfy the following commutation relations:
\begin{equation}
[[a, b], c] = [[b, c], a] = [[c, a], b] = 1.
\label{eq: triple commutativity}
\end{equation}
Similarly, the summation on the right-hand side of Eq.~\eqref{eq: W QD} is taken over group elements $a \in [g]$ and $b \in [h]$ that satisfy
\begin{equation}
[a^{-1}, b] [a, b] = [b, a] [b, a^{-1}].
\label{eq: commutation product}
\end{equation}
The second equality of Eq.~\eqref{eq: trefoil QD} follows from the relation $\alpha(g) = \theta_{[g], \alpha} \mathrm{id}$, which is an immediate consequence of Eq.~\eqref{eq: T patch QD} and Schur's lemma.
In the derivation of Eqs.~\eqref{eq: Borromean QD} and \eqref{eq: W QD}, we again used Eq.~\eqref{eq: representative}.
We note that $B_{([g], \alpha), ([h], \beta), ([k], \gamma)}$ is unity when $G$ is abelian.
The topological invariant $B_{([g], \alpha), ([h], \beta), ([k], \gamma)}$ was also computed by a different method in \cite{kulkarni2021topological}.

\section{Discussion}
\label{sec: Discussion}
In this paper, we proposed a general method to reconstruct the data of topological orders in 2+1 dimensions from symmetric transparent connectable patch operators in 1+1D systems with finite symmetries.
Our proposal is based on the observation that anyons in 2+1D topological orders are related to symmetric transparent connectable patch operators in 1+1 dimensions via the sandwich construction as illustrated in Fig.~\ref{fig: SymTFT patch}.
We demonstrated the validity of our proposal by explicitly computing the anyon data of Kitaev's quantum double topological order from patch operators of 1+1D systems with general finite group symmetry. 
This result supports the conjecture in \cite{Chatterjee:2022kxb} that the algebra of symmetric transparent patch operators determines a topological order in one higher dimension.

As an application of patch operators, we expect that the symmetric transparent connectable patch operators that we discussed in this paper would serve as order and disorder operators for gapped phases with finite symmetries.
Specifically, the expectation value of a patch operator $P_a^{ij}$ in the symmetric ground state would remain non-zero in the limit $|j - i| \gg 1$ if the anyon $a$ is condensed on the right boundary of the 2+1D bulk in Fig.~\ref{fig: SymTFT patch}, whereas it would go to zero in the limit $|j - i| \gg 1$ if the anyon $a$ is not condensed on the right boundary.
Thus, the expectation values of these patch operators would tell us which anyons are condensed and which are not.
This enables us to distinguish different gapped phases corresponding to different sets of condensed anyons on the right boundary.
Order and disorder operators are studied from this point of view in, e.g., \cite{Ji:2019jhk, Kong:2020cie, Chatterjee:2022tyg, Freed:2018cec, Moradi:2022lqp, Albert:2021vts}, see also \cite{bhardwaj231003784, bhardwaj231003786} for a general framework.
It would be interesting to see whether symmetric transparent connectable patch operators \eqref{eq: patch G} really serve as order and disorder operators in concrete lattice models with finite group symmetry $G$.
When $G$ is abelian, it was shown in \cite{Else2013} that all gapped phases are distinguished by the set of expectation values of the extended operators of the form \eqref{eq: patch G}.

A natural generalization of our study is to incorporate anomalies of finite group symmetries in 1+1D.
When finite group symmetries are anomalous, the corresponding 2+1D topological orders are those realized by the twisted quantum double model \cite{Hu:2012wx}.
It would be interesting to figure out symmetric transparent connectable patch operators in this case and reconstruct the twisted quantum double topological orders from them.
More generally, one can consider the case of more general fusion category symmetries \cite{BT2018, CLSWY2019, Thorngren:2019iar}, where it might be convenient to use matrix product operators in \cite{Lootens:2021tet, Lootens:2022avn} to represent patch operators and compute the anyon data of the corresponding topological orders.
Another interesting direction is to generalize our analysis to higher dimensions, where concrete descriptions of topological orders are less understood.

\begin{acknowledgments}

We would like to thank Arkya Chatterjee and Nathanan Tantivasadakarn for
helpful discussions.  We are grateful to the hospitality of the California
Institute of Technology where this work was initiated.  K.I. is supported by
FoPM, WINGS Program, the University of Tokyo, and by JSPS Research Fellowship
for Young Scientists.  X.-G.W. was partially supported by NSF grant DMR-2022428
and by the Simons Collaboration on Ultra-Quantum Matter, which is a grant from
the Simons Foundation (651446, XGW). 

\end{acknowledgments}

\appendix

\section{Ribbon operators on the rough boundary of Kitaev's quantum double model}
\label{sec: Ribbon operators on the boundary of Kitaev's quantum double model}
In this appendix, we briefly review ribbon operators of Kitaev's quantum double model following \cite{Bombin:2007qv, Beigi2011}.
In particular, we will see a relation between the ribbon operators on the rough boundary of Kitaev's quantum double model and the patch operators \eqref{eq: patch G} of 1+1D systems with finite group symmetry $G$.

\begin{figure*}
\subfloat[\label{fig: rough boundary}]{
\includegraphics[width = 6cm]{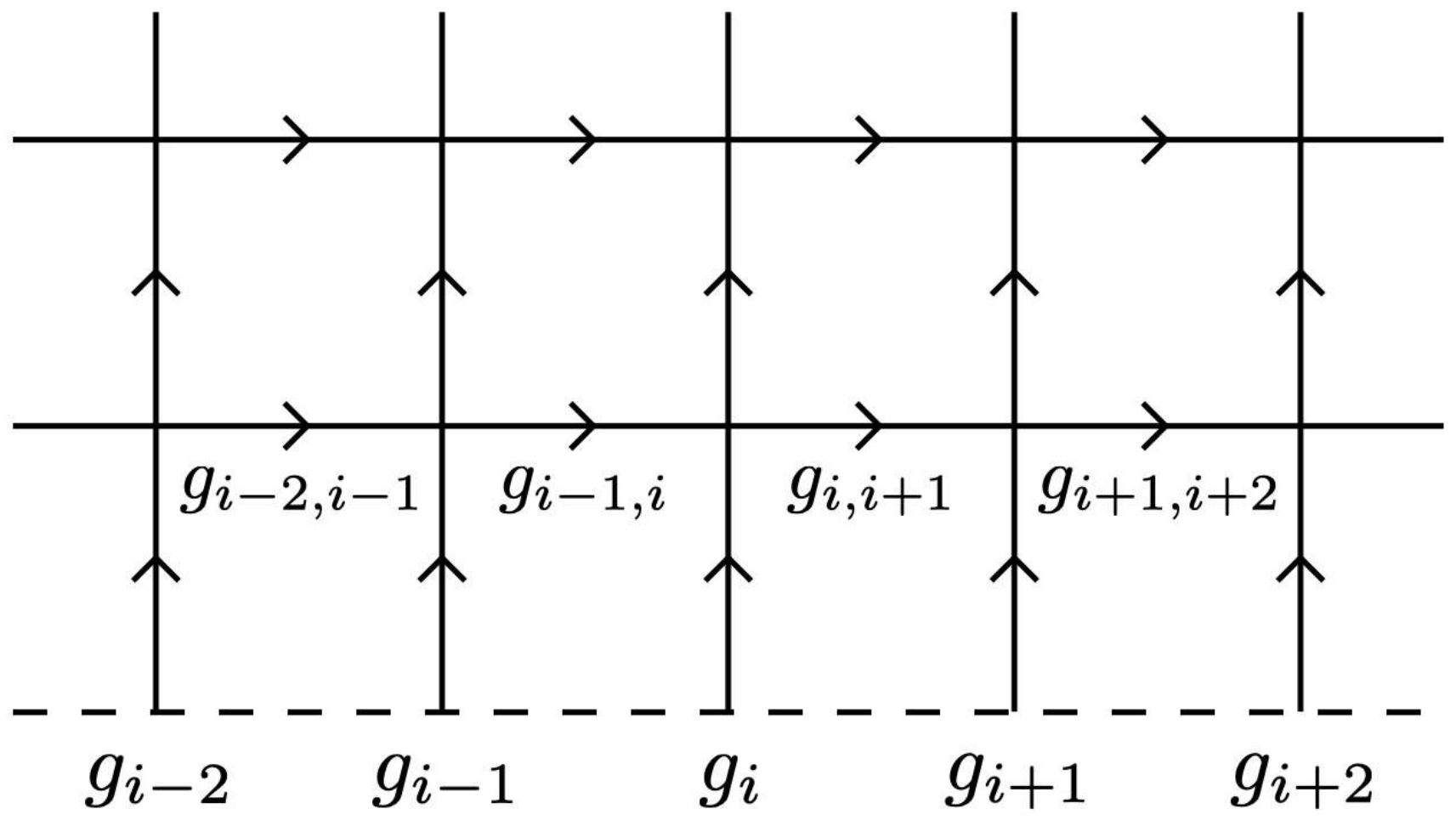}
}\quad \quad \quad
\subfloat[\label{fig: ribbon}]{
\includegraphics[width = 8cm]{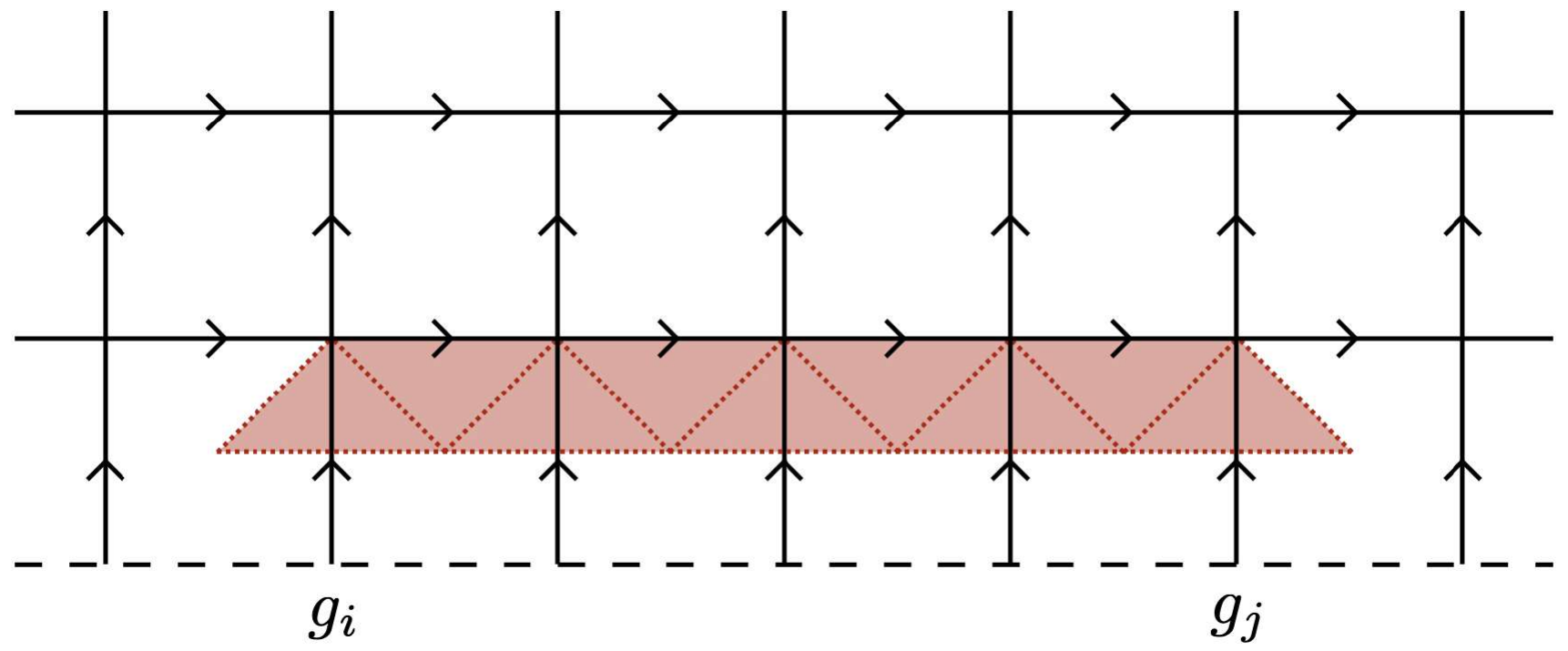}
}
\caption{(a) An oriented square lattice. The dashed line at the bottom represents a rough boundary. (b) A ribbon operator on the rough boundary.}
\end{figure*}

We consider Kitaev's quantum double model on a square lattice with a rough boundary.
Edges of the square lattice are oriented as shown in Fig.~\ref{fig: rough boundary}.
The state space $\mathcal{H}_e \cong \mathbb{C}^{|G|}$ on each edge $e$ is spanned by group elements of $G$.
The Hamiltonian of the model is given by
\begin{equation}
H = - \sum_{v:~\text{vertices}} A_v - \sum_{p:~\text{plaquettes}} B_p,
\end{equation}
where the vertex term $A_v$ and the plaquette term $B_p$ are defined as follows \cite{Kitaev:1997wr}:
\begin{equation}
A_v ~ \adjincludegraphics[valign = c, width = 1.7cm]{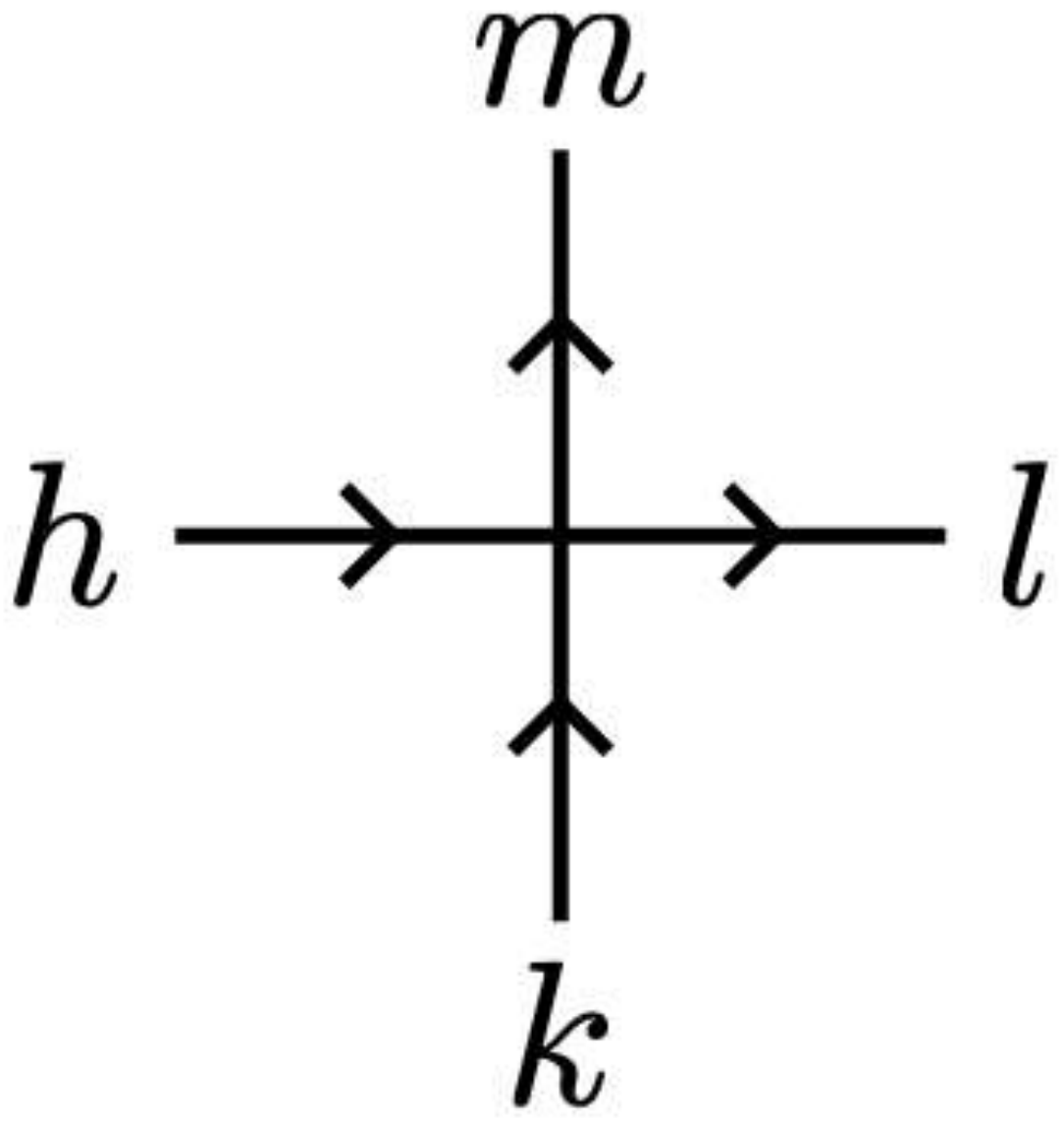} = \frac{1}{|G|} \sum_{g \in G} ~ \adjincludegraphics[valign = c, width = 2.25cm]{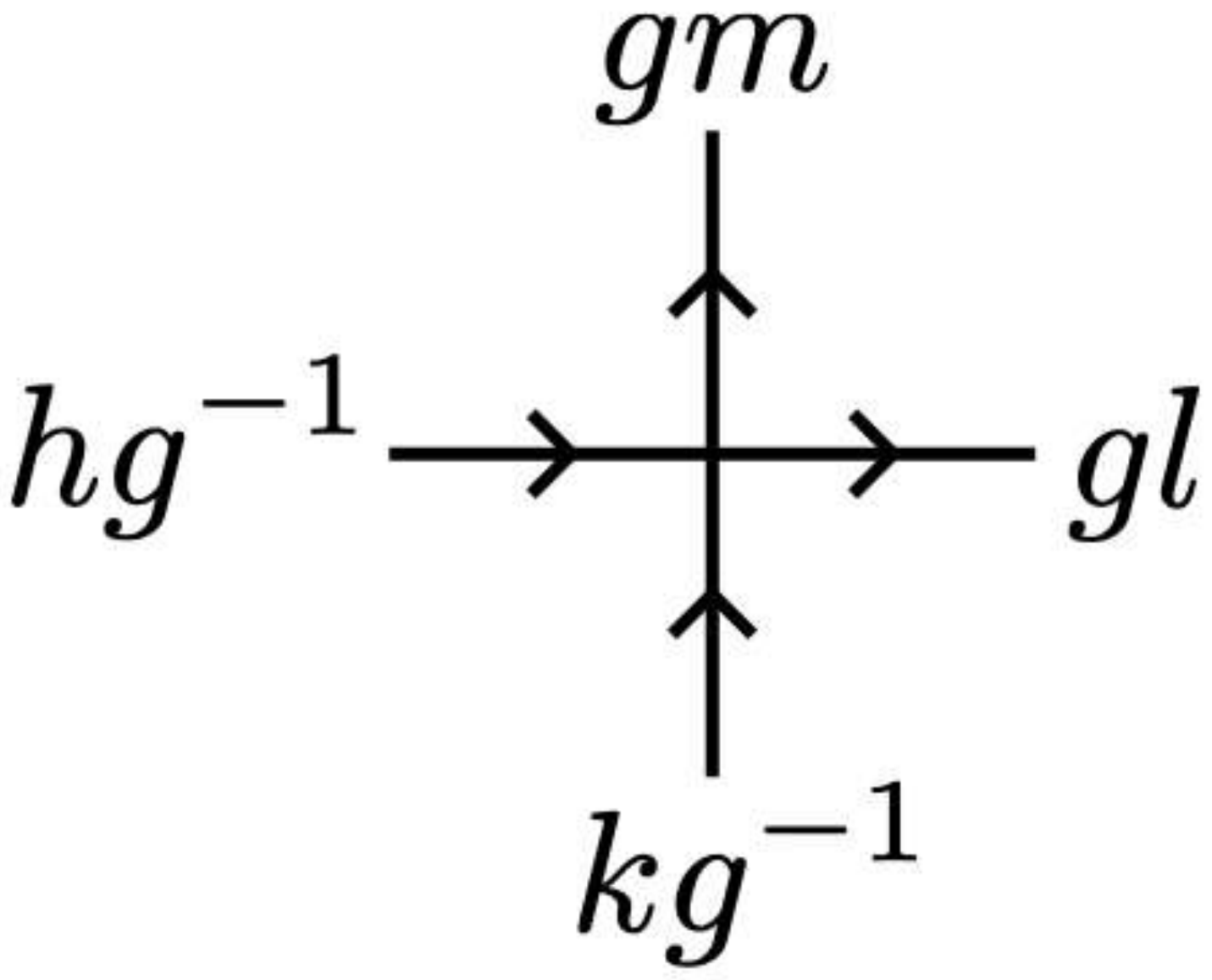},
\end{equation}
\begin{equation}
B_p ~ \adjincludegraphics[valign = c, width = 1.7cm]{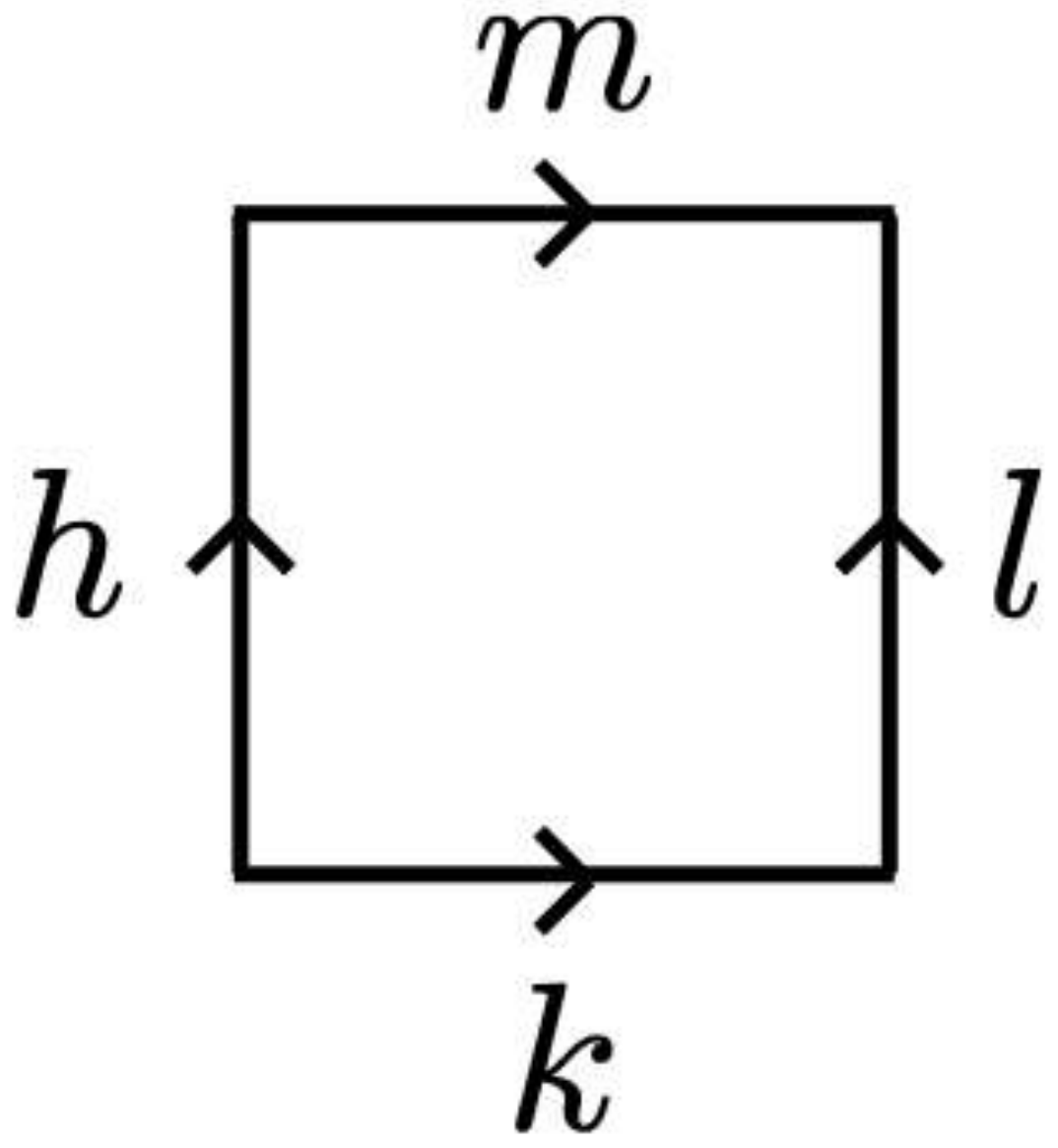} = \delta_{1, klm^{-1}h^{-1}} ~ \adjincludegraphics[valign = c, width = 1.7cm]{plaquette.pdf}.
\end{equation}
We note that the plaquette terms on the rough boundary constrain the configuration of dynamical variables in Fig.~\ref{fig: rough boundary} so that $g_{i, i+1} = g_i^{-1} g_{i+1}$ for all boundary edges.

Taking the above constraint into account, we can compute the action of a ribbon operator on an interval $[ij]$ on the rough boundary as \cite{Bombin:2007qv, Beigi2011}
\begin{equation}
\begin{aligned}
&\quad (F^{ij}_{[g], \alpha})_{(h_i, \mu_i), (h_j, \mu_j)} \ket{g_i, \cdots, g_j}\\
&= \frac{\mathrm{dim}\alpha}{|C(g)|} \sum_{n \in C(g)} \overline{\alpha}(n)_{\mu_i, \mu_j} \delta_{x^g_{h_i}n(x^g_{h_j})^{-1}, g_i^{-1} g_j} \ket{g_i^{\prime}, \cdots, g_j^{\prime}},
\end{aligned}
\label{eq: ribbon}
\end{equation}
where $h_i, h_j \in [g]$, $\ket{g_i, \cdots, g_j}$ is a basis state on the rough boundary (see Fig.~\ref{fig: rough boundary}), and $g_k^{\prime} := (g_i h_i g_i^{-1}) g_k$ for every site $k$ on interval $[ij]$.
As in the main text, $[g]$ and $C(g)$ denote the conjugacy class and centralizer of $g$ respectively, $\alpha$ is a unitary irreducible representation of $C(g)$, and $x^g_h$ is an arbitrary element of $G$ that satisfies $h = x^g_h g (x^g_h)^{-1}$ for $h \in [g]$.
The above ribbon operator is illustrated in Fig.~\ref{fig: ribbon}.
The right-hand side of Eq.~\eqref{eq: ribbon} is non-zero only when $(x^g_{h_i})^{-1} g_i^{-1} g_j x^g_{h_j}$ is in the centralizer of $g$.
This condition is satisfied if and only if $h_j = (g_i^{-1}g_j)^{-1} h_i g_i^{-1} g_j$.
Therefore, Eq.~\eqref{eq: ribbon} reduces to
\begin{equation}
\begin{aligned}
&\quad (F^{ij}_{[g], \alpha})_{(h_i, \mu_i), (h_j, \mu_j)} \ket{g_i, \cdots, g_j}\\
&= \frac{\mathrm{dim}\alpha}{|C(g)|} \delta_{h_j, (g_i^{-1}g_j)^{-1} h_i g_i^{-1} g_j} \overline{\alpha}\left((x^g_{h_i})^{-1} g_i^{-1} g_j x^g_{h_j}\right)_{\mu_i, \mu_j}\\
&\quad ~ \ket{(g_i h_i g_i^{-1}) g_i, \cdots, (g_i h_i g_i^{-1}) g_j}.
\end{aligned}
\end{equation}
Summing up the above ribbon operators for all $h_i, h_j \in [g]$ leads to
\begin{equation}
\begin{aligned}
&\quad \sum_{h_i, h_j \in [g]} (F_{[g], \alpha}^{ij})_{(h_i, \mu_i), (h_j, \mu_j)} \ket{g_i, \cdots, g_j}\\
&= \frac{\mathrm{dim}\alpha}{|C(g)|} \sum_{h \in [g]} \overline{\alpha}\left((x^g_{g_i^{-1}hg_i})^{-1} g_i^{-1} g_j x^g_{g_j^{-1}hg_j}\right)_{\mu_i, \mu_j} \ket{h g_i, \cdots, h g_j}\\
&= \frac{\mathrm{dim}\alpha}{|C(g)|} (P^{ij}_{[g], \alpha})_{\mu_i, \mu_j}\ket{g_i, \cdots, g_j},
\end{aligned}
\end{equation}
where $(P^{ij}_{[g], \alpha})_{\mu_i, \mu_j}$ is the patch operator \eqref{eq: patch G} of 1+1D systems with finite group symmetry $G$.
This equation shows the relation between ribbon operators of 2+1D Kitaev's quantum double model and symmetric transparent connectable patch operators in 1+1 dimensions.

\bibliography{bibliography,all,allnew,publst,publstnew}

\end{document}